\documentclass[12pt, letterpaper]{article}
\usepackage[utf8]{inputenc}
\usepackage[english]{babel}
\usepackage{authblk}
\usepackage[margin=1in]{geometry}
\usepackage{graphicx, caption, subcaption}
\usepackage{dcolumn}
\usepackage{bm}
\usepackage[font=small, labelfont=bf]{caption}
\usepackage[version=3]{mhchem}
\usepackage[tableposition=top]{caption}
\usepackage{hyperref}
\usepackage{multirow}
\usepackage{lineno}

\title{Composition of Terrestrial Exoplanet Atmospheres from Meteorite Outgassing Experiments}

\author[1]{Maggie A. Thompson*}
\affil[1]{Department of Astronomy and Astrophysics, University of California Santa Cruz, Santa Cruz, CA 95064}
\author[2]{Myriam Telus}
\affil[2]{Earth and Planetary Sciences, University of California Santa Cruz, Santa Cruz, CA 95064}
\author[3]{Laura Schaefer}
\affil[3]{Geological Sciences, School of Earth, Energy, and Environmental Sciences, Stanford University, Stanford, CA 94305}
\author[1]{Jonathan J. Fortney}
\author[4]{Toyanath Joshi}
\affil[4]{Department of Physics, University of California Santa Cruz, Santa Cruz, CA 95064}
\author[4]{David Lederman}

\begin{document}
\begin{titlepage}
\maketitle
*Email: maapthom@ucsc.edu
\end{titlepage}
\begin{abstract}

\noindent \textbf{Terrestrial exoplanets likely form initial atmospheres through outgassing during and after accretion, although there is currently no first-principles understanding of how to connect a planet's bulk composition to its early atmospheric properties. Important insights into this connection can be gained by assaying meteorites, representative samples of planetary building blocks. We perform laboratory outgassing experiments that use a mass spectrometer to measure the abundances of volatiles released when meteorite samples are heated to 1200 $^{\circ}$C.  We find that outgassing from three carbonaceous chondrite samples consistently produce H\textsubscript{2}O-rich (averaged $\sim$66 \%) atmospheres but with significant amounts of CO ($\sim$18 \%) and CO\textsubscript{2} ($\sim$15 \%) as well as smaller quantities of H\textsubscript{2} and H\textsubscript{2}S (up to 1\%).  These results provide experimental constraints on the initial chemical composition in theoretical models of terrestrial planet atmospheres, supplying abundances for principal gas species as a function of temperature.}

\end{abstract}

We are at the dawn of an exciting technological era in astronomy with new large-aperture telescopes and advanced instrumentation, both in space and on the ground, leading to major advances in exoplanet characterization. To optimize the use of these new facilities, we need suitable theoretical models to obtain a better understanding of the diversity of exoplanet atmospheres. Statistical studies using NASA's \textit{Kepler} mission data suggest that terrestrial and other low-mass planets are common around G, K and M stars \cite{Petigura2013a, Petigura2013b, Dressing2015}. Given the large number of current and anticipated low-mass exoplanet discoveries, the next phase in exoplanet science is to characterize the physics and chemistry of their atmospheres. For the foreseeable future, Solar System meteorites provide the only direct samples that can be rigorously studied in the laboratory to gain insight into the initial atmospheric compositions of these planets.  

Although gas giant planets, like Jupiter and Saturn, form primary atmospheres by capturing gases from the stellar nebula, atmosphere formation for low-mass planets is more complicated. While nebular ingassing can contribute to early atmosphere formation if a protoplanet accretes enough mass before the gas disk dissipates, several factors can result in loss of nebular volatiles early in the planet's history \cite{Sharp2017, Schlichting2018, Wu2018}. For instance, terrestrial planets' inability to retain significant primary atmospheres can be due to low planetary mass, large impactors and high EUV and X-ray flux from young host stars \cite{Lammer2018}. Instead, low-mass planets likely have secondary atmospheres which form via outgassing of volatiles during and after planetary accretion \cite{Elkins-Tanton2008}. The Solar System's terrestrial planets are believed to have formed by accretion of planetesimals that have compositions similar to chondritic meteorites, which are a likely source of atmospheric volatiles for such planets \cite{Ahrens1989, Lammer2018}.  As a result, an important step towards establishing the connection between terrestrial planets' bulk compositions and their atmospheres is to directly measure the outgassed volatiles from meteorites. 

While meteorites come in a wide variety with a range of volatile contents, they can be classified into three main types: chondrites, achondrites and irons. Chondrites are stony meteorites that come from undifferentiated planetesimals composed of aggregate material from the protoplanetary disk, while achondrites and iron meteorites have melted and derive from partially or fully differentiated planetesimals. Both chondrites and achondrites likely contributed to forming the Sun's terrestrial planets \cite{Lammer2018}. Our study focuses on CM-type carbonaceous chondrites because their compositions provide a strong match to that of the solar photosphere, second only to CI-type chondrites, so they serve as representative samples of the bulk composition of material in the solar nebula during planet formation. While planet formation alters planetesimals through various thermal and differentiation processes, carbonaceous chondrite-like material was a likely source of volatiles for the terrestrial planets, and CM chondrites are among the most volatile-rich of remnant materials, making these samples well-suited for studying initial outgassed atmospheres \cite{LoddersFegley, Wasson1988}. 

Planetary outgassing has been modeled both for the Solar System's terrestrial planets and for some low-mass exoplanets. Many prior studies have focused on outgassing during a planet's magma ocean phase and Earth's early degassing history, with several models investigating the outgassing composition from meteorites (e.g., \cite{Zahnle1988, Lammer2018, Gaillard2014, SchaeferFegly2010, Herbort2020}). Currently, however, there is limited experimental data to constrain these models and, in particular, none to inform meteorite outgassing studies. Prior experiments that heated meteorites were limited in several key ways due to restrictions in the number and type of samples used, the temperatures to which the samples where heated, and the number of gas species that were accurately measured (e.g., \cite{CourtSephton2009, GoodingMuenow1977, LangeAhrens1982, Burgess1991, Springmann2019}), while this study measures many of the dominant outgassing species across a temperature range relevant for terrestrial planet atmospheres for multiple meteorite samples. In addition, some studies shocked samples prior to analyzing their volatile species so they do not properly simulate outgassing conditions from bulk chondritic material (e.g., \cite{Tyburczy1986b}), while other studies focused on either a small subset of volatiles or trace metals and other volatile elements that are not major constituents of rocky planet atmospheres (e.g., \cite{Ikramuddin1977a, CourtSephton2009, Springmann2019}, see ``Comparison with Prior Studies'' in Methods). As a result, these studies are unsuitable for validating outgassing models for terrestrial planets.  Therefore, to inform the initial composition of terrestrial planet atmospheres, we have designed an experimental procedure to directly measure by mass spectrometry a large set of the major outgassed species (e.g., H\textsubscript{2}O, CO\textsubscript{2}) from diverse meteorites over a wide temperature range.

For this study, three chondrites are analyzed: Murchison, Jbilet Winselwan and Aguas Zarcas. Murchison was observed to fall in Australia in 1969 \cite{Murchison1969}; Jbilet Winselwan was collected in Western Sahara in 2013 \cite{Winselwan2015}; and Aguas Zarcas fell in Costa Rica in 2019 \cite{AguasZarcas2019}.  We minimized terrestrial contamination and weathering effects by ensuring that none of our samples have fusion crust, which is altered during atmospheric entry; by using two fall samples which are minimally altered by terrestrial contamination; and by significantly reducing the majority of adsorbed species on the samples (see Methods).

Our experimental set-up consists of a furnace connected to a residual gas analyzer (RGA) mass spectrometer and a vacuum system. This system heats samples at controlled rates (up to 1200 $^\circ$C) in a high-vacuum environment ($\sim 10^{-4}$ Pa ($10^{-9}$ bar) at lower temperatures and $\sim 10^{-3}$ Pa ($10^{-8}$ bar) at higher temperatures) and measures the partial pressures of up to 10 volatile species made up of hydrogen, carbon, nitrogen, oxygen and sulfur. For each of our experiments, a sample was heated in an open crucible (and thereby open to mass loss) from 200 to 1200 $^{\circ}$C (475-1475 K) at a rate of 3.3 $^{\circ}$C/minute and the partial pressures of the outgassing molecular species were continuously monitored using the RGA. The results from these experiments are expressed in three major forms. The first two are the instantaneous partial pressures ($p_i$, for species $i$) and mole fractions ($\chi_i = \frac{p_i}{p_{\text{Total}}}$, for species $i$ where $p_{\text{Total}} = \sum_i p_i$, and see Methods (e.g., Equation 3) for elemental mole fractions) of outgassed volatiles as a function of temperature. The third is relative abundances of outgassed volatiles from each sample, reported as partial pressures normalized to the total pressure of released gases summed over temperature, $P_{i,\text{Total}} = \frac{\sum_T p_i}{\sum_T p_{\text{Total}}}$, for species $i$, and for elemental abundances $P_{j,\text{Total}} = \frac{\sum_T p_j}{\sum_j (\sum_T p_j)}$, for element $j$ (i.e., H, C, O, N, and S).

The RGA measures the partial pressures of 10 selected species according to their molecular masses, assuming species are singly ionized (i.e., the mass-to-charge ratio equals the molecular mass): 2 amu (H\textsubscript{2}), 12 amu (carbon), 14 amu (nitrogen), etc. The signals for each of the species tracked during the outgassing experiments have been corrected for ion fragments and the possibility of terrestrial atmospheric adsorption and contamination using a set of linear equations. This approach also accounts for the background signal (Extended Data Figure 1). In addition, since the masses of several molecules overlap, we developed a method to address these degeneracies (see Methods). An alternative approach to correct for ion fragments using a least squares regression produces generally similar results (see Methods, Supplementary Table 3 and Extended Data Figure 2).

\section*{Results}

Tables 1 and 2 report the relative abundances of outgassed volatile species and elements from the three chondrites. We analyzed samples of Jbilet Winselwan twice under identical conditions to test the reproducibility of our experimental procedure, so its final reported relative abundances are given by the mean and 95\% confidence interval of the mean of the two analyses (see Methods and Extended Data Figure 3). As shown in Table 1, we find that H$_2$O has the largest relative abundance ($\sim$66$\pm$11 \%) for all the meteorite samples followed by CO ($\sim$18$\pm$8 \%), CO\textsubscript{2} ($\sim$15$\pm$5 \%), and H\textsubscript{2} ($\sim$1$\pm$1 \%) (see Extended Data Figure 4). The signal at 34 amu has a lower relative abundance while the signals at 12, 14, 16, 32, and 40 amu along with N\textsubscript{2} have relative abundances that are effectively 0 (see Methods). In terms of the elemental abundances in Table 2, hydrogen has the highest concentration ($\sim$48$\pm$5 \%) followed by oxygen ($\sim$41$\pm$2 \%), carbon ($\sim$12$\pm$4 \%), and sulfur ($\sim$0.03$\pm$0.02 \%), averaged across all three samples (uncertainties reported as the 95\% confidence interval).

We expect the three CM chondrite samples to have similar outgassed abundances given their similar bulk compositions which are within 20 mg/g for most volatile elements (Supplementary Table 2). Our experimental results confirm this prediction with the relative abundances for each species across the three samples being within 3$\sigma$ of each other.  Figure 1 shows the mole fractions of the measured volatile species as a function of temperature from each of the three meteorite samples. Several differences to note between the three meteorites are that Murchison has larger water but smaller CO outgassed abundances compared to Winselwan and Aguas Zarcas. In addition, Aguas Zarcas has a larger CO abundance but smaller CO\textsubscript{2} abundance than Murchison and Winselwan. Figure 2 shows the relative abundance ratios for the three samples as a function of temperature, which can inform the chemistry of the initial atmospheres. The mean outgassed carbon-to-oxygen, hydrogen-to-carbon, sulfur-to-oxygen and hydrogen-to-oxygen ratios summed over temperature for the three chondrite samples are 0.29$\pm$0.08, 4.15$\pm$1.80, 0.0008$\pm$0.0006, and 1.18$\pm$0.18, respectively (uncertainties reported as the 95\% confidence interval), with abundance ratios for the three chondrites within 2$\sigma$ of each other.  

The average C/O, H/C, S/O and H/O ratios for the initial bulk CM chondrite composition are 0.06$\pm$0.01, 7.09$\pm$1.18, 0.04$\pm$0.01, and 0.45$\pm$0.01, respectively \cite{Nittler2004, Alexander2012}. These initial bulk elemental abundances represent the volatile outgassing inventory for a planet that is outgassing predominantly CM chondrite-like material (Table 2). By comparing the outgassed and bulk CM chondrite abundance ratios, we find that, aside from the S/O ratios, the outgassed and bulk ratios are within an order of magnitude of each other. The C/O and H/O outgassed ratios are larger than the bulk ratios, which is likely due to the fact that many of the phases that host carbon and hydrogen (e.g., phyllosilicates, organics and carbonates) readily break down upon heating whereas a significant portion of the oxygen is in phases that do not easily decompose (e.g., forsterite and CaO). On the other hand, the H/C and S/O ratios are smaller than the bulk ratios, with the largest difference between the outgassed and bulk S/O ratios. This may be due to the fact that models predict that S\textsubscript{2} and SO\textsubscript{2} should also outgas in this temperature range but we only measure H\textsubscript{2}S due to the RGA's 10-species limit.

There are several major similarities between the results from our experiments (Figure 3 (b, d)) and those from the equilibrium calculations (Figure 3 (a, c)). For instance, water is the dominant outgassed species over almost the entire temperature range for both experimental and theoretical methods, with the average mole fraction being 0.6 and 0.4 for the experiments and theoretical models, respectively. In addition, CO and CO\textsubscript{2} constitute significant fractions of the vapor phase over the temperature range. In particular, for the Murchison sample, the experimental and theoretical outgassing trends for CO\textsubscript{2} and CO generally match with CO\textsubscript{2} outgassing more at the lowest ($\sim$300 $^{\circ}$C) and highest temperatures ($\sim$1100 $^{\circ}$C), and CO outgassing more at intermediate temperatures ($\sim$800-1000 $^{\circ}$C). In addition to the experimental results, we calculated `equilibrium-adjusted' abundances using  the equilibrium model to re-compute gas speciation based on the experimental abundances at intervals of 50 $^{\circ}$C (dashed curves Figure 3 (b, d)). Generally, the `equilibrium-adjusted' experimental H\textsubscript{2}O and CO\textsubscript{2} abundances provide a better match to the equilibrium model results, and the CO abundance provides a better match at higher temperatures. On the other hand, the other significant outgassing species (H\textsubscript{2} and H\textsubscript{2}S) did not exhibit an improved match compared to the experimental results, although the equilibrium-adjusted H\textsubscript{2} abundance is much larger compared to the experimental H\textsubscript{2}. Although our experiments monitor signals at 12 amu (carbon), 14 amu (nitrogen), 16 amu (CH\textsubscript{4}/O) and 40 amu (Ar), once we correct for ion fragments and atmosphere adsorption, we do not detect significant amounts of these species, also matching chemical equilibrium calculations (see Methods).

There are also significant differences between the experimental and theoretical results. For instance, although H\textsubscript{2}S mole fractions reach similar maxima of $\sim$1E-3, the peak is displaced from $\sim$600 $^{\circ}$C in the theoretical calculations to $\sim$950 $^{\circ}$C in the experiments. This offset may be due to the fact that, in carbonaceous chondrites, sulfur can occur as gypsum (a sulfate mineral, CaSO\textsubscript{4}H\textsubscript{2}O) which breaks down at 700 $^{\circ}$C and the corresponding phase change for sulfur gas may be kinetically inhibited, causing it to outgas at higher temperatures \cite{OBrienNielsen1959}. Preliminary X-ray diffraction (XRD) analyses carried out in our lab indicate that gypsum may be breaking down during the experiments (see Methods section on solid phases). However, the equilibrium models do not show gypsum being in a solid phase, which may explain the mismatch in outgassing trends of H\textsubscript{2}S. In addition, iron sulfide (FeS) and tochilinite (2Fe\textsubscript{0.9}S*1.67Fe(OH)\textsubscript{2}), which are known to be in carbonaceous chondrites, may decompose and contribute to the outgassed H\textsubscript{2}S \cite{Zhao2011, Gooding1987}. The `equilibrium-adjusted' H\textsubscript{2}S abundances are much lower than both the model and experimental results, further supporting H\textsubscript{2}S production being kinetically inhibited. Another significant difference is the experiments' absence of N\textsubscript{2} gas which does not match equilibrium models that show outgassing over the entire temperature range. The primary reason the experimental N\textsubscript{2} abundance is negligible is due to the atmospheric adsorption correction we apply to account for the possibility of contaminated N\textsubscript{2} gas from Earth’s atmosphere being adsorbed by the samples (see Methods). Without this correction, the measured N\textsubscript{2} outgassing varies from moderate levels consistent with theoretical models (i.e., average mole fractions of $\sim$5E-2) to negligible amounts, depending on the sample being investigated.

Several other differences between experimental outgassing and equilibrium model results lack complete explanations. For example, the second most abundant volatile species predicted to outgas over most temperatures in equilibrium models is H\textsubscript{2} with maxima mole fractions of $\sim$0.4, but our experimental results indicate an order of magnitude lower average H\textsubscript{2} mole fractions near $\sim$0.03. While the cause for much lower experimental H\textsubscript{2} abundance is uncertain, it may be due to the fact that our experiments do not allow sufficient time for some gas-gas reactions to take place that could raise the H\textsubscript{2} abundance, as equilibrium is reached between H\textsubscript{2}O and H\textsubscript{2}. The `equilibrium-adjusted' H\textsubscript{2} abundances are larger than the experimental results, pointing to a likely disequilibrium for gas phase reactions in the experiments. By comparing gas-gas reaction rates to the experiment's vacuum pumping rate, we conclude that these species likely do not have sufficient time to reach chemical equilibrium (see Methods). In addition, the H\textsubscript{2} background signals are large relative to the samples' signals, so over-subtraction of the background signal could also explain the lower abundance. Finally, our experiments do not detect a significant amount of outgassed species with mass number 32 amu after correcting for the possibility of contaminated terrestrial atmospheric adsorption of O\textsubscript{2} (see Methods). In the chemical equilibrium calculations, O\textsubscript{2} should only begin outgassing significantly at the highest temperatures ($\sim$1100 $^{\circ}$C). As described further in Methods, an alternative data analysis technique using least-squares fitting produces non-zero yields of CH\textsubscript{4}, which is not predicted to outgas based on equilibrium models (Extended Data Figure 2, Supplementary Table 3). It is possible that our original analysis (see Equations 4-15 in Methods) applies an overly conservative correction to CH\textsubscript{4}'s abundance, and further investigation is required to confirm if CH\textsubscript{4} is indeed outgassing from the samples.

Oxygen fugacity ($f_{O_2}$) represents the chemical potential of oxygen in a system which affects its gas chemistry and may be equated to the partial pressure of oxygen in a gas phase under low pressure and near-ideal gas conditions such as present in these experiments. Our instrument set-up does not allow equilibrium to be achieved, especially at lower temperatures, because many gas species do not have sufficient time to interact since their mean collision time is either comparable to or longer than the vacuum system's evacuation rate. Therefore, it is not surprising that the experimental $f_{O_2}$ curves from H$_2$O/H$_2$ and CO$_2$/CO do not agree in magnitude, except at the highest temperatures, revealing that the gas phase is likely not in equilibrium (see Figure 4).

\section*{Discussion}

The results from our outgassing experiments have several important implications for the initial atmospheric chemistry of low-mass exoplanets. While terrestrial planets experience a diversity of conditions during planet formation, these experiments represent an empirically-determined boundary condition of the less-theoretically-explored lower temperature/pressure paths that could be used to test outgassing models. Our experimental conditions approximately simulate the initial heating phase during planet formation, revealing the initial volatile species that would outgas assuming the bulk composition of material being outgassed is CM chondrite-like. As such low-mass planets form their initial atmospheres via outgassing during accretion, H\textsubscript{2}O-rich steam atmospheres form. These atmospheres will also likely contain significant amounts of CO and CO\textsubscript{2} and smaller amounts of H\textsubscript{2} and H\textsubscript{2}S. Until now, several common assumptions used for low-mass exoplanet atmospheric modeling include ad hoc abundances such as H\textsubscript{2}O-only or CO\textsubscript{2}-only, solar abundances (dominated by H\textsubscript{2} and He), or atmospheric compositions of Solar System planets \cite{Miller-Ricci2009, Fortney2013, Greene2016, Morley2017, Bower2019}. Our outgassing experiments suggest initial atmospheres may differ significantly from many of the common assumptions in atmosphere models, and provide the basis for more refined future models of terrestrial planets' initial atmospheres (see ``Comparison with Prior Studies" in Methods). 

In proposing that our results be used as initial conditions for terrestrial exoplanet atmosphere models, we note that while our experiments cover a wide range of temperatures, they were conducted in a low-pressure environment. Schaefer \& Fegley 2010 \cite{SchaeferFegly2010} suggest that varying pressure does not have a significant effect on the major gas composition of outgassed atmospheres from CM chondrite material. Chemical equilibrium or kinetics calculations can determine how these sets of initial compositions from outgassing experiments evolve as the pressure varies within a planet's atmosphere.

Theoretical models of terrestrial planet atmosphere formation usually lack experimental constraints. Based on the results from our experiments, the most direct way to improve such models of outgassing from CM chondritic-like building blocks would be to assume initial abundances at planetary surface boundaries averaging approximately 66\% H\textsubscript{2}O, 18\% CO and 15\% CO\textsubscript{2} over a temperature range similar to that used in this study. Depending on the capabilities of a particular model, temperature dependencies for species abundances could also be incorporated using our experimental results. Additional improvements could include the Langmuir coefficients for evaporation (and condensation) for the major outgassing species measured in our experiments to properly simulate chemical kinetics effects. Determining these coefficients often requires conducting Langmuir evaporation experiments \cite{SossiFegley2018}. In addition, reaction rates between the solid and gas phases relevant for chondrite outgassing would improve our understanding of when chemical equilibrium conditions are acceptable and when chemical kinetics effects are important. Subsequent experiments will determine whether other species predicted by chemical equilibrium models also outgas significantly. In addition, future experiments will measure volatiles from a wider variety of meteorites including ordinary and enstatite chondrites.

\newpage 

\noindent \textbf{Correspondence and requests for materials} should be addressed to M.\ Thompson.

\subsection*{Acknowledgements}
We thank A.\ K.\ Skemer for his helpful insights and K.\ Kim for performing the preliminary XRD experiments. M.\ Thompson acknowledges support from the ARCS Foundation Scholarship. M.\ Telus is supported by NASA EW grant 80NSSC18K0498 and NASA ECA grant 80NSSC20K1078. T.\ J. was supported from U.\ C. Santa Cruz start-up funds. 

\subsection*{Author Information} 
 M.\ Thompson performed the outgassing experiments and data analysis and wrote the manuscript. M.\ Thompson and M.\ Telus conceived the research. D.L., T.J., M.\ Telus and M.\ Thompson collaborated to configure the experimental set-up. M.\ Telus provided the meteorite samples used in the experiments and imparted essential guidance on the data analysis and interpretation of the results. D.L. provided the laboratory equipment for the experiments and helpful suggestions for the data analysis. T.J. helped with preparing the experiments and maintaining the instruments. L.S. provided the chemical equilibrium models and greatly contributed to interpreting the results and their implications. J.F. gave important insight into the scope of this work and its implications for exoplanet atmospheres. All authors contributed to editing the manuscript. \\
 
 \noindent The authors declare no competing interests.
 \clearpage 
 
\section*{Figures}

\begin{figure}[hbt!]
\centering\captionsetup[subfloat]{labelfont=bf}
\vspace{-0.5cm}
    \centering
    \begin{subfigure}{0.49\textwidth}
        \centering
        \includegraphics[width=\textwidth]{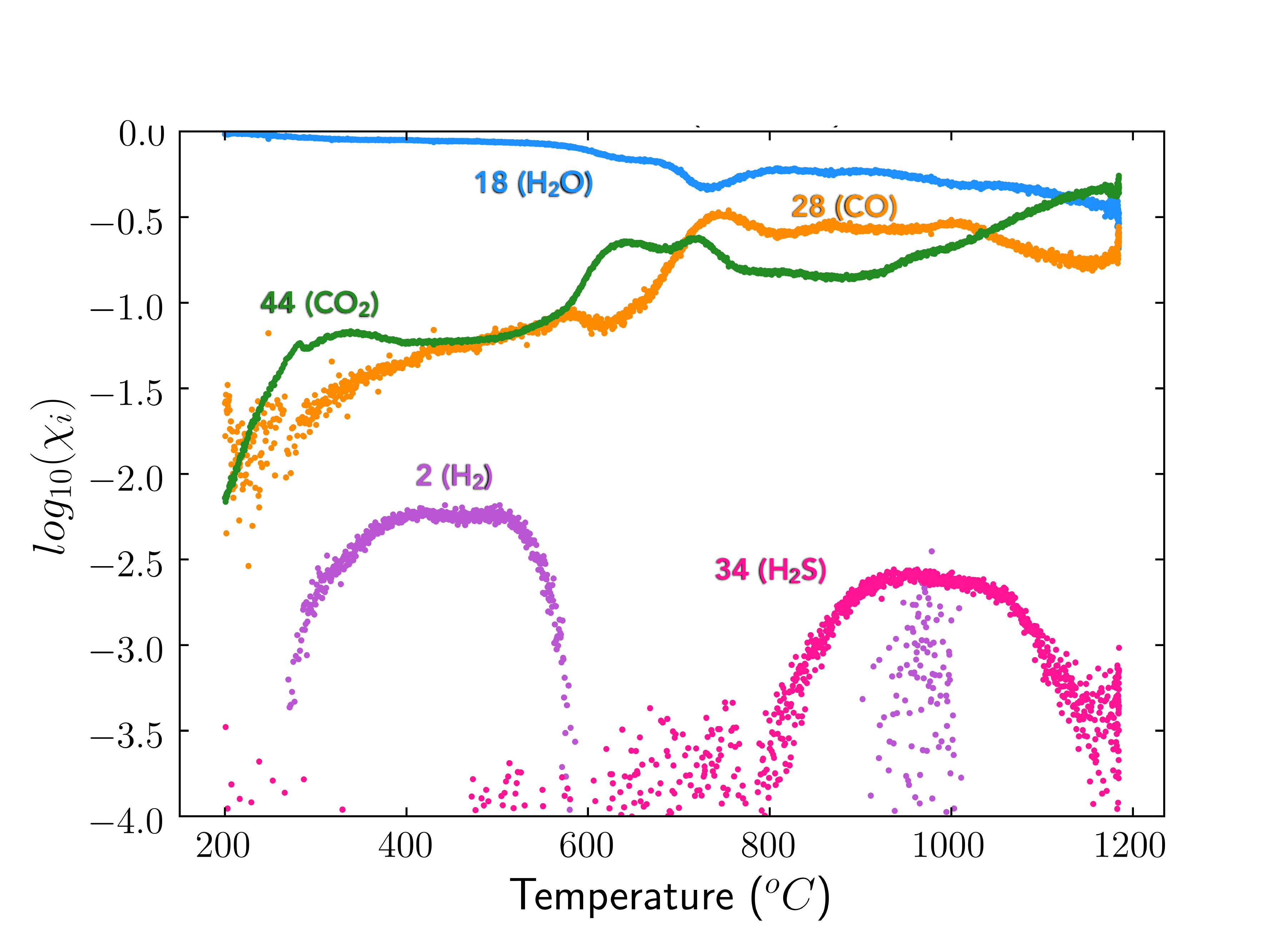}
        \caption[Murchison]%
        {{\small Murchison}}    
        \label{fig:Murchison logX} \vspace{-0.5cm}
    \end{subfigure}
    \hfill
    \begin{subfigure}{0.49\textwidth}  
        \centering 
        \includegraphics[width=\textwidth]{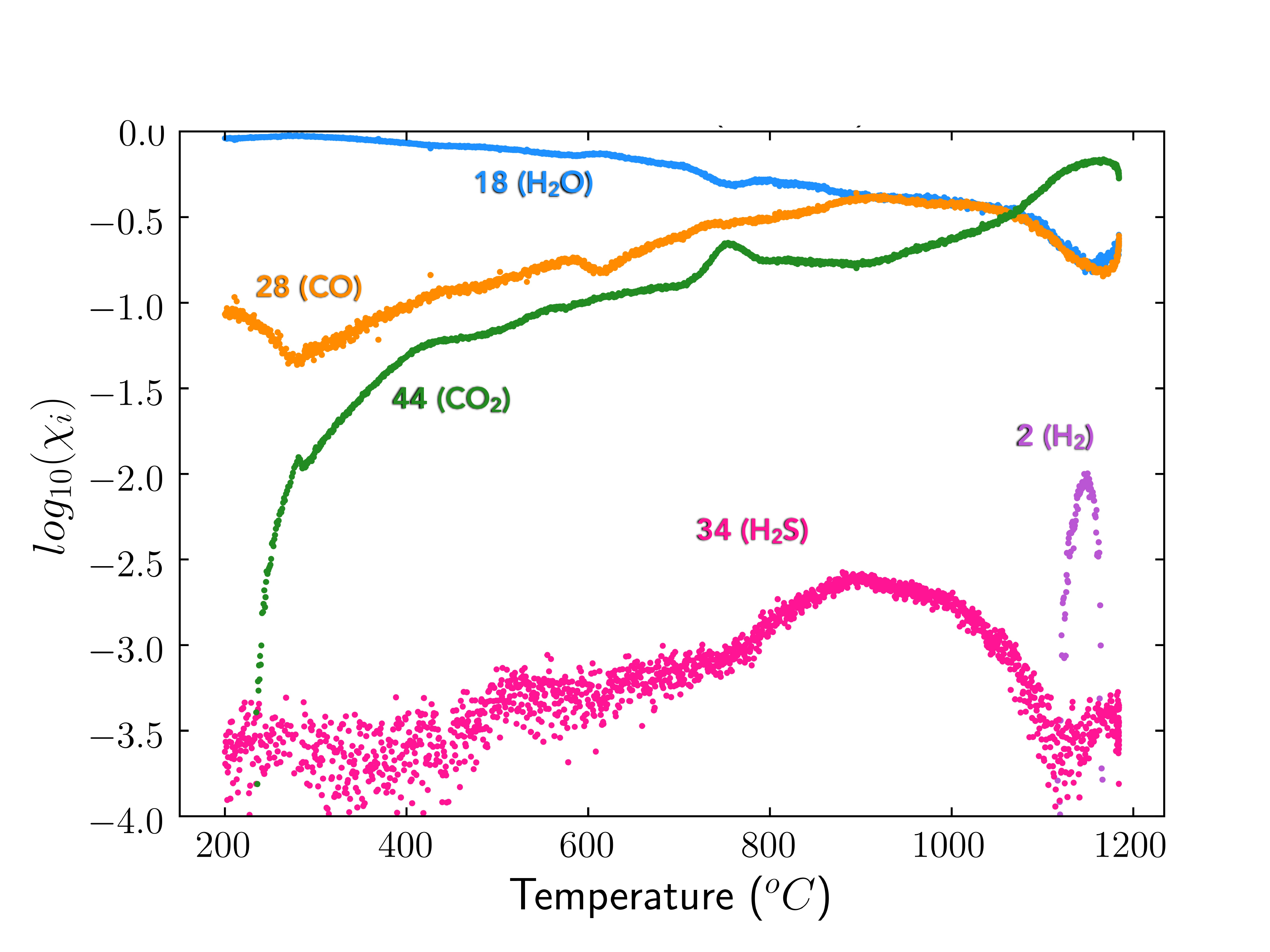}
        \caption[]%
        {{\small Jbilet Winselwan (Experiment 1)}}  
        \label{fig:Winselwan1 logX} \vspace{-0.5cm}
    \end{subfigure}
    \vskip\baselineskip
    \begin{subfigure}[b]{0.49\textwidth}
        \centering 
        \includegraphics[width=\textwidth]{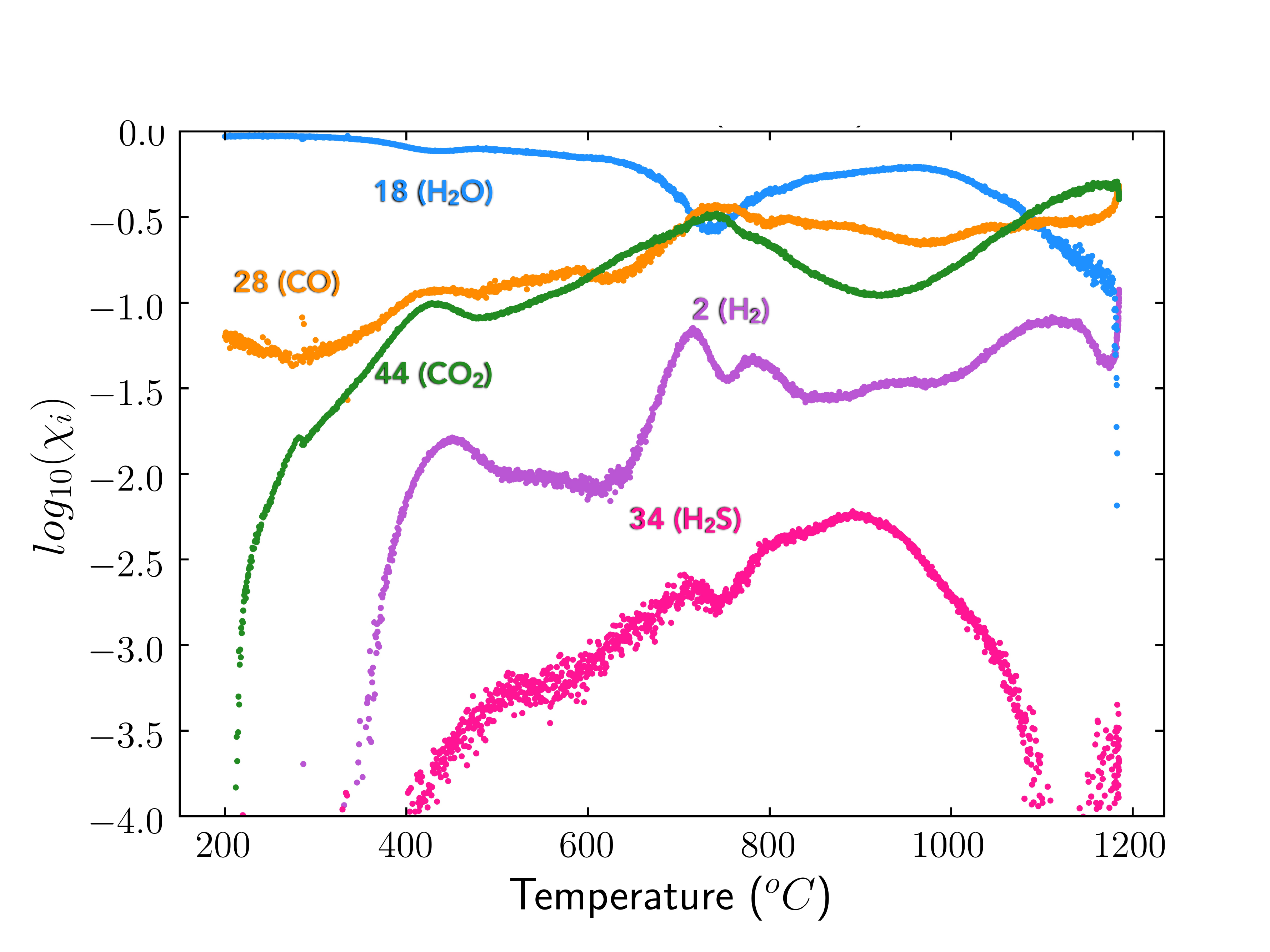}
        \caption[]%
        {{\small Jbilet Winselwan (Experiment 2)}}  
        \label{fig:Winselwan2 logX} 
    \end{subfigure}
    \hfill
    \begin{subfigure}[b]{0.49\textwidth}
        \centering 
        \includegraphics[width=\textwidth]{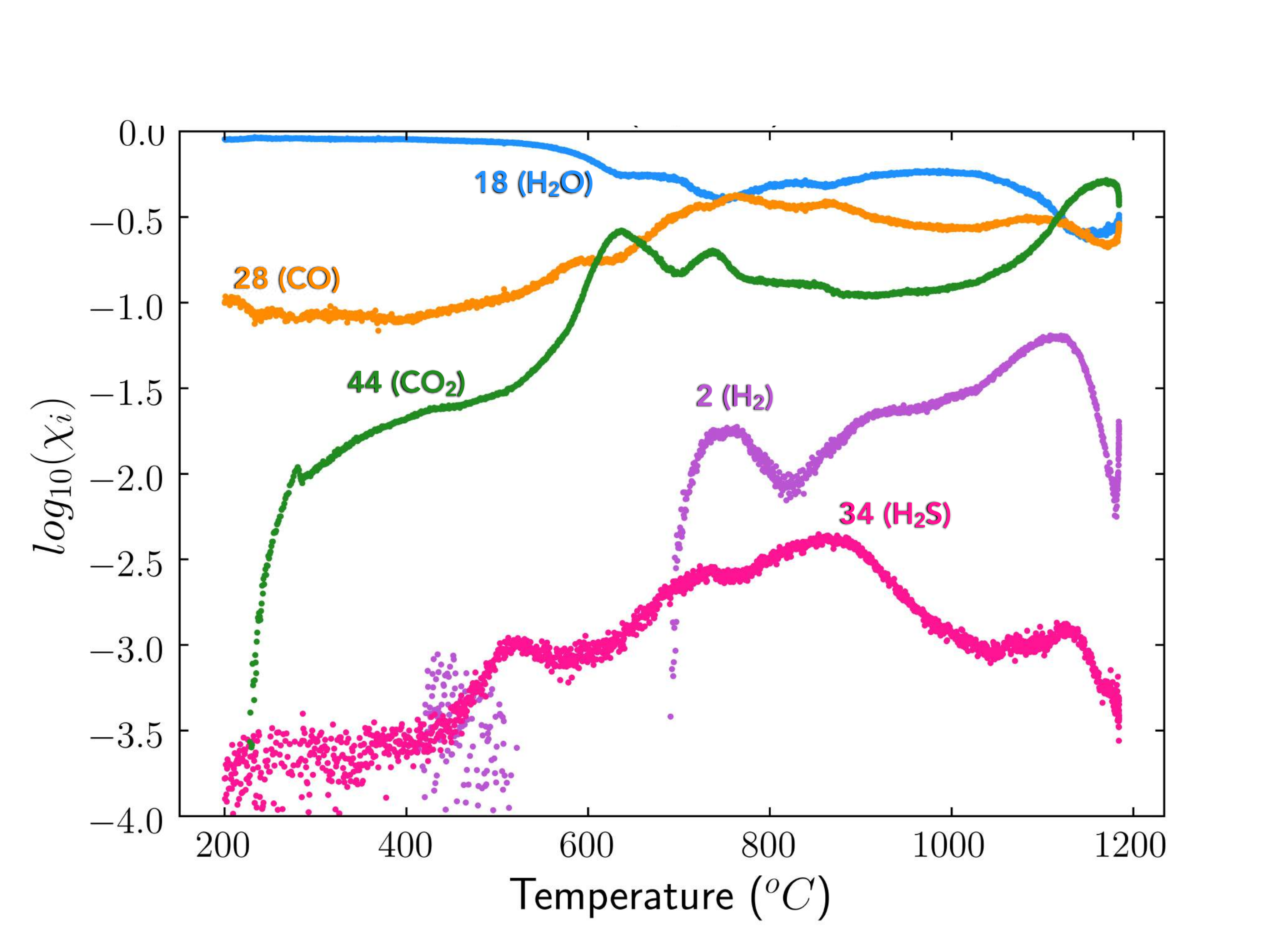}
        \caption[]%
        {{\small Aguas Zarcas}}    
        \label{fig: AZ logX} 
    \end{subfigure}
    \caption{\textbf{Mole fractions of the measured species outgassed as a function of temperature for each chondrite sample}. The results are for 3 mg samples of (a) Murchison, (b) and (c) Jbilet Winselwan, and (d) Aguas Zarcas. We analyzed two 3 mg samples of Jbilet Winselwan under identical conditions to test reproducibility and show the results in (b) and (c). H\textsubscript{2} has the largest variation between the two experiments with Jbilet Winselwan. Across the three samples, some species exhibit major variations in their relative abundances over specific temperature intervals. For instance, CO and CO\textsubscript{2}'s abundances increase around 650 - 750 $^{\circ}$C. Although the mole fraction of H\textsubscript{2}S varies considerably over the entire heating range, it peaks near $\sim$900-1000 $^{\circ}$C and then decreases at higher temperatures for all three chondrites. For most samples, there is a prominent increase in H\textsubscript{2}'s abundance near $\sim$1100  $^{\circ}$C. } 
    \label{fig:heatingresults}
\end{figure}
 
\begin{figure}[hbt!]
    \includegraphics[scale=0.4]{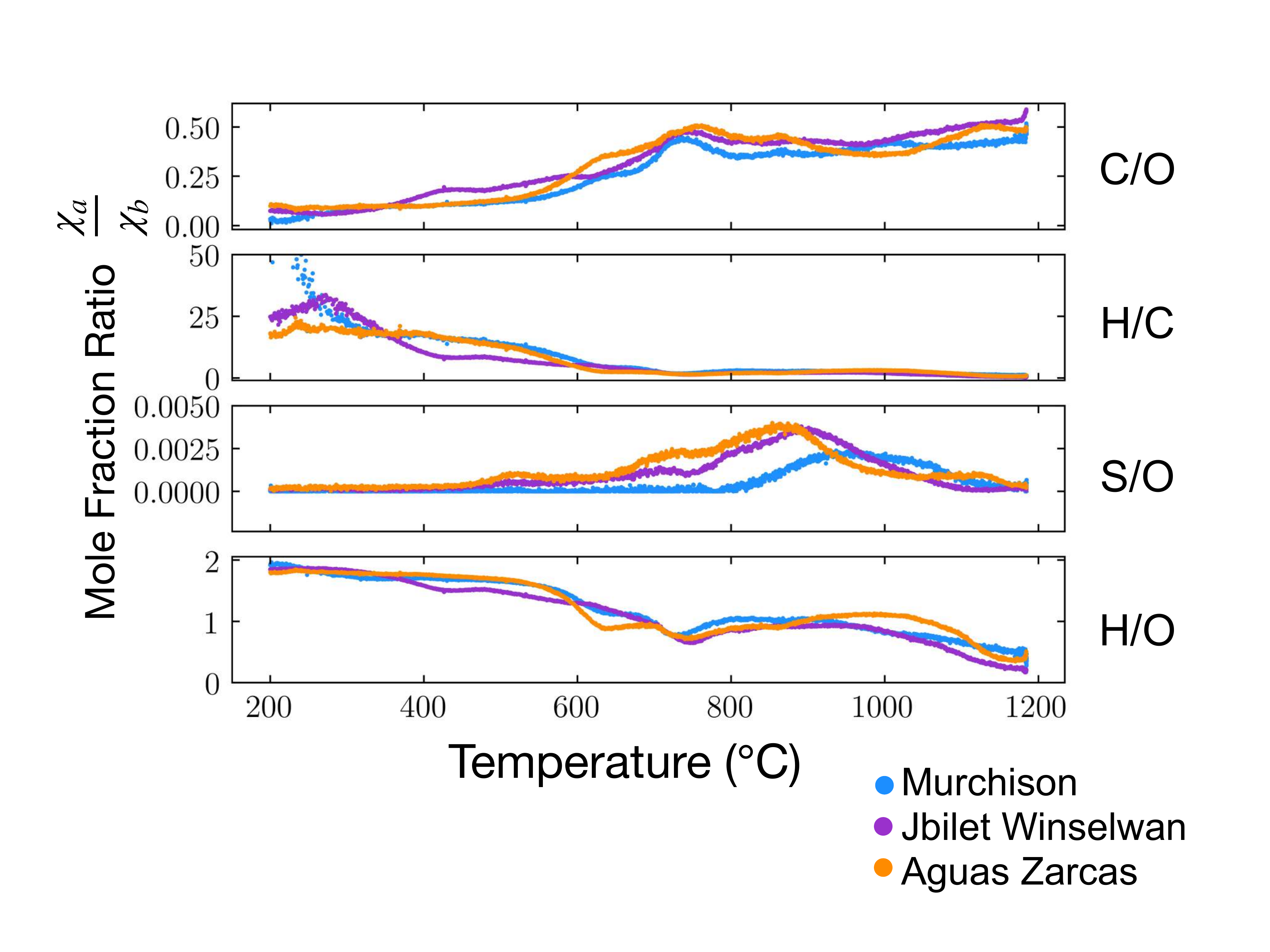}
    \centering
    \caption{\textbf{Ratios of mole fractions of outgassed bulk elements hydrogen, carbon, oxygen, and sulfur as a function of temperature for the three chondrite samples.} From top to bottom the ratios are: carbon/oxygen, hydrogen/carbon, sulfur/oxygen and hydrogen/oxygen. Blue, purple and orange curves represent elements outgassed from Murchison, Winselwan, and Aguas Zarcas, respectively.}
    \label{fig:image2}
\end{figure}

 \begin{figure}[hbt!]
    \centering
    \begin{subfigure}{0.49\textwidth}
        \centering
        \includegraphics[width=\textwidth]{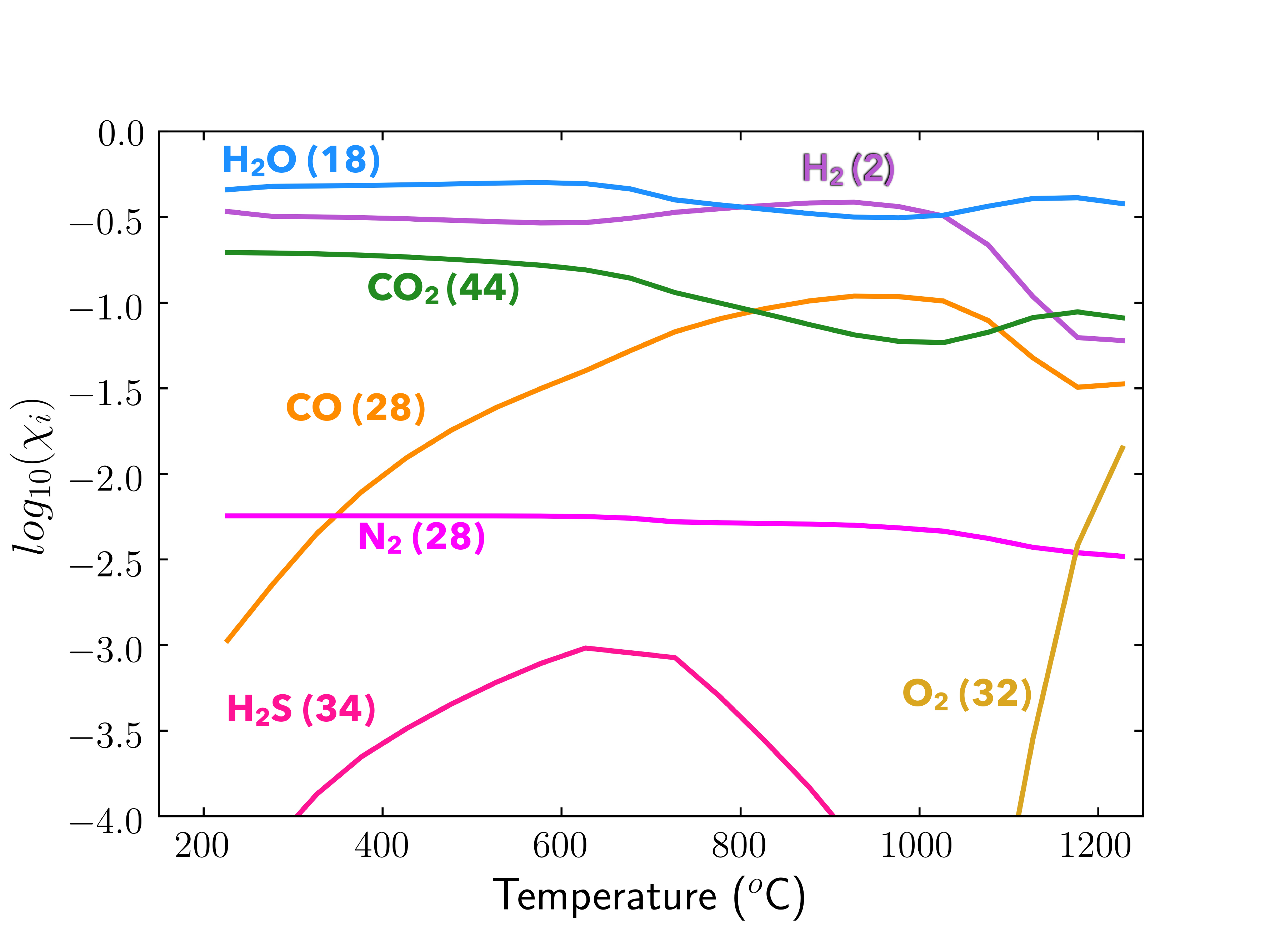}
        \caption[]%
        {{\small Theoretical Equilibrium Results for Average CM Chondrite Composition}}    
        \label{fig:CMAvgTheory}
    \end{subfigure}
    \hfill
    \begin{subfigure}{0.49\textwidth}  
        \centering 
        \includegraphics[width=\textwidth]{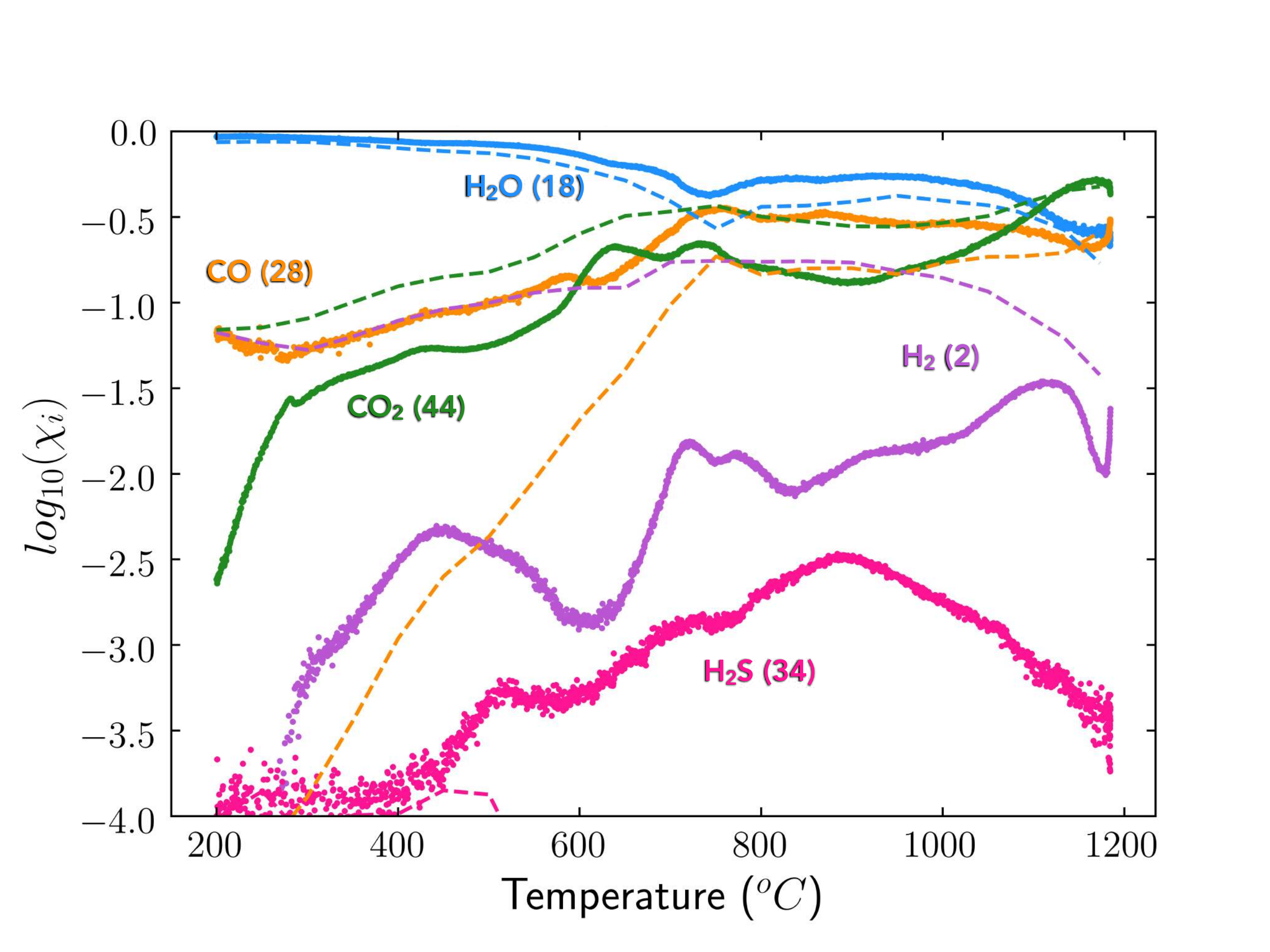}
        \caption[]%
        {{\small Experimental Results for Average of 3 CM Chondrite Samples}}    
        \label{fig:CMAvgExperiment}
    \end{subfigure}
    \hfill
    \begin{subfigure}{0.49\textwidth}
        \centering
        \includegraphics[width=\textwidth]{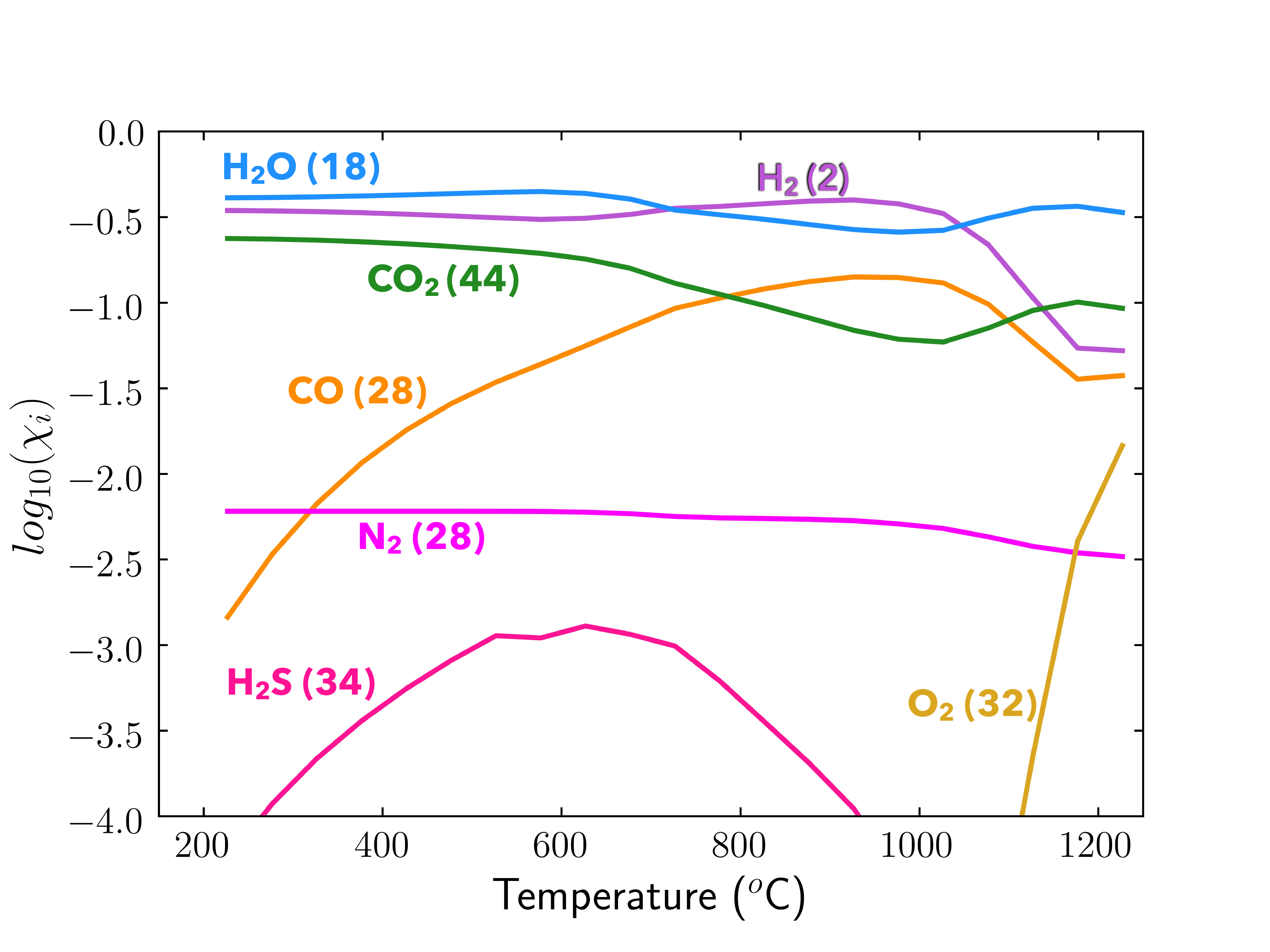}
        \caption[]%
        {{\small Theoretical Equilibrium Results\\ for Murchison}}    
        \label{fig:MurchisonTheory}
    \end{subfigure}
    \hfill
    \begin{subfigure}{0.49\textwidth}
        \centering
        \includegraphics[width=\textwidth]{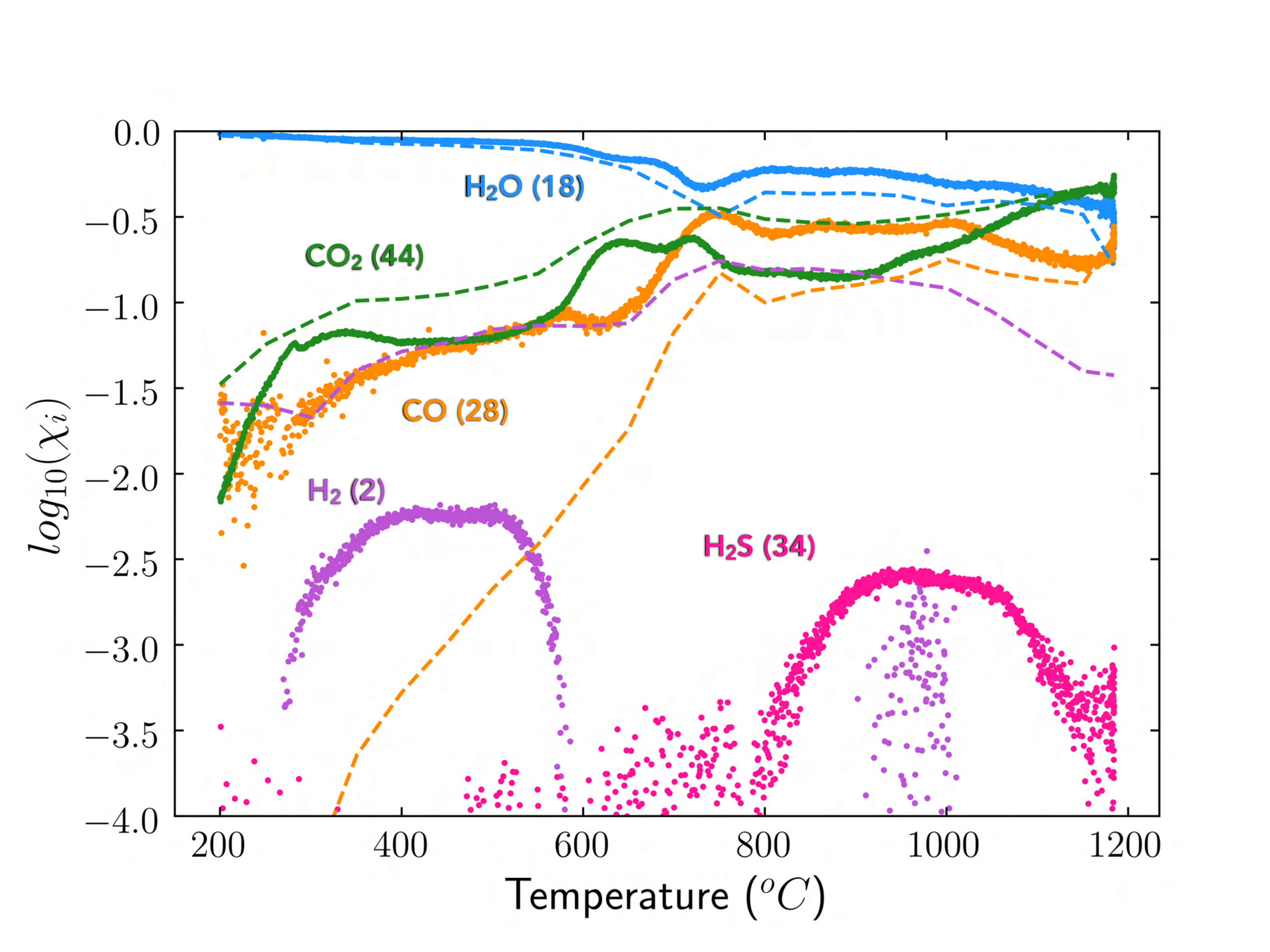}
        \caption[]%
        {{\small Experimental Results for Murchison Sample}}    
        \label{fig:MurchisonExperiment}
    \end{subfigure}
\caption{\textbf{Comparison between equilibrium calculations (left) and experimental results (right) under the same pressure and temperature conditions.} Figures (a) and (b) illustrate outgassing abundances calculated assuming chemical equilibrium for an average CM chondrite bulk composition at 1E-3 Pa (a) and experimental outgassing results for the average of the three CM chondrite samples measured at 1E-3 Pa (b). In (b), each species' curve is dominated by the sample that has the most abundant amount of that species at a given temperature. Figures (c) and (d) show the results for outgassing from a Murchison composition using chemical equilibrium calculations (c) and experimental outgassing results from the Murchison sample (d). The dashed curves in (b) and (d) show `equilibrium-adjusted' experimental abundances in which the equilibrium model was used to recalculate gas speciation using the experimental abundances at intervals of 50 $^{\circ}$C. The mass (in amu) of each species is in parentheses. See Extended Data Figure 5 for other volatile species that theoretically degas with mole fractions above $1\times10^{-4}$ according to chemical equilibrium calculations but are not measured in the experiments. } 
\label{fig:TheoryvsExperiment}
\end{figure}
 
 \begin{figure}[hbt!]
    \centering
    \includegraphics[width=0.9\textwidth]{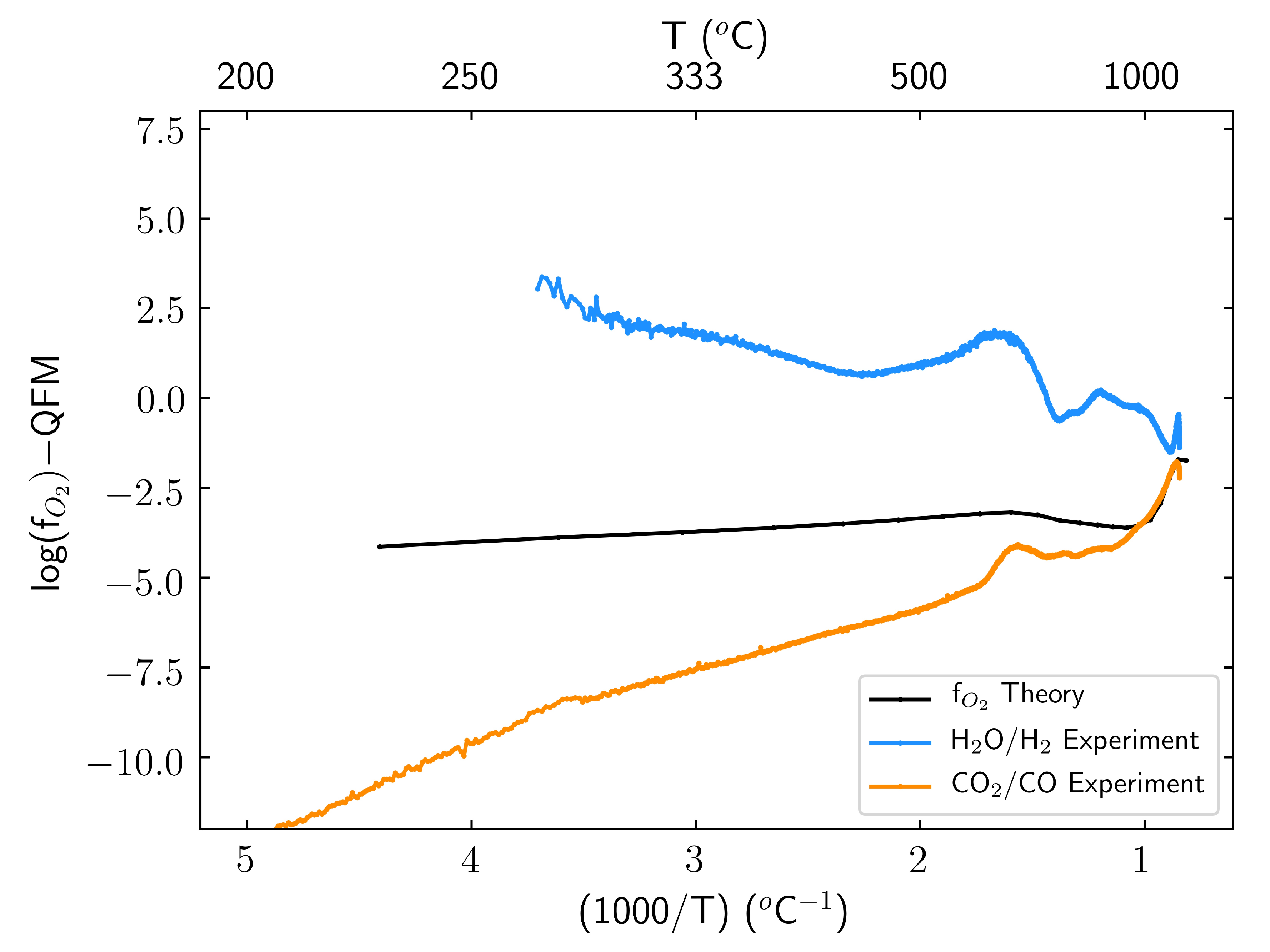}
\caption{\textbf{Oxygen fugacities relative to the quartz-fayalite-magnetite (QFM) buffer from theory and experiments.} Oxygen fugacity of an average bulk CM chondrite composition as a function of temperature from chemical equilibrium calculations (black curve, labeled Theory) and the oxygen fugacity of the average of the three CM chondrites measured experimentally (blue and orange curves, labeled Experiment). Two abundance ratios were used to calculate $f_{O_2}$: H\textsubscript{2}O/H\textsubscript{2} (blue curve) and CO\textsubscript{2}/CO (orange curve). We cannot determine $f_{O_2}$ directly from the O\textsubscript{2} abundance because after correcting for terrestrial atmospheric adsorption its abundance goes to zero (see Methods).} 
    \label{fig:Ofugacity}
\end{figure}
\clearpage

\section*{Tables}

\begin{table}[hbt!]
\setlength{\tabcolsep}{9pt}
\centering
\caption{\textbf{Total mass of volatile species released (in g) and relative abundances (in \%) of outgassed species summed over temperature for the three CM chondrite samples.} Abundances are in partial pressures normalized to the total pressure of all released gases measured summed over temperature and are reported as percentages. The species corresponding to each mass number is in parentheses. The uncertainties reported for Jbilet Winselwan are the 95\% confidence intervals of the means. Since some species have overlapping mass numbers (e.g., S and O\textsubscript{2}), we provide a detailed description of the calculations made in determining these relative abundances in Methods. \label{tab:meas}}
\begin{tabular}{c c c c}
\hline
  & Murchison & Jbilet Winselwan & Aguas Zarcas \\
 \hline
 \vspace{0.1cm}
 Total Gas Mass & 5.00$\times10^{-4}$ g & 5.50$\pm1.50\times10^{-4}$ g & 8.00$\times10^{-4}$ g \\
 2 amu (H$_2$) & 0.20 & 1.27$\pm$15.64 & 1.38 \\
 12 amu (C) & 0.0 & 0.0 & 0.0 \\
 14 amu (N) & 0.0 & 0.0 & 0.0 \\
 16 amu (CH$_4$/O) & 0.0 & 0.0 & 0.0 \\
 18 amu (H$_2$O) & 71.62 & 61.30$\pm$33.73 & 65.02 \\
 28 amu (CO) & 13.45 & 19.90$\pm$15.74 & 20.50 \\
 28 amu (N$_2$) & 0.0 & 0.0 & 0.0 \\
 32 amu (S/O$_2$) & 0.0 & 0.0 & 0.0 \\
 34 amu (H$_2$S) & 0.05 & 0.11$\pm$0.40 & 0.11 \\
 40 amu (Ar) & 0.0 & 0.0 & 0.0 \\ 
 44 amu (CO$_2$) & 14.67 & 17.43$\pm$34.03 & 12.98 \\
 \hline
\end{tabular}

\vspace{0.4cm}

\setlength{\tabcolsep}{5pt}
\centering
\caption{\textbf{Relative outgassed atomic abundances (in \%) summed over temperature of hydrogen, carbon, oxygen, nitrogen and sulfur for the three samples.} As in Table 1, abundances are in partial pressure normalized to the total pressure of all released gases measured and are reported as percentages. The uncertainties reported for Jbilet Winselwan are the 95\% confidence intervals on the means. Comparison between the initial (pre-degassing) normalized atomic abundances for average CM chondrites and the outgassed normalized atomic abundances are shown, both reported as percentages (bottom two rows). These atomic abundances are normalized to the sum of the elements measured in the experiments, i.e., H, C, O, N, S. The uncertainties for the pre-degassing normalized atomic abundances are (1$\sigma$) standard deviations. The uncertainties for the outgassed quantities are expressed as the 95\% confidence intervals of the means. The initial bulk atomic abundances of CM chondrites come from the literature \cite{Nittler2004, Alexander2012}. \label{tab:meas2}}
\begin{tabular}{c c c c c c}
\hline
 Sample & Total H & Total C & Total O & Total N & Total S\\ 
 \hline
  Murchison & 50.20 & 9.82 & 39.96 & 0.0 & 0.02 \\ 
  Jbilet Winselwan & 44.95$\pm$35.69 & 13.39$\pm$17.86 & 41.62$\pm$17.97 & 0.0 & 0.04$\pm$0.14 \\ 
  Aguas Zarcas & 47.83 & 12.04 & 40.09 & 0.0 & 0.04 \\
  \hline
 Initial* Bulk CM Abundance & 28.81$\pm$0.44 & 4.07$\pm$0.67  & 64.37$\pm$0.12 & 0.18$\pm$0.05 & 2.58$\pm$0.70 \\
 Outgassed CM Abundance & 47.66$\pm$5.33 & 11.75$\pm$3.65 & 40.55$\pm$1.88 &  0.0  & 0.03$\pm$0.02 \\
 \hline
 \noindent *Pre-degassing
\end{tabular}
\end{table}

\section*{Extended Data Figures}

\renewcommand{\figurename}{Extended Data Figure}
\setcounter{figure}{0}
\begin{figure}[hbt!]
    \begin{subfigure}{0.49\textwidth}
        \centering
        \includegraphics[width=\textwidth]{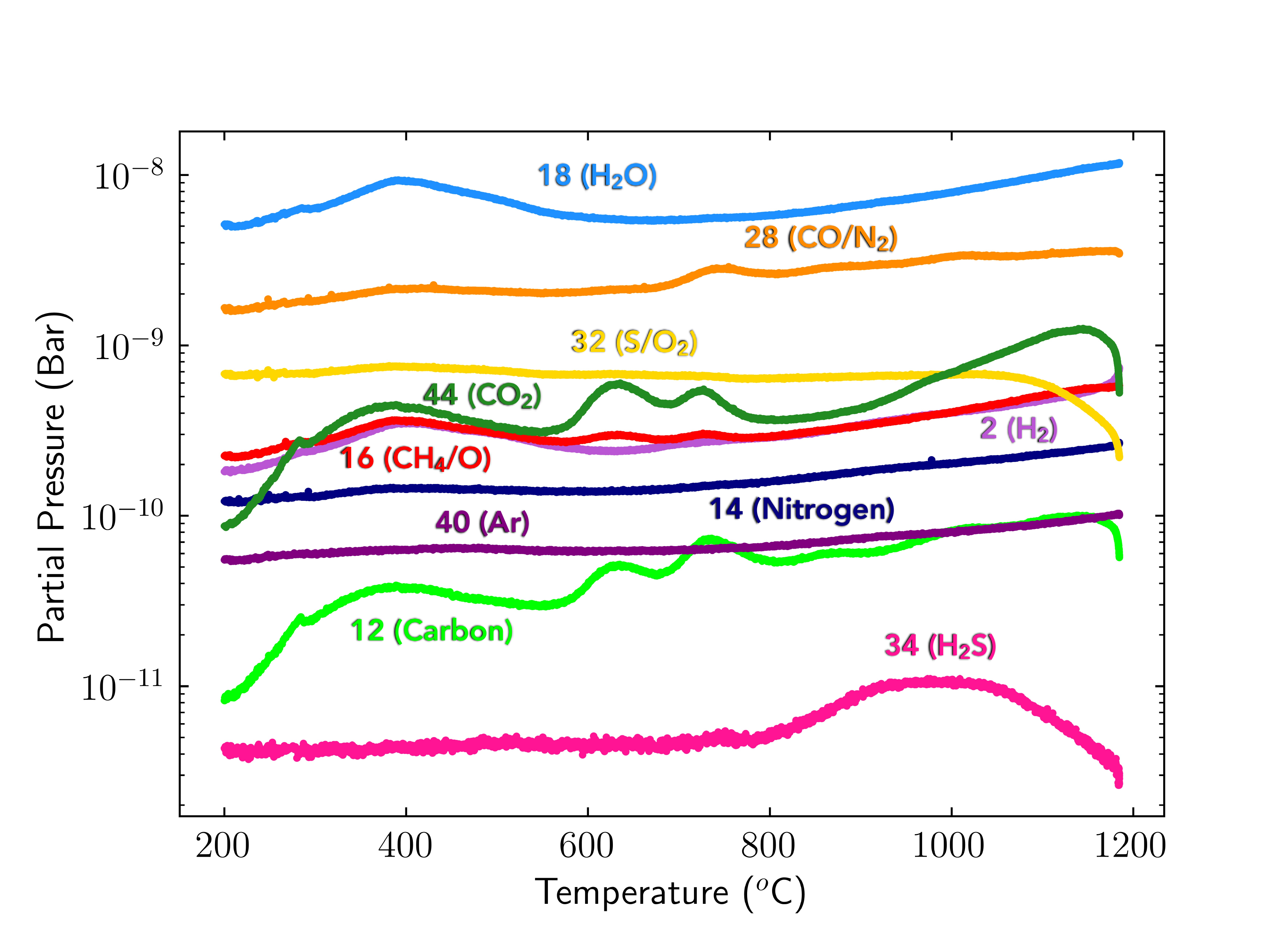}
        \caption{\small Murchison Raw Measurement}  
        \label{fig:e1a} 
    \end{subfigure}
    \hfill
    \begin{subfigure}{0.45\textwidth}  
        \centering 
        \includegraphics[width=\textwidth]{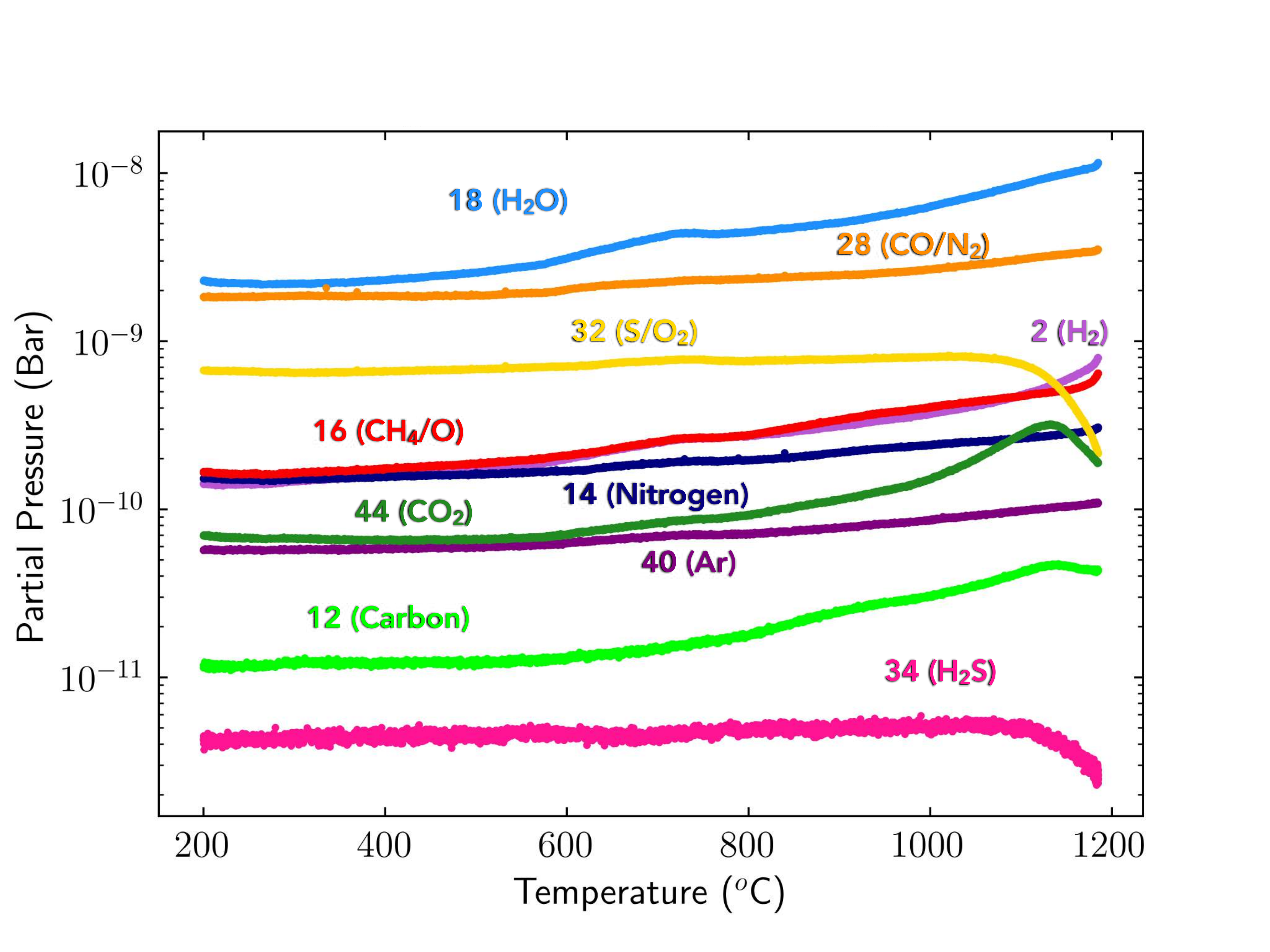}
        \caption{\small Background Raw Measurement}
        \label{fig:e1b}  
    \end{subfigure}
    \hfill
    \begin{subfigure}{0.45\textwidth}   
        \centering 
        \includegraphics[width=\textwidth]{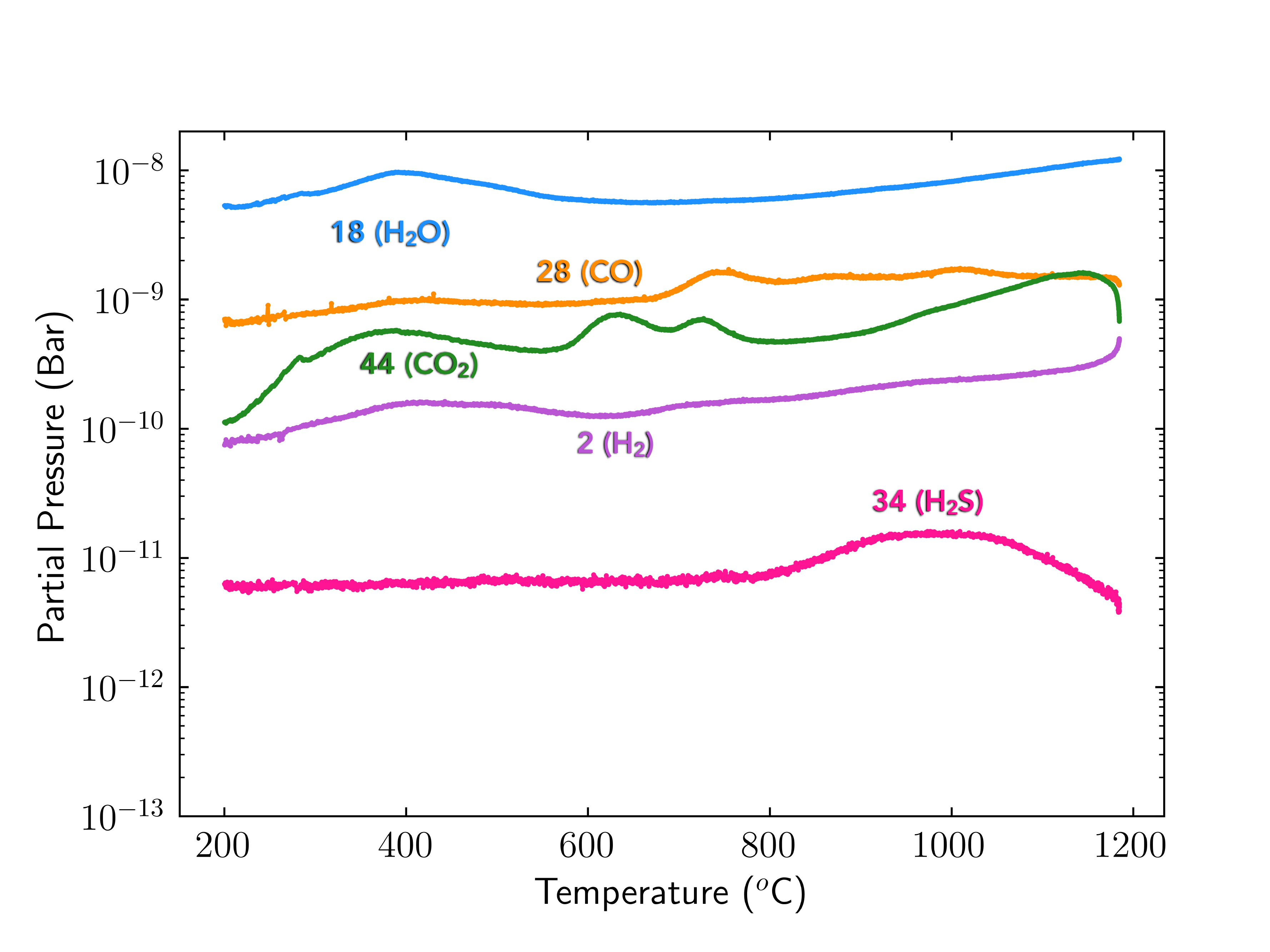}
        \caption{\small Murchison Measurement Corrected for \\Ion Fragments and Atmospheric Adsorption}   
        \label{fig:e1c}
    \end{subfigure}
    \hfill
    \begin{subfigure}{0.45\textwidth}   
        \centering 
        \includegraphics[width=\textwidth]{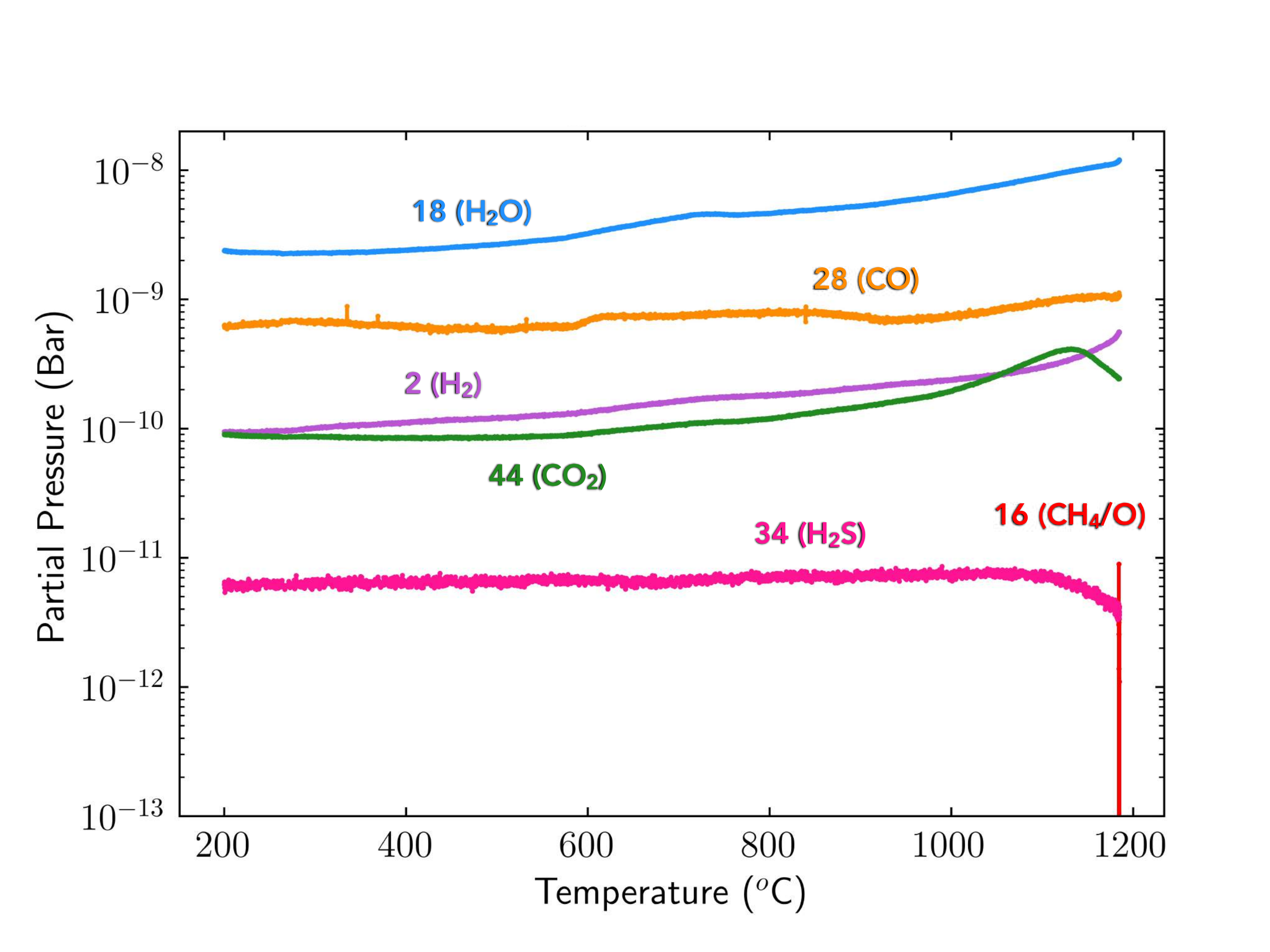}
        \caption{\small Background Measurement Corrected for \\Ion Fragments and Atmospheric Adsorption}   
        \label{fig:e1d}
    \end{subfigure}
    \hfill
    \begin{subfigure}{\textwidth}   
    \centering 
        \includegraphics[width=0.45\textwidth]{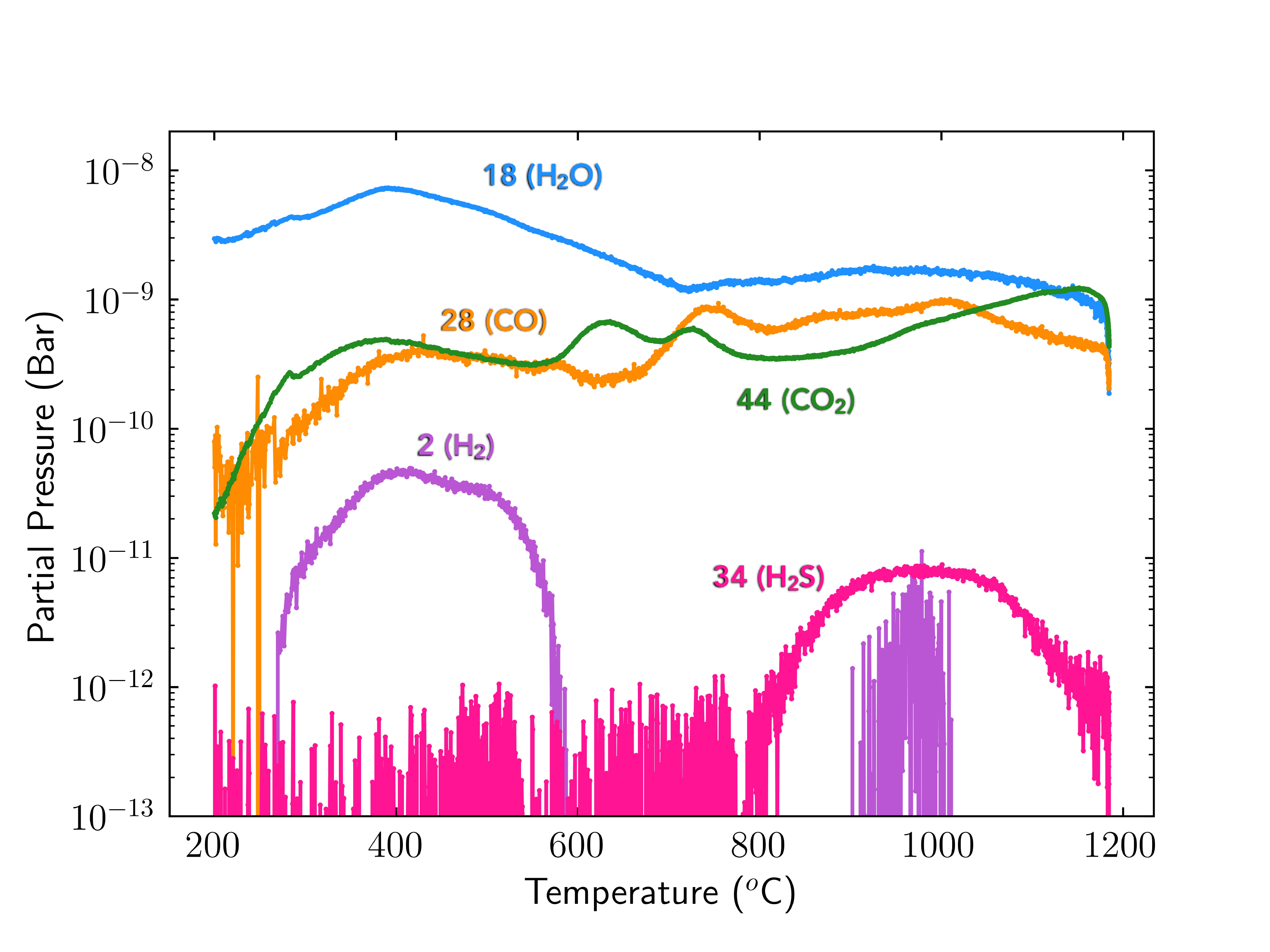}
        \caption{\small Murchison Measurement Corrected for Ion Fragments and Atmospheric Adsorption and Background-Subtracted (i.e., data from (d) subtracted from data from (c)).} 
        \label{fig:e1e}
    \end{subfigure}
\caption{\textbf{Data Calibration Steps}. Each figure illustrates the partial pressures (bars) for the molecular species measured from 200 $^{\circ}$C to 1200 $^{\circ}$C. Each sample's data is calibrated by first correcting for ion fragments and atmospheric adsorption and then background subtracting.}
\label{fig:edimage1}
\end{figure}

\begin{figure}[hbt!]
    \begin{subfigure}{0.5\textwidth}
        \centering
        \includegraphics[width=\textwidth]{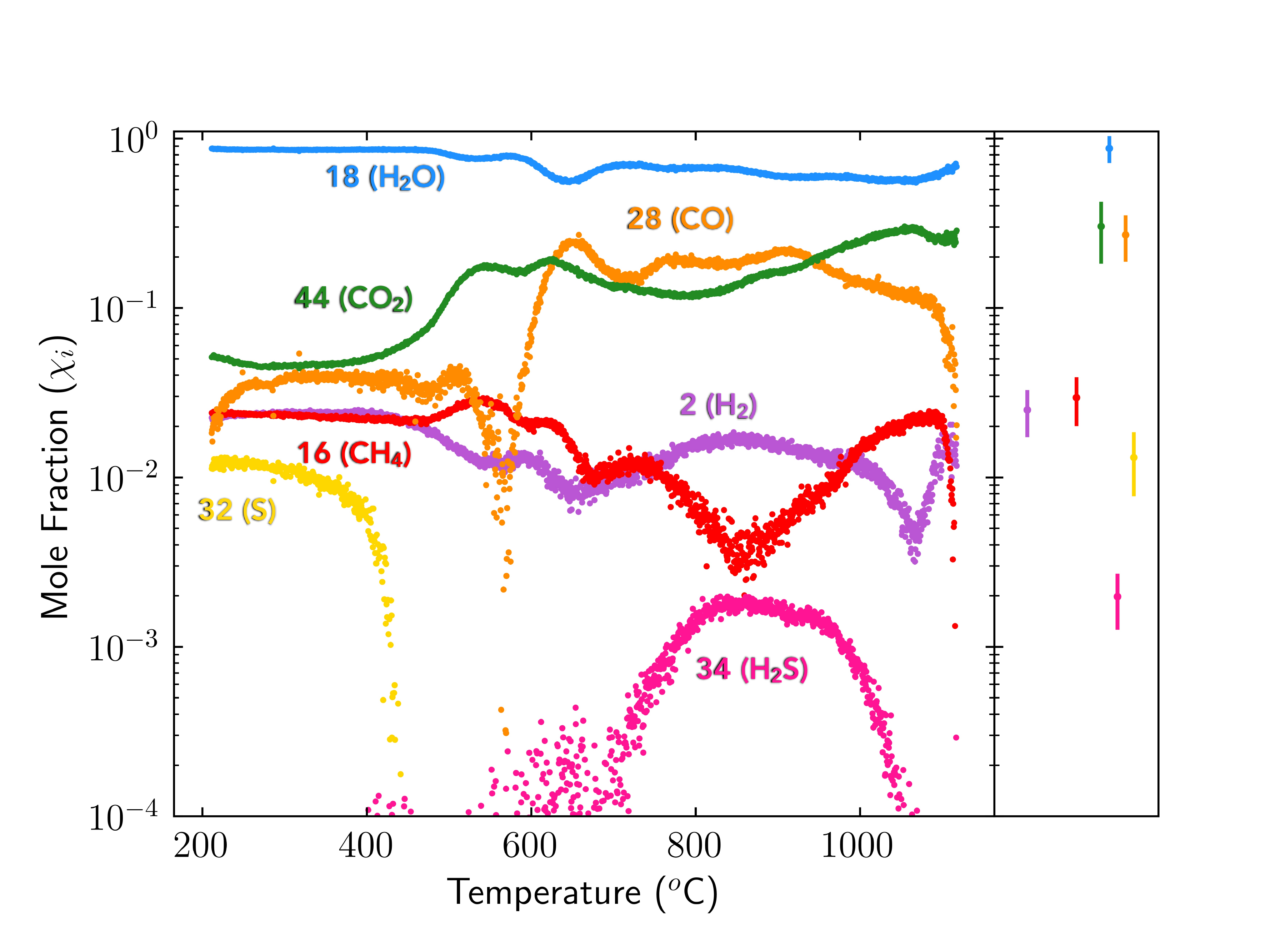}
        \caption{\small Murchison}  
        \label{fig:e4a} 
    \end{subfigure}
    \hfill
    \begin{subfigure}{0.5\textwidth}  
        \centering 
        \includegraphics[width=\textwidth]{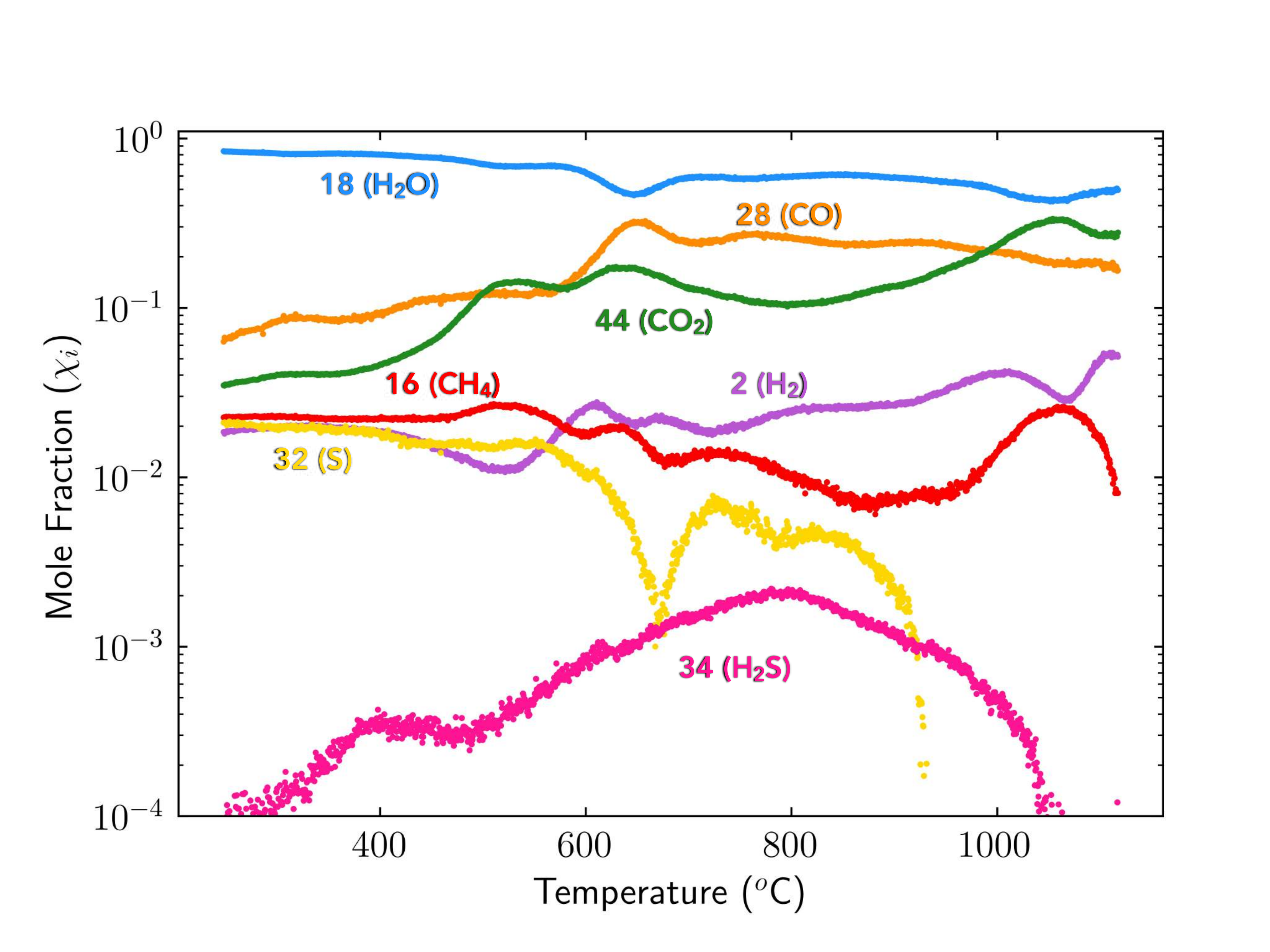}
        \caption{\small Average of 3 CM Chondrites}
        \label{fig:e4b}  
    \end{subfigure}
\caption{\textbf{Results of analyzing ion fragments using a non-linear least squares regression}. The outgassing abundances in (a) are for the Murchison sample with the panel on the right side showing the average standard deviation determined from the Monte Carlo simulation for each of the species measured. The abundances in (b) are the average of the three CM chondrites.}
\label{fig:edimage2}
\end{figure}

\begin{figure}[hbt!]
    \centering
    \includegraphics[width=\textwidth]{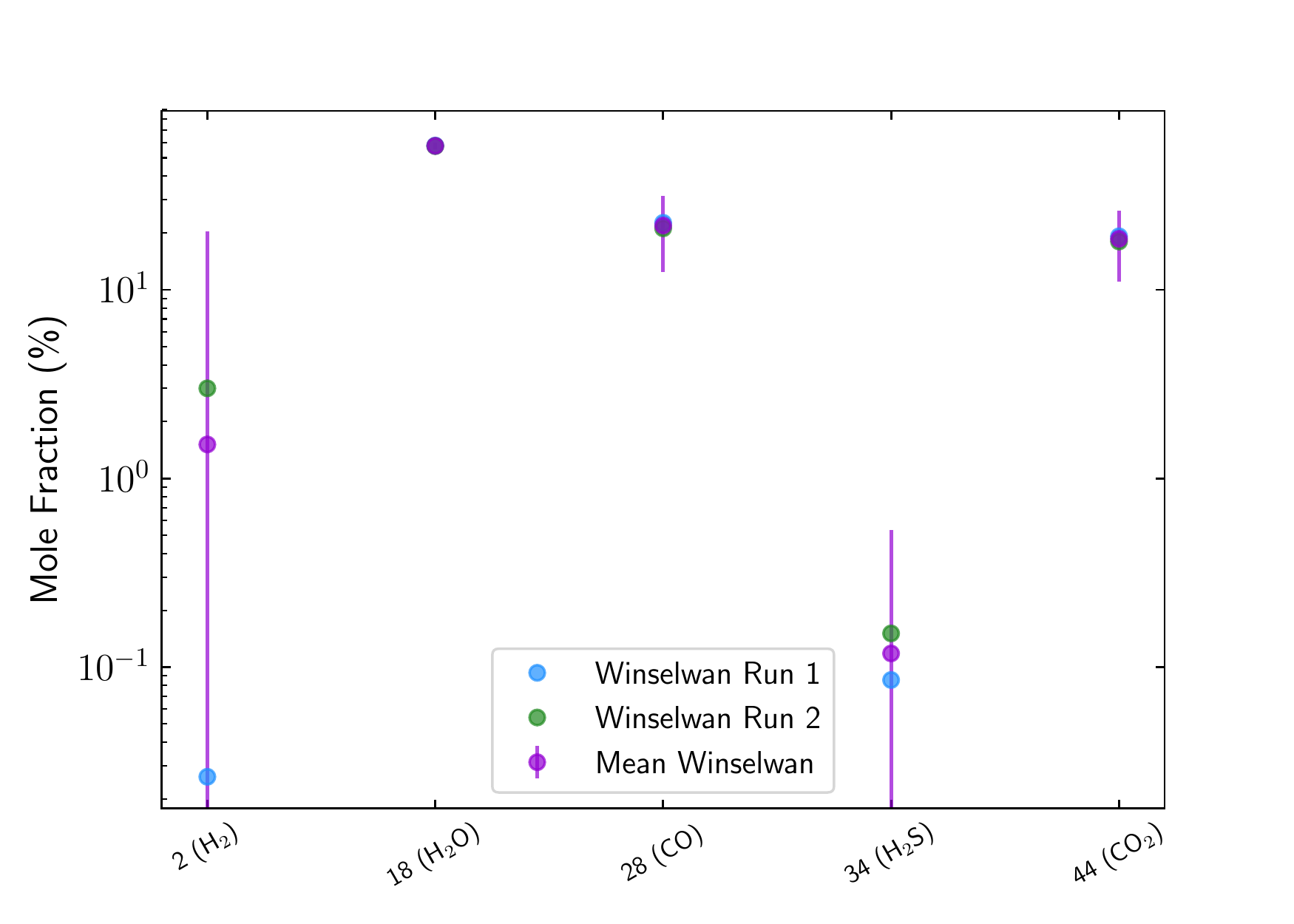}
\caption{\textbf{Comparison between the yields of major volatiles released from Jbilet Winselwan samples during two identical experiments.} The mole fraction summed over temperature for each volatile species is normalized to the total mole fraction of released gases summed over temperature and expressed as a percentage. The uncertainty on the mean relative abundance for each volatile species is the 95\% confidence interval of the mean. The volatile yields are fairly reproducible between the two experiments, especially for the most dominant outgassed species (H\textsubscript{2}O, CO, CO\textsubscript{2}).} 
\label{edimage3}
\end{figure}

\begin{figure}[hbt!]
    \includegraphics[width=\textwidth]{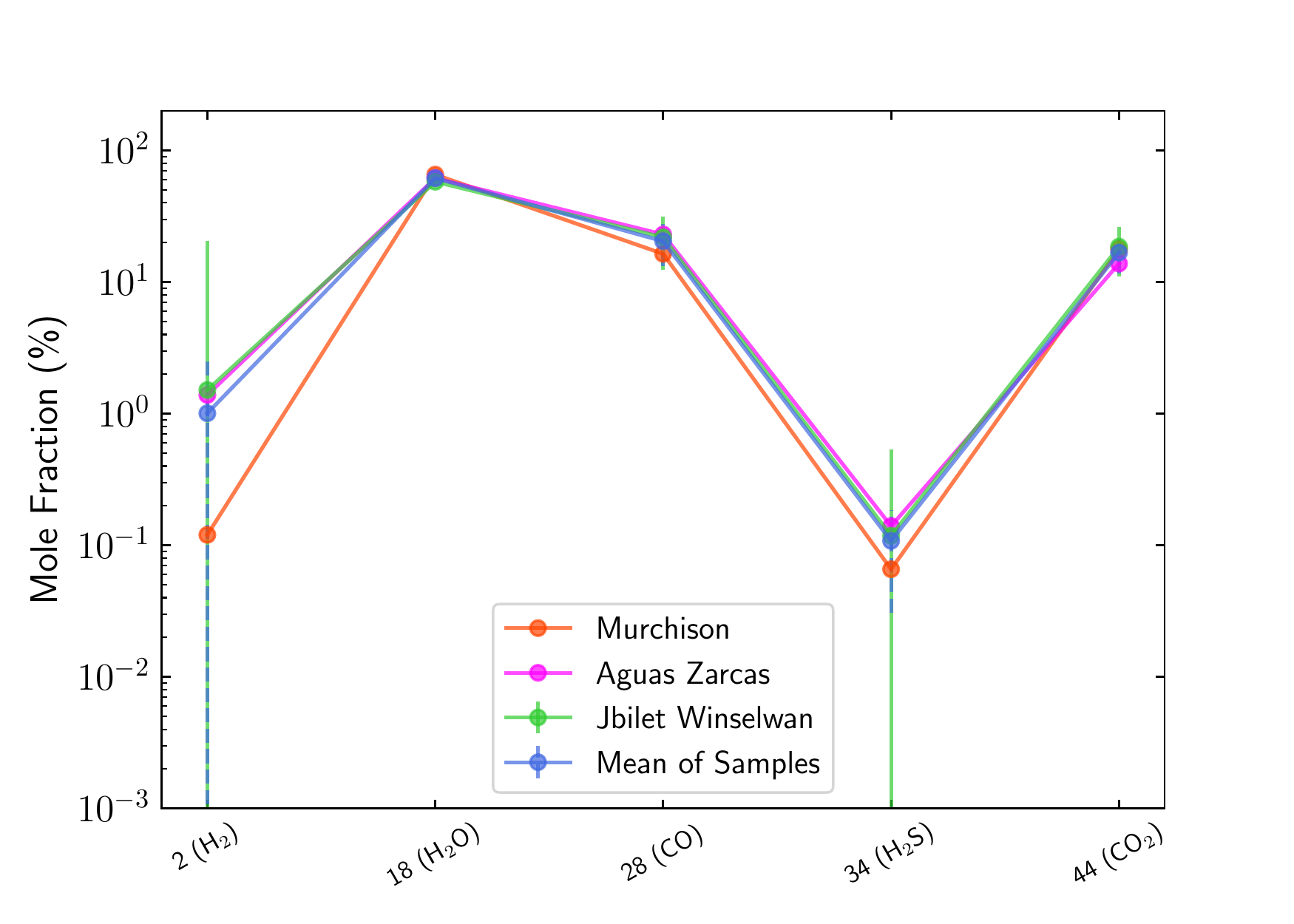}
    \centering
    \caption{\textbf{Comparison between the yields of major volatiles released from the samples.} The mole fraction summed over temperature for each volatile species is normalized to the total mole fraction of released gases summed over temperature and expressed as a percentage. The data for Winselwan is the mean of the two individual experiments conducted with the uncertainty reported as the 95\% confidence interval of the mean (see Methods and Extended Data Figure 3). The mean relative abundance of all three samples for each volatile species is also shown with the uncertainty reported as the 95\% confidence interval of the mean. All three samples have similar outgassing abundances for the most dominant outgassing species (H\textsubscript{2}O, CO, and CO\textsubscript{2}). While H\textsubscript{2} and H\textsubscript{2}S have larger variations up to an order of magnitude, the relative abundances for each species across the three samples are within 2$\sigma$ of each other.}
    \label{fig:edimage4}
\end{figure}

\begin{figure}[hbt!]
    \begin{subfigure}{0.5\textwidth}
        \centering
        \includegraphics[width=\textwidth]{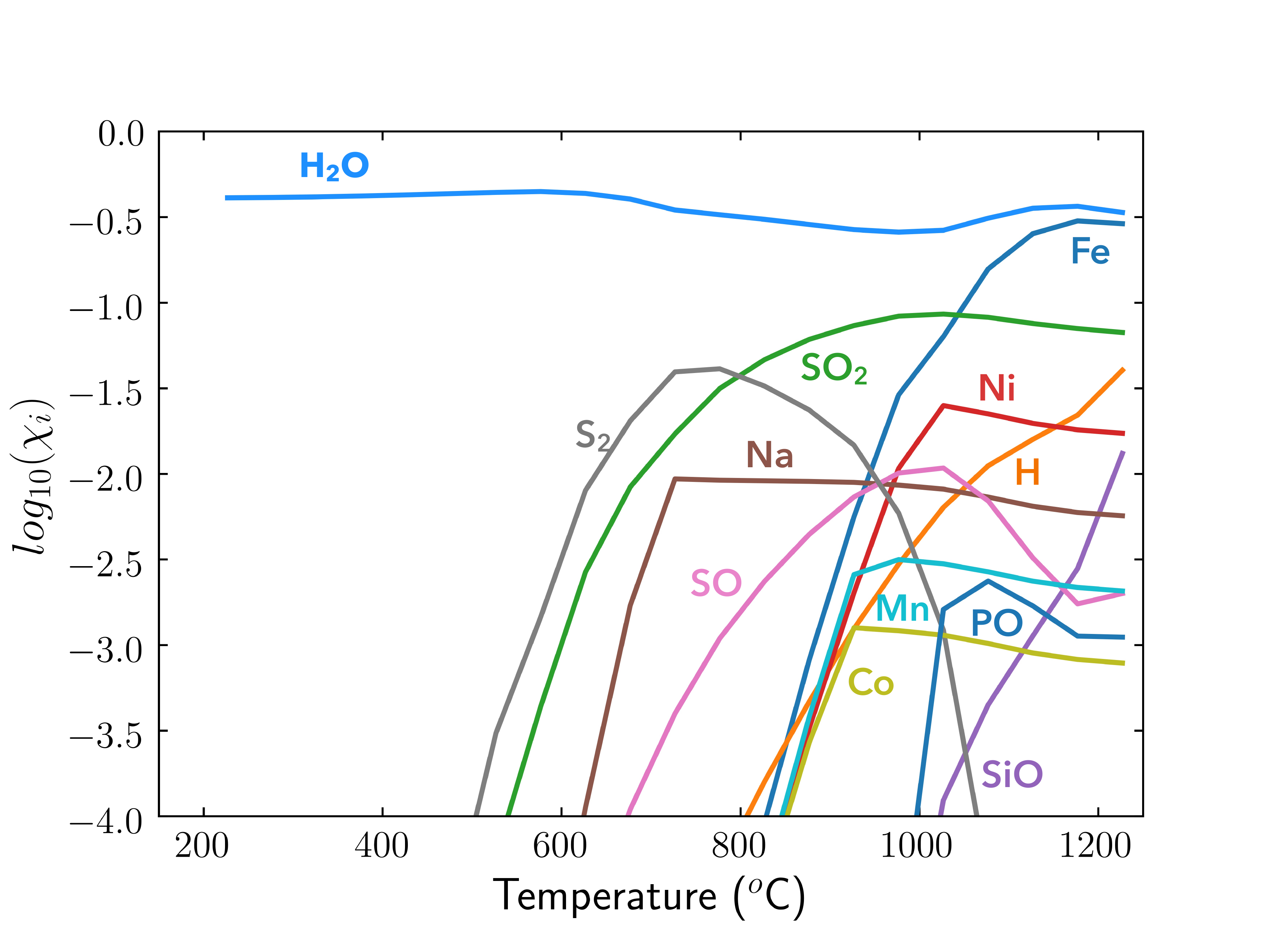}
        \caption{\small Murchison}  
        \label{fig:e5a} 
    \end{subfigure}
    \hfill
    \begin{subfigure}{0.5\textwidth}  
        \centering 
        \includegraphics[width=\textwidth]{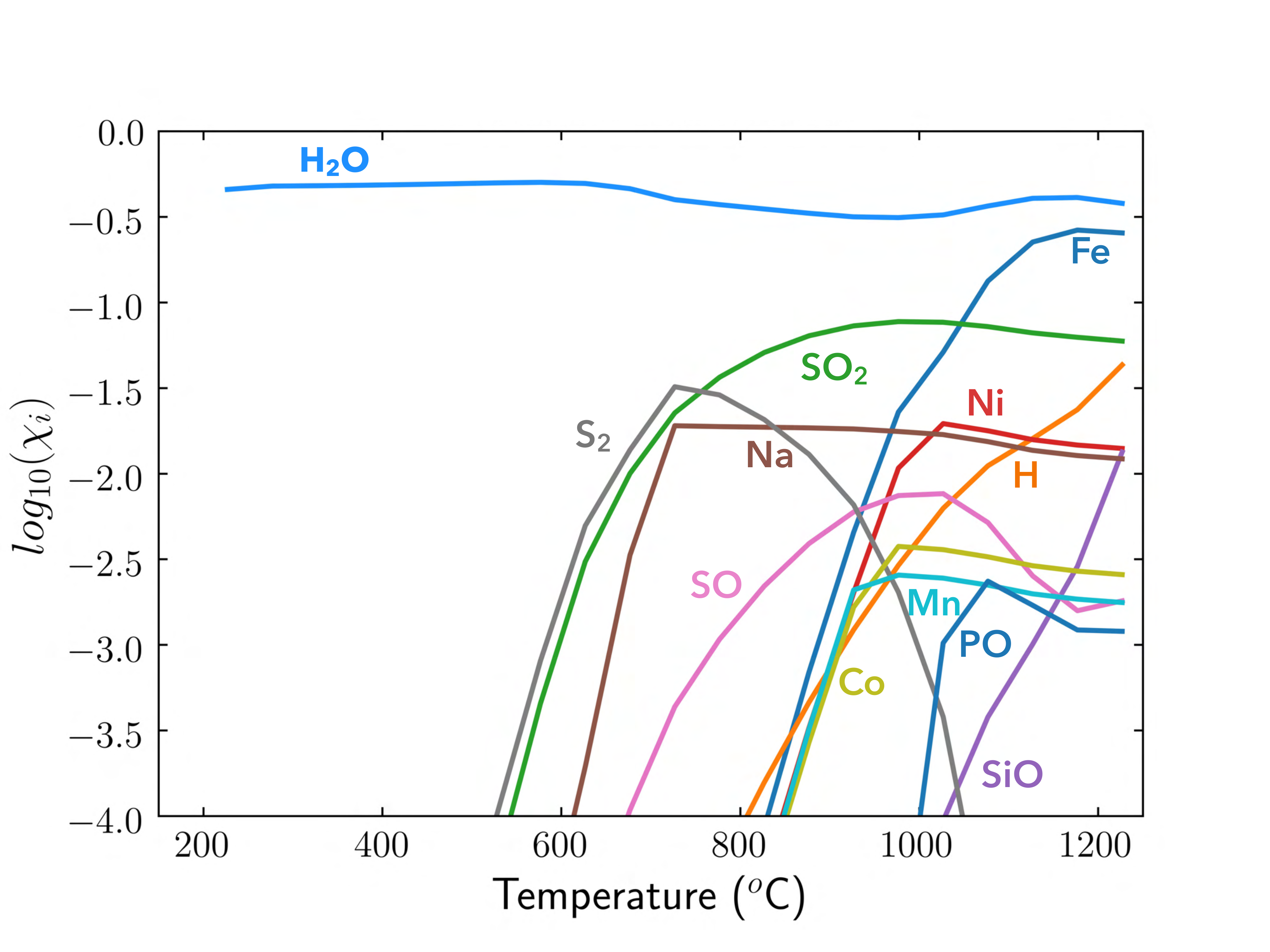}
        \caption{\small Average CM Chondrite Composition}
        \label{fig:e5b}  
    \end{subfigure}
\caption{\textbf{Additional Outgassing Species from Chemical Equilibrium Calculations} Outgassing abundances for additional species not measured in the experiments calculated assuming chemical equilibrium for Murchison (a) and an average CM chondrite bulk composition (b) at 1E-3 Pa. The outgassing of H\textsubscript{2}O is also shown as a reference.}
\label{fig:edimage5}
\end{figure}

\clearpage

\section*{Methods}

\subsection*{Sample Preparation \& Experimental Procedure}

CM chondrite samples were powdered with an agate mortar and pestle and sieved so that only material between 20 and 106 $\mu$m in diameter was analyzed to ensure homogeneous samples. Powdered samples were stored in a desiccator under vacuum to minimize terrestrial contamination. For each heating experiment, $\sim$3 mg of powdered sample was evenly distributed into a 6.5 $\times$ 4.0 mm$^2$ alumina crucible, as shown in Supplementary Figure 1. This sample size was chosen because larger sample sizes saturate the RGA.

Prior to assaying each sample, we first heated an empty small crucible and the larger 50 $\times$ 20 $\times$ 20 mm$^3$ combustion boat in the tube furnace to allow impurities, particles in the tube and adsorbed volatiles to degas which otherwise could interfere with our measurements. The heating procedure to bake-out the tube and sample containers consisted of five steps: (1) heating from room temperature to 200 $^{\circ}$C over 40 minutes, (2) holding at 200 $^{\circ}$C for 30 minutes, (3) heating from 200 $^{\circ}$C to 1200 $^{\circ}$C over 5 hours, (4) holding at 1200 $^{\circ}$C for 5 hours, and (5) cooling the system to room temperature over 5 hours. After one of the bake-outs, we calibrated the background signal by performing a similar heating procedure on empty sample containers except for step (4) in which the time at 1200 $^{\circ}$C is reduced to 10 minutes.  

The RGA mass spectrometer used in this study operates inside the vacuum chamber and ionizes gas molecules according to their molecular masses (up to 100 atomic mass units (amu)) and measures their partial pressures. Since an RGA is commonly used for detecting low-levels of contamination in vacuum systems, its sensitivity to trace amounts of gas makes it ideal for carrying out this study \cite{RGAm}. The experimental procedure for each sample is identical to that used to determine the background signal. We chose the heating rate of 3.3 $^{\circ}$C/min because it is similar to those in prior studies using mass spectrometers to monitor released gases from meteorite samples (e.g., \cite{GoodingMuenow1977}, \cite{Okumura}). We also note that the combination of very low pressure and high vacuum pumping rate precludes many gas-gas reactions or volatile phase changes in the experimental system. At the high-temperature end of the experiments, some gas-gas reaction rates approach or exceed the vacuum pumping rate suggesting that some gas species, but not all, may approach equilibrium. 

\subsection*{Data Calibration: Ion Fragmentation, Terrestrial Atmospheric Adsorption \& Background Subtraction Corrections}

The RGA's ionizer can cause different types of ions to be produced from a single species of gas molecule due to processes such as molecular fragmentation \cite{RGAm}. The mass spectrum of each molecule has contributions from all ion fragments formed from that molecule and they define the molecule's fragmentation pattern. In residual gas analysis, standard fragment patterns of common atoms and molecules are well established.  Supplementary Table 1 explains, for a given gas species whose mass number is analyzed during the experiments, the percentage each of its known ion fragments contributes to the intensity relative to the major peak due to that gas species itself. These fragment patterns were determined from the NIST Mass Spectrometry Data Center \cite{NIST}.  To correct for ion fragments for each species, we subtract its partial pressure from the partial pressures of other species that contribute to its mass signal weighted by the percentage of the other species's contribution (see Equations 4-15). For example, the partial pressure of H\textsubscript{2} is given by:

\begin{equation}
    p_{H_2} = p_{2\text{amu}} - (0.02 * p_{H{_2}O}).
\end{equation}

We also correct for the possibility of terrestrial atmospheric adsorption onto the samples. We assume that the signal at 40 amu is due entirely to atmospheric argon adsorbed onto the samples (see the section on degeneracies below). Given the composition of Earth's atmosphere (78\% N\textsubscript{2}, 21\% O\textsubscript{2}, 1\% Ar), we determine the amount of atmospheric N\textsubscript{2} and O\textsubscript{2}.  After correcting for ion fragments, we subtract the atmospheric N\textsubscript{2} and O\textsubscript{2} contributions from the signals due to N\textsubscript{2} and 32 amu (see Equations 8, 12 and 14). We also subtract the ion fragments of atmospheric N\textsubscript{2} and O\textsubscript{2} from the signals at 14 and 16 amu (see Equations 6 and 9).

The step-wise heating procedure allows us to disentangle terrestrial weathering and contamination from the actual volatile composition of our samples \cite{GradyWright2003}. In the heating procedure, we hold each sample at 200 $^{\circ}$C for 30 minutes which helps eliminate any adsorbed water or nitrogen that is not intrinsic to the sample.  Although we conduct each heating experiment under high-vacuum conditions ($\sim10^{-4}$ Pa), slight contamination may still be possible. Therefore, to properly calibrate the background signal, we conducted an additional experiment following the same procedure used for the empty sample containers (see Extended Data Figure 1). The partial pressures during this background measurement are corrected for ion fragments and terrestrial atmospheric adsorption and then serve as the background pressures which are subtracted from the ion fragment-corrected and atmospheric adsorption-corrected partial pressures during sample heating to determine the fully calibrated (i.e., ion fragment-corrected, atmospheric adsorption-corrected and background-subtracted) partial pressures (i.e., for species i, $p_i$ = $p_{i,\text{heating}}$ - $p_{i,\text{background}}$). The total background pressure averages $\sim$6E-4 Pa, and the dominant background species, H\textsubscript{2}O, has an average partial pressure of $\sim$5E-4 Pa, both of which are $\sim$1.5 times lower than their corresponding sample values. The total pressure of the system at each temperature step is given by:
\begin{equation}
    p_{\text{Total}} = \sum_{i} p_{i,\text{heating}} - \sum_{i} p_{i,\text{background}} 
\end{equation} 

Supplementary Figure 2 shows the variations in total pressure with temperature for the samples. To calculate the mole fraction ($\chi_i$) of a gas species (atomic or molecular) at each temperature step, we divide its background-subtracted and ion fragment-corrected partial pressure by the total pressure, $\chi_i = p_i / p_{\text{Total}}$. For the elemental mole fractions of hydrogen, carbon, oxygen, nitrogen, and sulfur at each temperature step, we sum the mole fraction of each gas species containing the element of interest multiplied by the number of atoms of that element in the species, and divide by a normalization factor. For example, for hydrogen: 
\begin{equation}
    \chi_{\text{H}} = \frac{2 \chi_{H_2O} + 4 \chi_{CH_4} + 2 \chi_{H_2} + 2 \chi_{H_2S}}{\text{Norm}}
\end{equation}
where Norm is the normalization factor that ensures that the elemental mole fractions sum to unity and is given by $\chi_H + \chi_C + \chi_O + \chi_N + \chi_S$.
The reported relative abundance of a given species $i$ is its partial pressure summed over temperature and normalized to the total pressure of the released gases also summed over temperature ($P_{\text{i, Total}} = \sum_T p_{i} / \sum_T p_{\text{Total}}$), and is expressed as a percent (see Extended Data Figure 4). The relative abundance of a given element $j$ is its partial pressure, determined the same way as Equation 3 except using partial pressures instead of mole fractions, summed over temperature and normalized to the sum of the pressures of all elements measured in the experiments also summed over temperature ($P_{\text{j, Total}} = \sum_T p_{j} / \sum_T (\sum_j p_{j})$).

\subsection*{Calculations to Determine Gas Species' Partial Pressures}

Several of the mass numbers analyzed for this study correspond to multiple gas species (e.g., 28 amu corresponds to CO and N\textsubscript{2}). Bulk composition measurements of the samples, measurements of the other masses, and melting/evaporation temperatures for the different species allow us to disentangle which species dominate the signal and, in some cases, distinguish between different gas species' signals that correspond to the same mass number. Equations 4-15 below show the calculations to determine the partial pressures of different volatile species by accounting for ionization fragmentation, disentangling some of the species with overlapping mass numbers and correcting for atmospheric adsorption:

\begin{equation}
    p_{H_2} = p_{\text{2 amu}} - 0.02 p_{H_2O}
\end{equation}
\begin{equation}
    p_{H_2O} = 1.04 p_{\text{18 amu}}
\end{equation}
\begin{equation}
    p_{CH_4} = 1.25 p_{\text{16 amu}} - 0.10 p_{CO_2} - 0.015 p_{H_2O} - 4.96 p_{\text{40 amu}}
\end{equation}
\begin{equation}
    p_{N_{2} \text{, pre-atmosphere correction}} = (p_{\text{14 amu}} - 0.21 p_{CH_4})/0.14
\end{equation}
\begin{equation}
    p_{N_2} = 1.14 p_{N_{2} \text{, pre-atmosphere correction}} - (83.96 p_{\text{40 amu}})
\end{equation}
\begin{equation}
    p_{N} = p_{\text{14 amu}} - 0.21 p_{CH_4} - 0.14 p_{N_{2}}
\end{equation}
\begin{equation}
    p_{CO} = 1.07 (p_{\text{28 amu}} - p_{N_{2} \text{, pre-atmosphere correction}}) - 0.10 p_{CO_2}
\end{equation}
\begin{equation}
    p_C = p_{\text{12 amu}} - 0.09 p_{CO_2} - 0.05 p_{CO} - 0.04 p_{CH_4}
\end{equation}
\begin{equation}
    p_{S} \text{ or } p_{O\textsubscript{2}} = (1.22 p_{\text{32 amu}}) - (22.53 p_{\text{40 amu}})
\end{equation}
\begin{equation}
    p_{H_2S} = 1.45 p_{\text{34 amu}}
\end{equation}
\begin{equation}
    p_{Ar} = p_{\text{40 amu}} \text{ (All due to atmospheric adsorption)}
\end{equation}
\begin{equation}
    p_{CO_2} = 1.29 p_{\text{44 amu}}
\end{equation}

In Equation 6, we account for the fact that the signal at 16 amu can be due to ion fragments of CO\textsubscript{2} and H\textsubscript{2}O. We also account for the fact that contaminated O\textsubscript{2} due to atmospheric adsorption has an ion fragment at 16 amu (see ``16 and 32 amu" section in ``Degeneracies'' section). Although 16 amu can also be due to ion fragments of CO, we do not account for them because 16 amu only contributes 2\% to CO. Equations 7-10 explain how we disentangle the signals due to CO and N\textsubscript{2} given that they have the same mass number (28 amu). In Equation 7, we first account for the fact that the signal at 14 amu can be due to ion fragments of CH\textsubscript{4}. We assume that the remaining signal at 14 amu is due entirely to atomic nitrogen which is an ion fragment of N\textsubscript{2}, and we use it to determine the partial pressure of N\textsubscript{2}. In Equation 8, we account for contaminated N\textsubscript{2} due to atmospheric adsorption by assuming all of the 40 amu signal is due to terrestrial atmospheric argon and using its signal and the known composition of Earth's atmosphere to determine the amount of contaminated N\textsubscript{2}. In Equation 9, we determine the partial pressure of atomic nitrogen which is zero since we assume all of it was an ion fragment of N\textsubscript{2}. In Equation 10, we determine the signal due to CO by subtracting the total amount of N\textsubscript{2} from the signal at 28 amu and also accounting for the fact that this signal can be an ion fragment of CO\textsubscript{2}.  Equation 11 determines the partial pressure of atomic carbon and accounts for the various ion fragments at 12 amu. The signal at 32 amu can be due to either O\textsubscript{2} or atomic sulfur. Equation 12 accounts for the contamination from atmospheric adsorbed O\textsubscript{2}, and once this correction is applied the signal at 32 amu becomes negligible. Although the signal at 32 amu can be an ion fragment of H\textsubscript{2}S, we do not account for it since we are not certain that this signal is due to sulfur or O\textsubscript{2}. Equation 13 determines the partial pressure of H\textsubscript{2}S, and Equation 14 assumes that the signal at 40 amu is entirely due to terrestrial atmospheric argon. Finally, Equation 15 determines the partial pressure of CO\textsubscript{2}. 

Equations 5, 6, 8, 10, 12, 13, and 15 determine the partial pressures of molecules subject to fragmentation.  An additional factor that should be taken into account when correcting for ion fragments is adding fragments back to those species that are subject to molecular fragmentation. While this may cause a slight over-correction, it does not significantly affect the results, as the average difference between the relative abundances summed over temperature with or without adding fragments back in is $\sim$1\%. The differences between the relative abundances summed over temperature with or without adding fragments back in are also within the uncertainties (expressed as 95\% confidence intervals of the means) for each species and atomic abundance. As examples, to determine H\textsubscript{2}O and CH\textsubscript{4}'s partial pressures, we add back the contributions from their fragments:

\begin{equation}
    p_{H_2O} = (1.0p_{\text{18 amu}}) + (0.02p_{\text{18 amu}}) + (0.02p_{\text{18 mu}}) = 1.04p_{\text{18 amu}}
\end{equation}

\begin{equation}
    \begin{aligned}
        p_{CH_4} & = ((1.0p_{\text{16 amu}}) + (0.21p_{\text{16 amu}}) + (0.04p_{\text{16 mu}}))  - (0.10p_{CO_2}) - (0.02p_{H_2O}) - (4.96p_{\text{40 amu}})\\ 
        & = 1.25 p_{\text{16 amu}} - (0.10p_{CO_2}) - (0.02p_{H_2O}) - (4.96p_{\text{40 amu}})
    \end{aligned}
\end{equation}

\subsection*{Reproducibility of Experimental Results}

In order to test the reproducibility of our experiment and to confirm that it precisely measures the outgassed species from various samples, we analyzed samples of Jbilet Winselwan twice under identical conditions. Jbilet Winselwan's final reported relative abundances are given by the mean and the 95\% confidence interval of the mean calculated from a t-distribution of the two trials (see Table 1 in main article). As Extended Data Figure 4 illustrates, the relative abundances of the three most abundant outgassed species, H\textsubscript{2}O, CO and CO\textsubscript{2}, between the two experiments agree with each other within 6 \% with 95\% confidence intervals less than 35\%.  The other species' abundances between the two experiments have variations of up to $\sim$3 \% and 95 \% confidence intervals less than 16 \%.  As Tables 1 and 2 in the main article illustrate, the species with the largest uncertainties are CO\textsubscript{2}, H\textsubscript{2}O, and the total amount of H (95 \% confidence intervals of $\sim$34 \%, 34 \%, and 36 \%, respectively). These large confidence intervals are due to the the small sample size. All other species have confidence intervals less than 18 \% with difference between the two measurements less than 3 \%. \\

\subsection*{Calculating Oxygen Fugacity}
Although our experiments simulate a non-equilibrium open system, we can compare our results to what is expected at equilibrium. For determining the oxygen fugacity of the system as shown in Figure 4, there are several ways to calculate $f_{O_2}$ from the relative abundances of various gas species including ratios of H\textsubscript{2}/H\textsubscript{2}O and  CO/CO\textsubscript{2}. Since our experimental O\textsubscript{2} abundance is zero, we cannot use O\textsubscript{2} alone to compute $f_{O_2}$. Assuming the system is in equilibrium, to calculate $f_{O_2}$ as a function of temperature from H\textsubscript{2} and H\textsubscript{2}O we use Equations 18-20: 

\begin{equation}
    H_{2}O = H_2 + 0.5O_2
\end{equation}

\begin{equation}
    \log_{10}(K_1) = \frac{-12794}{T} + 2.7768
\end{equation}

\begin{equation}
    f_{O_2} = (K_{1} \frac{\chi_{H_{2}O}}{\chi_{H_{2}}})^{2}
\end{equation}

\noindent Similarly, to calculate $f_{O_2}$ from CO and CO\textsubscript{2} we use Equations 21-23: 

\begin{equation}
    CO\textsubscript{2} = CO + 0.5O_2
\end{equation}

\begin{equation}
    \log_{10}(K_2) = \frac{-14787}{T} + 4.5472
\end{equation}

\begin{equation}
    f_{O_2} = (K_{2} \frac{\chi_{CO_2}}{\chi_{CO}})^{2}
\end{equation}

\noindent $K_1$ and $K_2$ are the equilibrium constants that are functions of temperature $T$ (in Kelvin) and taken from the IVTANTHERMO database (see \cite{SchaeferFegley2017} for details). Figure 4 in the article shows $f_{O_2}$ calculated from chemical equilibrium (black curve) and $f_{O_2}$ determined using H\textsubscript{2}/H\textsubscript{2}O and CO/CO\textsubscript{2} from our experiments (blue and orange curves, respectively). The $f_{O_2}$ are plotted relative to the quartz-fayalite-magnetite (QFM) mineral buffer \cite{Fegley2013}. Under equilibrium conditions, we expect the $f_{O_2}$ values calculated from H$_2$O/H$_2$ and CO$_2$/CO to match, but we do not find this with our experimental data. The fact that the H$_2$O/H$_2$ trend is larger than the theoretical $f_{O_2}$ suggests that the abundance of H$_2$O compared to H$_2$ is larger than that predicted under chemical equilibrium conditions. This larger abundance may be due to kinetic barriers that result in hydrogen initially outgassing as H$_2$O, but our experiments do not allow sufficient time for reactions to take place that would transform it to H$_2$.  

\subsection*{Comparison with Model Assumptions} 

A key distinction between our experimental results and equilibrium model calculations is that our experiments simulate \textit{initial} (or instantaneous) outgassing compositions, not the long-term outgassing abundances once equilibrium has been achieved. Also, in our experiments the meteorite composition changes as the temperature increases and volatiles are removed whereas the equilibrium calculations assume a closed system in which the volatiles are not removed.  Nevertheless, the preliminary outgassed abundances determined experimentally may have important implications for the subsequent evolution of outgassed atmospheres that eventually achieve chemical equilibrium, as initial outgassed species' abundances control what is available to subsequently evolve within an atmosphere. For example, our experiments find that H\textsubscript{2}S outgasses at higher temperatures than predicted in equilibrium models, which means that if a planet does not reach 900-1000 $^{\circ}$C, H\textsubscript{2}S may not have a significant atmospheric abundance. 

The experimental results for this work are the \textit{instantaneous} outgassing compositions because those are the more appropriate ones to compare to thermochemical equilibrium models rather than the \textit{cumulative} outgassing compositions. Both the instantaneous measurements and equilibrium model results represent contained assemblages, although the composition in the experiments is evolving. On the other hand, the \textit{cumulative} outgassed abundances do not represent such an assemblage, since the gases are removed at each measurement and do not react with material outgassed at higher temperatures. However, the cumulative outgassed compositions are a useful way to determine the extent to which volatiles have been released, so we have calculated the cumulative outgassing composition. Supplementary Figure 4 illustrates the cumulative outgassing compositions as a function of temperature for each of the three chondrite samples as well as the average of the three samples.  We see a leveling-off behavior at high temperatures for nearly all of the outgassing species across each of the three CM chondrite samples. Some of this leveling-off behavior is due to the fact that a volatile’s outgassing is decreasing (e.g., H\textsubscript{2}S), but there could still be other gas species being produced that we do not track in these experiments. Compared to the other volatiles measured, CO\textsubscript{2} and H\textsubscript{2}'s cumulative outgassing trends do not level off as significantly at the higher temperatures.  

\subsection*{Least Squares Regression Technique for Ion Fragments and Species Degeneracies}

Our chosen technique to correct for ion fragmentation and, when possible, break the degeneracies between volatile species that have the same mass number involves making logical assumptions regarding which gas species likely dominates a given mass signal and using a set of arithmetic corrections (see ``Calculations to Determine Gas Species' Partial Pressures" section above). However, a non-linear least squares regression is an alternate method to account for ion fragments and disentangle gas species with overlapping mass numbers. This technique involves performing a least squares regression on the normalized mass spectrum library (Supplementary Table 1, column 4) and constraining the outgassed abundances to be positive and less than appropriate upper bounds. For all of the species, the upper bounds are twice the maximum value of that species mass number. Extended Data Figure 2 shows the fully-calibrated outgassing abundances calculated using this method and Supplementary Table 3 compares the average partial pressures for each gas species calculated using our original analysis to those from the least squares technique. In order to calculate the uncertainties on the fitted parameters from the least squares analysis, we ran a Monte Carlo simulation on a sparser data array. The least squares analysis results are consistent with the original results within 2$\sigma$ for most species.  While the results for the most dominant outgassing species (i.e., H\textsubscript{2}O, CO, CO\textsubscript{2}) are similar to our original analysis, the least squares result finds non-negligible amounts of methane and atomic sulfur. Outgassing of these species is not predicted from chemical equilibrium calculations, which suggests that this technique may not fully account for ion fragments compared to our original calculations or, in the case of sulfur, they are leftover fragments from species we are not currently measuring in our experiments (e.g., S\textsubscript{2}, SO\textsubscript{2}, etc.).\\

\subsection*{Degeneracies between Gas Species and Mass Numbers}

For mass numbers that could correspond to multiple volatile species, we describe in the following subsections additional details on how we either determined which species dominates the signal or disentangled multiple species' signals.\\

\noindent \textbf{\underline{16 and 32 amu:}} 16 amu is the molecular weight of CH\textsubscript{4} and the atomic weight of oxygen. As shown in Supplementary Table 1, mass number 16 is affected by ionization fragmentation, being fragments of CO\textsubscript{2}, CO, H\textsubscript{2}O and O\textsubscript{2}. Ion fragments at 16 amu contribute to 22\% of O\textsubscript{2}'s signal (assuming the signal at 16 amu is due to atomic oxygen), 9.8\% of CO\textsubscript{2}'s signal, 2.2\% of CO's signal, and 1.5\% of H\textsubscript{2}O's signal. We assume that the majority of the signal at 16 amu outgassing from the samples is due to CH\textsubscript{4} because prior meteorite ablation studies have detected small amounts CH\textsubscript{4} from carbonaceous chondrites and atomic oxygen is not expected to outgas significantly \cite{CourtSephton2009_2}. 32 amu is the atomic weight of sulfur and the molecular weight of O\textsubscript{2} and methanol (CH\textsubscript{3}OH). If mass number 32 amu is due to atomic sulfur, then it can be an ion fragment of hydrogen sulfide (H\textsubscript{2}S), contributing to 45\% of H\textsubscript{2}S's signal. According to theoretical calculations, atomic sulfur is not predicted to outgas significantly and O\textsubscript{2} is only predicted to begin outgassing around 1100 $^{\circ}$C.

As indicated below, our data does not allow a definitive determination of which species dominate the signals at 16 and 32 amu. For the 32 amu signal, we correct for the possibility of atmospheric adsorption of O\textsubscript{2} onto the samples by assuming the 40 amu signal is entirely due to atmospheric argon and using the known ratio of O\textsubscript{2}/Ar in Earth's atmosphere to subtract the atmospheric contribution to the signal at 32 amu (Equation 12). After correcting for atmospheric adsorption, we do not detect a significant outgassing signal at 32 amu for any of the samples. The signal at 32 amu is likely not due to methanol because it is not predicted to outgas significantly across the entire temperature range.

For the 16 amu signal, even after correcting for ion fragments of CO\textsubscript{2} and H\textsubscript{2}O, the signal is still significant at lower temperatures (up to $\sim$600 $^{\circ}$C) for all three samples. However, since O\textsubscript{2} has an ion fragment at 16 amu, we also have to account for the possibility of atmospheric adsorption of O\textsubscript{2}'s ion fragment at 16 amu so we subtract 22 \% of the atmospheric adsorbed O\textsubscript{2} from the 16 amu signal. When we apply this correction, the signal at 16 amu becomes negligible across all temperatures for all samples. If the signal at 32 amu is predominantly due to sulfur not O\textsubscript{2}, then the signal at 16 amu would be significant and likely due to CH\textsubscript{4}. However, because we cannot definitively resolve which species dominate the signals at 32 and 16 amu, we must conservatively apply the atmospheric adsorption correction for both masses. In order to disentangle CO and N\textsubscript{2}, as described in the next subsection, we assume the ion fragment of CO at 16 amu is negligible, which is reasonable because it only contributes to 2 \% of CO's signal.

If we assume the signal at 16 amu is due entirely to atomic oxygen resulting from ion fragments of O\textsubscript{2}, we can use it to disentangle the abundances of sulfur and O\textsubscript{2}. To calculate O\textsubscript{2}'s signal from the 16 amu signal, we first correct for the fact that 16 amu can contribute to ion fragments of CO\textsubscript{2} and H\textsubscript{2}O (we assume the ion fragment of CO at 16 amu is negligible), and then use the remaining signal to calculate the abundance of O\textsubscript{2}, knowing that 16 amu contributes to 22\% of O\textsubscript{2}'s mass spectrum (Equation 24). We then calculate the signal due to sulfur, by subtracting O\textsubscript{2}'s signal from the 32 amu signal and correcting for ion fragmentation of H\textsubscript{2}S (Equation 25). Finally, we correct for atmospheric adsorption of O\textsubscript{2} (Equation 26). The resulting signal due to 16 amu is given by Equation 27. Once we disentangle the signals from O\textsubscript{2} and sulfur, we find that sulfur and O\textsubscript{2} abundances are negligible at all temperatures including higher temperatures where predictions indicate that O\textsubscript{2} should begin outgassing (Supplementary Figure 3). In addition when the sulfur and O\textsubscript{2} components are separated, the abundance at 16 amu is also negligible. Ultimately, further work is required to determine which species dominate the signals at 16 and 32 amu.

\begin{equation}
    p_{O_{2} \text{, pre-atmosphere correction}} = 1.22((1.25 p_{\text{16 amu}} - 0.10 p_{CO_2} - 0.02 p_{H_2O})/0.22)
\end{equation}

\begin{equation}
    p_{S} = p_{\textsubscript{32 amu}} - p_{O_{2} \text{, pre-atmosphere correction}} - 0.45 p_{H_2S}
\end{equation}

\begin{equation}
    p_{O_2} = p_{O_{2} \text{, pre-atmosphere correction}} - (22.53 p_{40 \text{amu}})
\end{equation}

\begin{equation}
    p_{CH_4} = 1.25 p_{\text{16 amu}} - 0.10 p_{CO_2} - 0.02 p_{H_2O} - (0.22 p_{O_{2}})
\end{equation}

\noindent \textbf{\underline{28 amu:}} This is the molecular weight of CO, N\textsubscript{2} and ethylene (C\textsubscript{2}H\textsubscript{4}). Mass number 28 amu can also be an ion fragment of CO\textsubscript{2}, contributing to 10\% of CO\textsubscript{2}'s signal (Supplementary Table 1). To disentangle the signals due to CO and N\textsubscript{2}, we assume the signal at 14 amu is predominantly due to ion fragments of N\textsubscript{2} which is valid because atomic nitrogen is not expected to outgas. We correct for the fact that 14 amu is also an ion fragment of CH\textsubscript{4}, and then we use the resulting signal at 14 amu to calculate the signal due to N\textsubscript{2}, knowing that 14 amu contributes to 14 \% of N\textsubscript{2}'s mass spectrum (Equation 7). We then determine the signal due to CO by subtracting the N\textsubscript{2} signal from the 28 amu signal and correcting for ion fragmentation of CO\textsubscript{2} (Equation 10). We correct for atmospheric adsorption of N\textsubscript{2} by assuming the signal at 40 amu is entirely due to atmospheric argon and using the known ratio of N\textsubscript{2}/Ar in Earth's atmosphere to subtract the atmospheric contribution to N\textsubscript{2} (Equation 8). Even after disentangling the signal at 28 amu into the contributions from N\textsubscript{2} and CO and correcting for the effects of ionization fragmentation, the abundance of CO is still very high, being the second most abundant species (Figure 1 and Extended Data Figure 4). After correcting for atmospheric adsorption, N\textsubscript{2}'s outgassed abundance is not significant for any of the three samples. Since CM chondrites have a higher bulk abundance of oxygen (432 mg/g) compared to hydrogen (14 mg/g) while prior theoretical and experimental studies do not predict significant amounts of C\textsubscript{2}H\textsubscript{4} to outgas, the 28 amu signal is more likely to be CO than C\textsubscript{2}H\textsubscript{4} (see Supplementary Table 2). Further investigation is required to definitively rule out C\textsubscript{2}H\textsubscript{4} contributing to the signal at 28 amu, so our experimental results should be considered an upper limit on the CO abundances.  As Figure 3 illustrates, this result agrees fairly well with chemical equilibrium calculations. The fact that the oxygen fugacity calculated from the abundance ratios of CO\textsubscript{2} and CO is lower than $f_{O_2}$ under chemical equilibrium at lower temperatures suggests that there is more CO than CO\textsubscript{2} in our experiments than would be expected if at equilibrium. Once our experiment reached higher temperatures ($\sim$900 $^{\circ}$C), the experimental $f_{O_2}$ determined by CO\textsubscript{2}/CO matches the theoretical chemical equilibrium value.

\noindent \underline{\textbf{40 amu:}} This is the atomic weight of argon and the molecular weight of sodium hydroxide (NaOH), potassium hydride (KH), and methyl cyanide (CH\textsubscript{3}CN). Mass number 40 should not contribute to the signals of any other gas species due to the ionization fragmentation process. In terms of the average bulk composition of CM chondrites, oxygen has the highest abundance (432 mg/g) followed by carbon (22 mg/g), hydrogen (14 mg/g) and finally sodium (4.1 mg/g), nitrogen (1.52 mg/g) and potassium (0.4 mg/g) (Supplementary Table 2). Although Argon has an even smaller bulk abundance than these species, it is relatively abundant in Earth's atmosphere ([Ar]/[O\textsubscript{2}] for air is 0.05). Atmospheric $^{40}$Ar is known to contaminate prior meteorite experiments (e.g., \cite{Huss1996}). In addition, NaOH, KH, and CH\textsubscript{3}CN are not predicted to outgas significantly from CM chondrites at these temperatures. Therefore, we conclude that the 40 amu signal is due to atmospheric Ar. As described earlier, we use this signal to determine the atmospheric contributions of N\textsubscript{2} and O\textsubscript{2}. Future investigation is required to determine if any of the 40 amu signal is due to outgassing from the samples rather than atmospheric adsorption of Ar.    

\subsection*{Solid Phases}

Twenty four solid phases are stable in the theoretical equilibrium calculations: CaAl\textsubscript{2}Si\textsubscript{2}O\textsubscript{8}, Mg\textsubscript{2}SiO\textsubscript{4}, MgCaSi\textsubscript{2}O\textsubscript{6}, MgSiO\textsubscript{3}, MgTiO\textsubscript{3}, FeCr\textsubscript{2}O\textsubscript{4}, FeTiO\textsubscript{3}, FeSiO\textsubscript{3}, Fe\textsubscript{2}SiO\textsubscript{4}, Ca\textsubscript{3}(PO\textsubscript{4})\textsubscript{2}, Ni, Co, Fe\textsubscript{0.947}O, Mn\textsubscript{2}SiO\textsubscript{4}, MgAl\textsubscript{2}O\textsubscript{4}, Ni\textsubscript{3}S\textsubscript{2}, FeS, KAlSi\textsubscript{3}O\textsubscript{8}, NaAlSiO\textsubscript{4}, Ca\textsubscript{5}P\textsubscript{3}O\textsubscript{12}F, Co\textsubscript{9}S\textsubscript{8}, Na\textsubscript{8}Al\textsubscript{6}Si\textsubscript{6}O\textsubscript{24}C\textsubscript{l2}, Fe\textsubscript{3}O\textsubscript{4}, NaAlSi\textsubscript{3}O\textsubscript{8}.  While many of these phases are only stable over a narrow temperature range, the phases that were stable across nearly the entire temperature range include Mg\textsubscript{2}SiO\textsubscript{4}, MgCaSi\textsubscript{2}O\textsubscript{6}, FeCr\textsubscript{2}O\textsubscript{4}, FeTiO\textsubscript{3}, Fe\textsubscript{2}SiO\textsubscript{4}, Ca\textsubscript{3}(PO\textsubscript{4})\textsubscript{2}. 

Preliminary X-ray diffraction (XRD) analyses were performed on the sample residues and unheated samples. For each XRD measurement, the $\sim$3 mg sample was spread in a thin layer over a silicon sample holder and continuously rotated 360$^{\circ}$ for two hours while data was collected, covering angles 0 to 70$^{\circ}$. Comparing the solid phases from the equilibrium calculations to what we detect in the samples from our preliminary XRD analysis, we find that almost all of these phases may be present in the unheated samples and the post-heated residues but, for the post-heated residues, most of the phases have reduced signals despite the unheated and post-heated sample masses being nearly the same. Notable exceptions include Ca\textsubscript{3}(PO\textsubscript{4})\textsubscript{2} and Co which were not definitively detected in the unheated samples and post-heated residues, and CaAl\textsubscript{2}Si\textsubscript{2}O\textsubscript{8} and Na\textsubscript{8}Al\textsubscript{6}Si\textsubscript{6}O\textsubscript{24}C\textsubscript{l2} which were not detected in most of the post-heated residues. For example, troilite (FeS) is present in the unheated samples but has a much weaker signal in the post-heated residues, matching the equilibrium calculations that have FeS being a stable phase up until $\sim$775 $^{\circ}$C. Our XRD analysis suggests that gypsum (CaSO\textsubscript{4}(H\textsubscript{2}O)\textsubscript{2}) may be breaking down during the experiments.  However, in the equilibrium calculations, gypsum is never stable, and this difference may be due to an issue with the data for gypsum that is used in the equilibrium models or uncertainties in the bulk composition used for the equilibrium calculations. Further XRD analyses are required to confirm these preliminary results.

\subsection*{Outgassed Gas Species' Masses}

The average molar mass of the mixture across 200 to 1200 $^{\circ}$C of volatile species $i$ is determined by the equation: $\bar{M} = \sum_{i} M_i * \sum_T \chi_{i}$, where $M_i$ is the molar mass of species $i$. To calculate the mass fraction of a species, $w_i$: 
\begin{equation}
    w_i = 100*((\sum_T \chi_{i}) * M_i) / \bar{M}.
\end{equation}
To determine the outgassed mass of a certain element or species ($\text{Mass}_{i}$), the mass fraction is multiplied by the total outgassed mass ($\text{Mass}_{\text{Total}}$) which is determined by measuring the mass change of the sample before and after heating:
\begin{equation}
    \text{Mass}_{i} = w_i \times \text{Mass}_{\text{Total}}.
\end{equation}

Each sample was weighed before and after heating to determine the mass loss of each volatile species as a result of outgassing.  The total gas released during the experiments based on mass loss measurements is similar between Murchison and Winselwan but higher for Aguas Zarcas. For all three chondrites, the mass released is mostly in H\textsubscript{2}O, CO, and CO\textsubscript{2}. Comparing the initial bulk abundance of an element for CM chondrites to the outgassed abundance informs the degree to which the samples have outgassed relative to complete vaporization of the samples (Table 2 in main article). On average, the samples have higher outgassed abundances of hydrogen and carbon but lower outgassed abundances of oxygen, nitrogen and sulfur compared to the initial bulk abundances. These differences between the initial bulk abundances for an average CM chondrite composition and the outgassed abundances suggests that the samples have not outgassed fully relative to complete vaporization and could also reflect heterogeneities in the meteorite samples themselves.

\subsection*{Comparison with Prior Studies}

Planetary outgassing has been modeled both for the Solar System's terrestrial planets and for some low-mass exoplanets. For instance, studies find that Earth's early degassing produced a steam atmosphere during planetary accretion and a reducing atmosphere of H\textsubscript{2} and/or CH\textsubscript{4} near the end of accretion (e.g., \cite{Zahnle1988, LangeAhrens1982, AbeMatsui1985, Hasimoto2007}).  The major factors controlling speciation during Earth's early degassing included the water content of accreting planetesimals as well as temperature and pressure conditions during the atmosphere's degassing history. For the Zahnle et al.\ 1988 model of Earth's steam atmosphere during accretion, water is the only atmospheric species considered, while the Hashimoto et al.\ 2007 model of Earth's reducing atmosphere assumed accretion of only a specific type of chondritic material with varying amounts of water (\cite{Zahnle1988, Hasimoto2007}). Outgassing models for a planet's magma ocean phase suggest that the degassed atmospheric composition depends on the concentration of volatiles in the accreted body and the pressure at which degassing occurs \cite{Lammer2018, Gaillard2014}. Many of these studies assume chemical equilibrium conditions and lack experimental data to validate some of their assumptions. For example, Gaillard \& Scaillet 2014 considered outgassed species composed of a limited set of elements (H, C, O, S, Fe) to investigate volcanic outgassing of basaltic material on planetary atmospheres. However, they do not include other potentially important elements (e.g., F, Na, Cl, K) nor do they apply experimental data to validate using a simplified set of elements.  Finally, prior research used chemical equilibrium calculations assuming meteorite abundances to determine planet atmospheric compositions \cite{SchaeferFegley2007, SchaeferFegly2010, Lupu2014}. These planetary outgassing models have been applied to low-mass exoplanet atmosphere studies to help interpret current observational data (e.g., \cite{MbarekKempton2016, Dorn2018}).

As noted above, there is limited experimental data to inform these theoretical outgassing models and, in particular, none to fully inform meteorite outgassing work. Prior meteorite heating experiments have used a variety of instrumental techniques including mass spectrometry, infrared spectroscopy and shock devolatilization (e.g., \cite{CourtSephton2009, GoodingMuenow1977, Muenow1995, LangeAhrens1982, Tyburczy1986b, Gerasimov1998, Ikramuddin1977a, Burgess1991, Springmann2019}). However, studies that heated meteorites were limited in several key ways due to restrictions in the number and type of samples used, the temperatures to which the samples where heated, and the number of gas species that were accurately measured. For example, some prior studies focused on the contribution from meteorites on impact-induced atmosphere formation which often involved shocking samples prior to analyzing their volatile contents, and therefore do not properly simulate conditions expected for outgassing from a planet (e.g., \cite{CourtSephton2009, LangeAhrens1982, Tyburczy1986b}). In addition, these experiments only measured a small subset of volatile species, namely H\textsubscript{2}O and CO\textsubscript{2} (\cite{CourtSephton2009, Tyburczy1986b}). It is important to note that the prior studies that focused on shock-induced devolatilization experiments did not continuously monitor the composition of degassed species and focused on higher pressures ($10^{-4}-10^4$ bars) than those in our experiments ($\sim10^{-8}$ bars). These prior works cannot be quantitatively compared to theoretical outgassing models because they either monitored the evolving composition of only a few gas species as a function of temperature or instead inferred loss of volatiles by comparing samples before and after heating. Other studies focused on trace metals (e.g., Co, Zn, In) and moderately volatile and volatile elements (e.g., Se, Ga, As), which are not major constituents of the atmospheres of temperate rocky planets (\cite{Springmann2019, Ikramuddin1977a}). As a result, prior studies are unsuitable for validating outgassing models for low-mass planets.  To fill this gap in the understanding of meteorite outgassing compositions, we designed an experimental procedure to analyze the abundances of a wide range of degassed components: H\textsubscript{2}, C, N, CH\textsubscript{4}/O, H\textsubscript{2}O, CO, N\textsubscript{2}, S/O\textsubscript{2}, H\textsubscript{2}S, Ar, CO\textsubscript{2}, which informs the initial compositions of outgassed atmospheres assuming the outgassing material is CM chondrite-like. 

Comparing our results to other prior meteorite heating experiments, we find that our detection of significant outgassing of H\textsubscript{2}S from Murchison (beginning at 800 $^{\circ}$C) is consistent with the stepped combustion experiments of Murchison from Burgess et al.\ 1991 that found the highest outgassing yield of sulfur occurring at 800 $^{\circ}$C \cite{Burgess1991}. Court \& Sephton 2009 rapidly heated CM2 chondrites to 1000 $^{\circ}$C and using FTIR found that outgassed H\textsubscript{2}O and CO\textsubscript{2} yields relative to the initial sample masses were $\sim$9 \% and 5 \%, respectively; they did not detect significant amounts of CO and CH\textsubscript{4}. These H\textsubscript{2}O and CO\textsubscript{2} yields are similar to those measured in our experiments, which reached higher temperatures over a much longer period of time: $\sim$9 \% for H\textsubscript{2}O and $\sim$6 \% for CO\textsubscript{2}, where both of these values are determined by taking the outgassed mass of the volatile species divided by the initial sample mass \cite{CourtSephton2009}. 

Mbarek \& Kempton 2016 \cite{MbarekKempton2016} used the theoretical outgassing composition of chondritic meteorites from Schaefer \& Fegley 2010 \cite{SchaeferFegly2010} as their initial condition and then performed Gibbs free energy minimization to determine what condensate cloud species may form in super-Earth atmospheres. This work explores similar atmospheric temperatures to those measured in our experiments ($\sim$350-1500 K) and finds that the C/O and H/O ratios have a strong influence on cloud chemistry in exoplanet atmospheres. They claim that if a planet's bulk composition is made of CM chondrite-like material, its outgassed atmosphere will have C/O and H/O ratios of 0.18 and 1.39, respectively \cite{MbarekKempton2016}. From our experimental outgassing abundances, we find similar C/O and H/O ratios, 0.29$\pm$0.08 and 1.18$\pm$0.18, respectively. Our experimental C/O ratio is between Mbarek \& Kempton's values for CM, CI and CV chondrites, whereas our H/O ratio is closest to their values for CM and CI chondrites. They predict that the atmospheres of super-Earth exoplanets with bulk compositions similar to CM chondrites may form KCl and ZnS clouds, but slightly more oxidizing conditions (e.g., CV chondrites) may hinder the formation of cloud condensates \cite{MbarekKempton2016}.

In summary, our results provide a comprehensive experimental comparison to prior theoretical chemical equilibrium models \cite{SchaeferFegley2007, SchaeferFegly2010} that aim to study the outgassing compositions of chondritic meteorites and their implications for terrestrial planets' early atmospheres. Additional experiments on a wider range of chondritic meteorites, including ordinary and enstatite chondrites, will allow for a more complete comparison with prior theoretical work and will reveal more insight into the possible atmospheric composition of early Earth as well as various exoplanets.  \\

\noindent \textbf{Data Availability:} The data that support the findings of this study and corresponding plots in the paper are available from \href{https://github.com/maggieapril3/CMChondritesOutgassingData}{\text{https://github.com/maggieapril3/CMChondritesOutgassingData}} or from the corresponding author upon request. Figures 1-4 and Extended Data Figures 1-5 and Supplementary Figures 2-4 have associated raw data that is available from \\ \href{https://github.com/maggieapril3/CMChondritesOutgassingData}{\text{https://github.com/maggieapril3/CMChondritesOutgassingData}} or from the corresponding author. The thermochemical equilibrium models used in Figures 3 and 4 are available from L.S. upon request.\\

\noindent \textbf{Code Availability:} The code used to calibrate and analyze the data used in this study is also available from \href{https://github.com/maggieapril3/CMChondritesOutgassingData}{\text{https://github.com/maggieapril3/CMChondritesOutgassingData}}.

\section*{Supplementary Information}

\renewcommand{\figurename}{Supplementary Figure}
\setcounter{figure}{0}
\begin{figure}[hbt!]
    \includegraphics[scale=0.25]{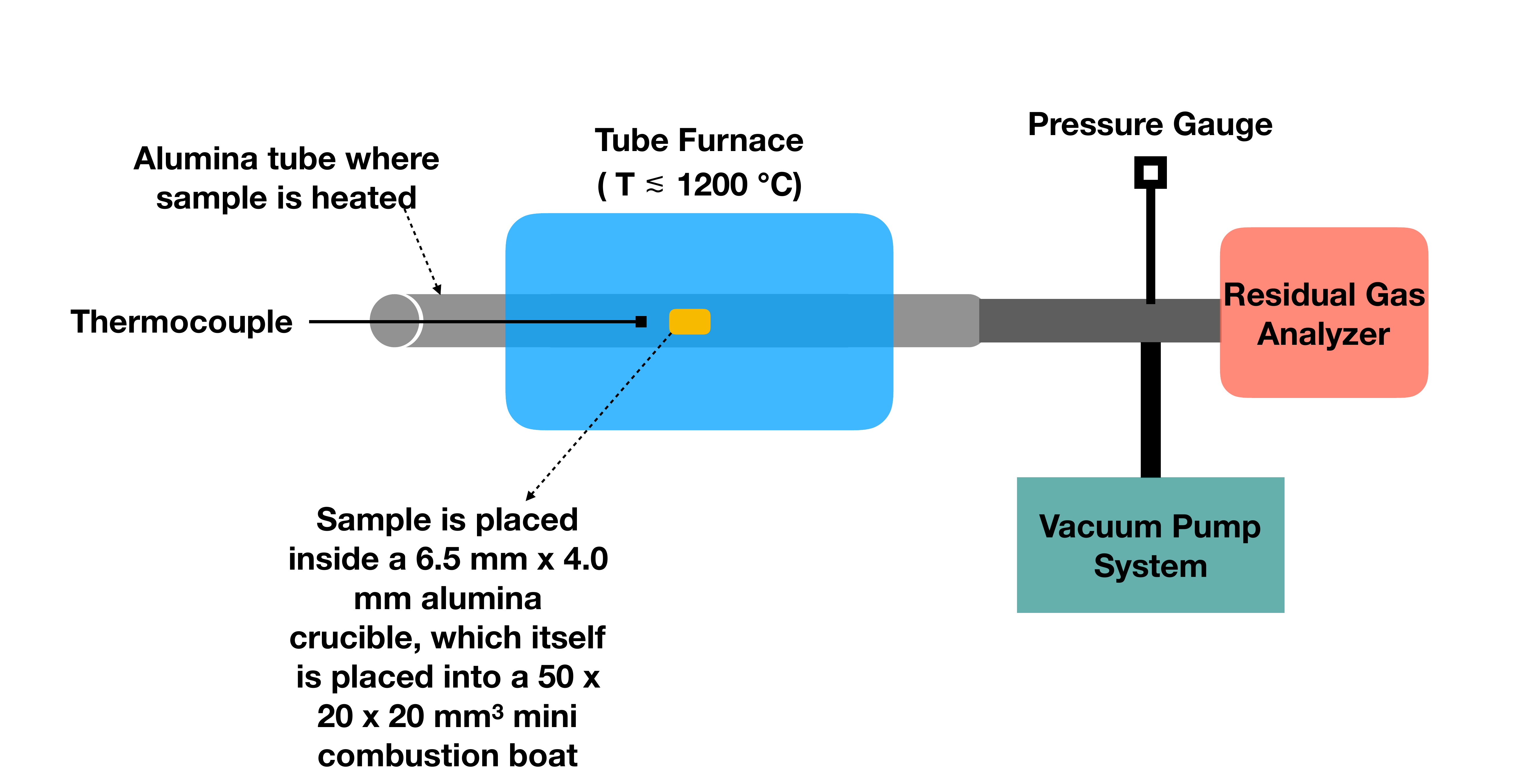}
    \centering
    \caption{\textbf{Schematic of Instrument Set-Up.} Each powdered sample is placed inside a small alumina crucible which itself is placed inside a alumina mini combustion boat. The boat is inserted into an alumina tube to the center of the furnace that can reach temperatures up to 1200 $^\circ$C. The furnace is connected to a turbomolecular pump which maintains the entire system at a high-level vacuum, and to a residual gas analyzer which measures the partial pressures of up to 10 species continuously throughout the experiment. A thermocouple inside the tube measures the temperature as a function of time. The thermocouple is placed within 50 mm of the sample containers and both are within the furnace's 13 cm hotspot to ensure accurate temperature measurements.}
\label{fig:supimage1}
\end{figure}

\renewcommand{\tablename}{Supplementary Table}
\setcounter{table}{0}
\begin{table}[hbt!]
\setlength{\tabcolsep}{5pt}
\centering
\caption{\textbf{Table of the mass spectrum for each gas species in our experiments.} For the ten masses measured during the experiments, several could correspond to gas molecules that, when ionized by the RGA, produce fragments that contribute to the signal of other masses measured. For each of the gas species that correspond to one of the ten measured masses, we used the mass spectrum from National Institute of Standards and Technology (NIST)'s Mass Spectrometry Data Center to determine the possible ion fragments \cite{NIST2}. We also include an additional ion fragment for water at 2 amu from MKS \cite{MKS2}. For each species in the table, we list the mass numbers of its known ion fragments and their signal intensities relative to the major peak due to the gas species itself (i.e., the signal percentage of the gas species is 100\%). In the last column, we provide the normalized signal intensities (i.e., all of the signals sum to 100\%). We only list ion fragments whose mass numbers correspond to those measured in our outgassing experiments. For atoms and species that either do not suffer from ion fragments or whose fragments correspond to masses that we do not measure, we assume all of its signal is concentrated at its mass number.}  \label{tab:fragment}
\begin{tabular}{c c c c}
\hline
 Gas Species & Mass Number (amu) &  \% of Signal Relative to Major Peak & Normalized \% of Signal \\
 \hline 
    H\textsubscript{2}: & 2 & 100 & 100\\
    \hline
    C: & 12 & 100 & 100 \\
    \hline
    N: & 14 & 100 & 100 \\
    \hline
    CH\textsubscript{4}: & 16 & 100 & 79.99 \\
    & 14 & 20.7 & 16.56 \\
    & 12 & 4.31 & 3.448 \\
    \hline
    O: & 16 & 100 & 100 \\
    \hline
    H\textsubscript{2}O: & 18 & 100 & 96.62 \\
     & 16 & 1.5 & 1.449 \\
     & 2 & 2 & 1.932 \\
    \hline
    CO: & 28 & 100 & 93.11 \\
    & 16 & 2.2 & 2.048 \\
    & 12 & 5.2 & 4.842 \\
    \hline
    N\textsubscript{2}: & 28 & 100 & 87.72 \\
    & 14 & 14 & 12.28 \\
    \hline
    S: & 32 & 100 & 100 \\
    \hline
    O\textsubscript{2}: & 32 & 100 & 81.97 \\
    & 16 & 22 & 18.03 \\
    \hline
    H\textsubscript{2}S: & 34 & 100 & 68.97 \\
    & 32 & 45 & 31.03 \\
    \hline 
    Ar: & 40 & 100 & 100 \\
    \hline 
    CO\textsubscript{2}: & 44 & 100 & 77.58 \\
    & 28 & 10.2 & 7.913 \\
    & 16 & 9.8 & 7.603 \\
    & 12 & 8.9 & 6.905 \\
 \hline
\end{tabular}
\end{table}

\begin{table}[ht]
\caption{\textbf{Previously determined average bulk composition of CM chondrites and Murchison from literature} \textsuperscript{a}\cite{Alexander2012_2}, \textsuperscript{b}\cite{Nittler2004_2}, \textsuperscript{c}\cite{Fuchs1973_2}. The uncertainties are the 1$\sigma$ standard deviations.} \label{tab:bulk}
\setlength{\tabcolsep}{3pt}
\centering
\begin{tabular}{c c c}
\hline
 Element & Average CM Chondrite & Murchison\\ 
 \hline
  H & 11.5$\pm$0.18\textsuperscript{a} mg/g & 10.7$\pm$0.002\textsuperscript{a} mg/g \\
  C & 19.5$\pm$3.24\textsuperscript{a} mg/g & 20.8\textsuperscript{a} mg/g \\
  N & 996.5$\pm$280\textsuperscript{a} $\mu$g/g & 1050\textsuperscript{a} $\mu$g/g \\
  O & 412.0$\pm$0.75\textsuperscript{a,b} mg/g & 410\textsuperscript{b} mg/g \\
  S & 33$\pm$9.0\textsuperscript{b} mg/g & 14\textsuperscript{c} mg/g \\
  K & 400\textsuperscript{b} $\mu$g/g & 280\textsuperscript{c} $\mu$g/g  \\
  Na & 4.1\textsuperscript{b} mg/g & 4.2\textsuperscript{c} mg/g \\
 \hline
\end{tabular}
\end{table}

\begin{table}[ht]
\caption{\textbf{Comparison of primary algebraic data analysis and Monte Carlo non-linear least squares (MC) data analysis for Murchison.} The second and third columns show the average partial pressure (in bars) for each species. The fourth column shows the standard deviation of the average partial pressure for each species analyzed using the Monte Carlo technique. These partial pressures are corrected for ion fragments and atmospheric adsorption but have not been background subtracted.} \label{tab:MCvsoriginal}
\setlength{\tabcolsep}{3pt}
\centering
\begin{tabular}{c c c c}
\hline
 Species & Primary Analysis & MC Analysis & Standard Deviation of MC Analysis\\ 
 \hline
  H\textsubscript{2} & 1.9E-10  & 3.4E-10 & 1.1E-10 \\
  C & 0.0  & 1.3E-13 & 1.4E-25 \\
  N & 0.0 & 0.0 & 0.0 \\
  CH\textsubscript{4} & 0.0 & 3.5E-10 & 9.7E-11 \\
  O & -- & 0.0 & 0.0 \\
  H\textsubscript{2}O & 7.6E-09 & 7.3E-09 & 1.9E-09 \\
  N\textsubscript{2} & 0.0 & 0.0 & 0.0  \\
  CO & 1.2E-09 & 2.6E-09 & 6.3E-10 \\
  S & 0.0 & 6.4E-10 & 1.0E-10 \\
  O\textsubscript{2} & 0.0 & 0.0 & 0.0 \\
  H\textsubscript{2}S & 8.5E-12 & 5.8E-12 & 2.2E-12  \\
  Ar & 0.0 & 0.0 & 0.0 \\
  CO\textsubscript{2} & 6.9E-10 & 5.3E-10 & 2.9E-10 \\
 \hline
\end{tabular}
\end{table}

\begin{figure}[hbt!]
    \includegraphics[width=\textwidth]{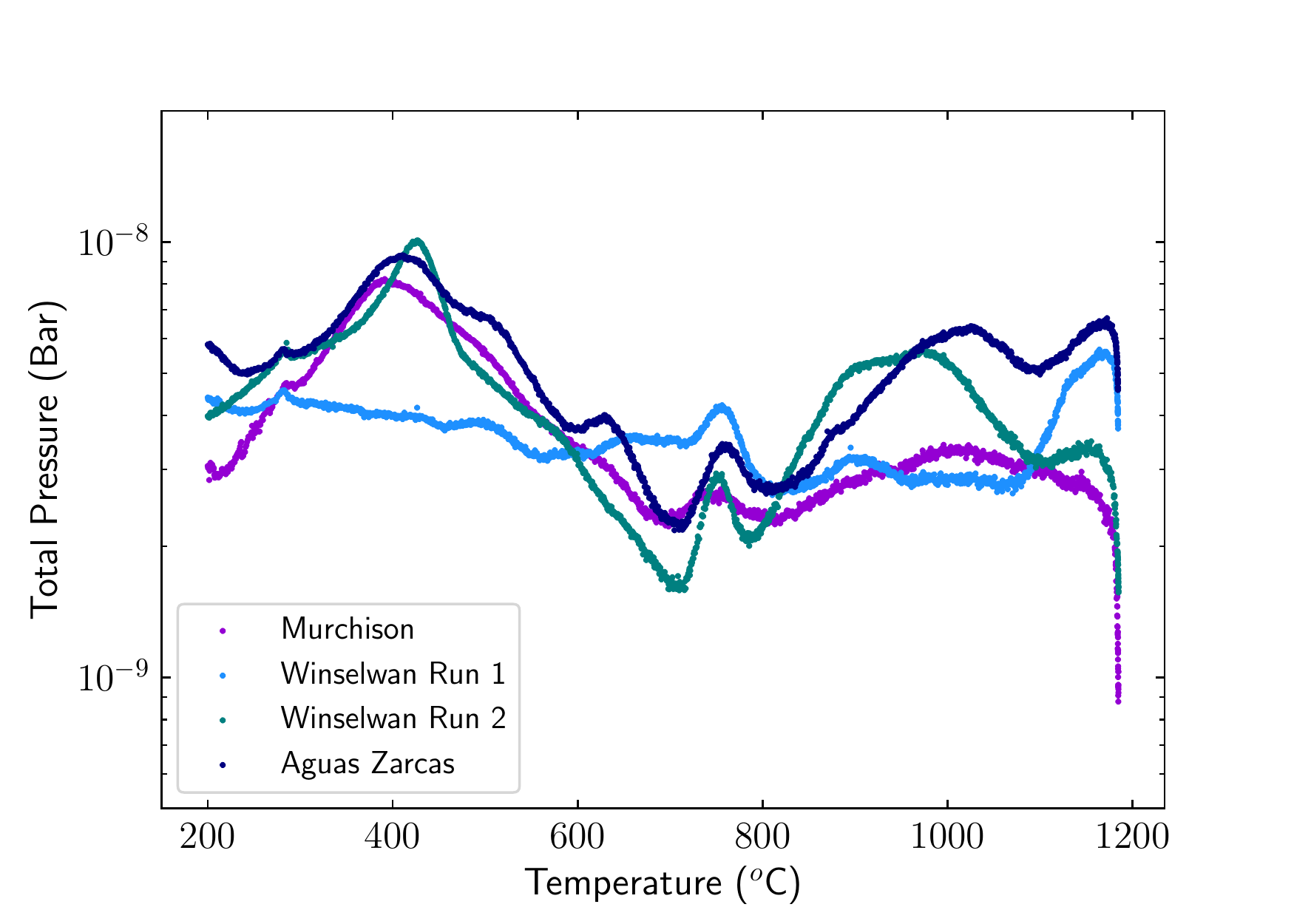}
    \centering
    \caption{\textbf{Total pressure of measured volatiles released from the samples as a function of temperature.} Variations in total pressure with temperature suggest that the amount of outgassing varies throughout the experiment. The average difference between the maximum and minimum total pressure is 6E-9 bars. Most samples show an increase in total pressure near 400 $^{\circ}$C.}
    \label{fig:supimage3}
\end{figure}

\begin{figure}[hbt!]
    \begin{subfigure}{0.49\textwidth}
        \centering
        \includegraphics[width=\textwidth]{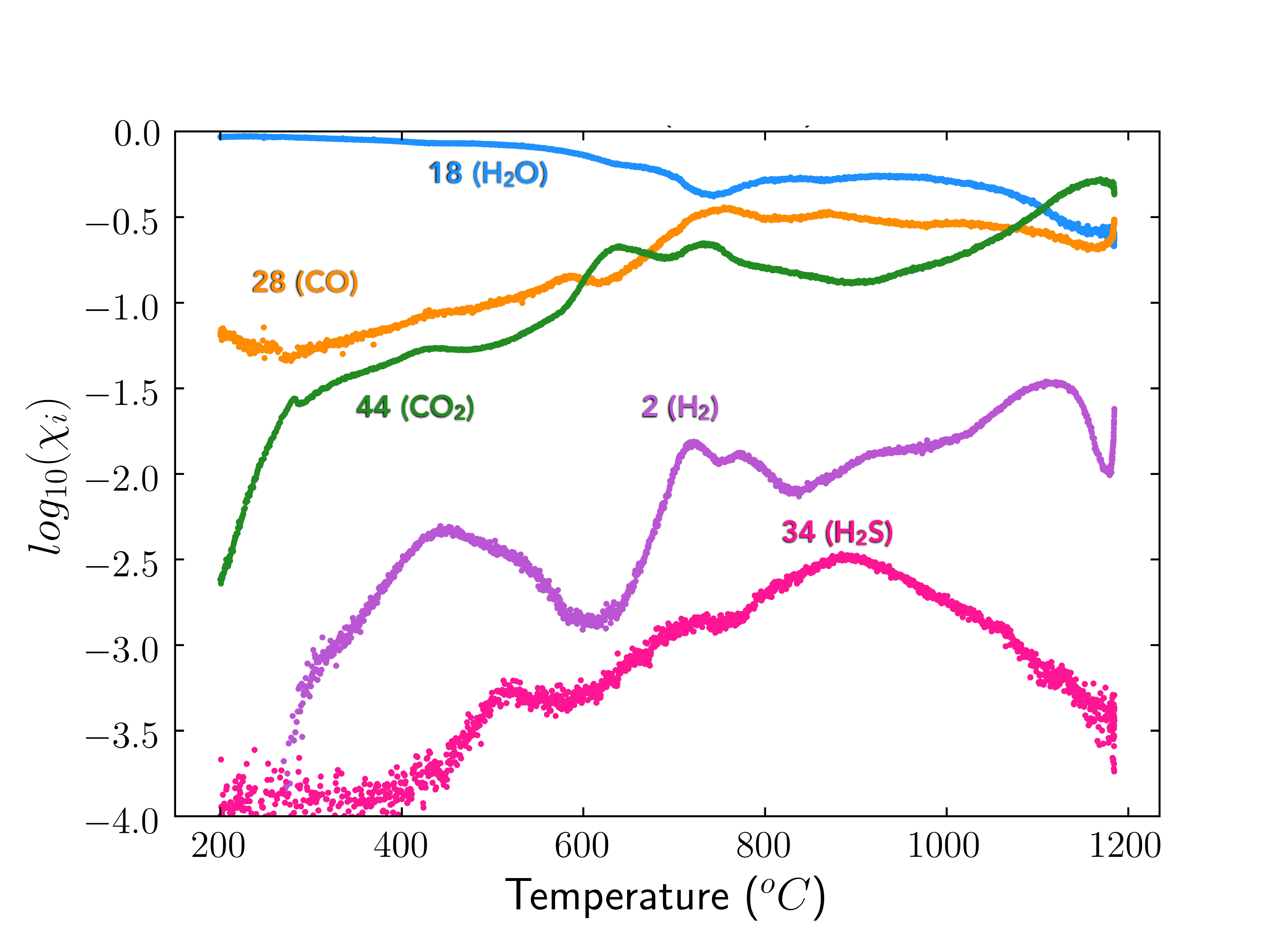}
        \caption{\small Average of 3 CM Chondrite Samples with the Signal at 32 amu Not Separated into Sulfur and O\textsubscript{2} Components} 
        \label{fig:s6a} 
    \end{subfigure}
    \hfill
    \begin{subfigure}{0.49\textwidth}  
        \centering 
        \includegraphics[width=\textwidth]{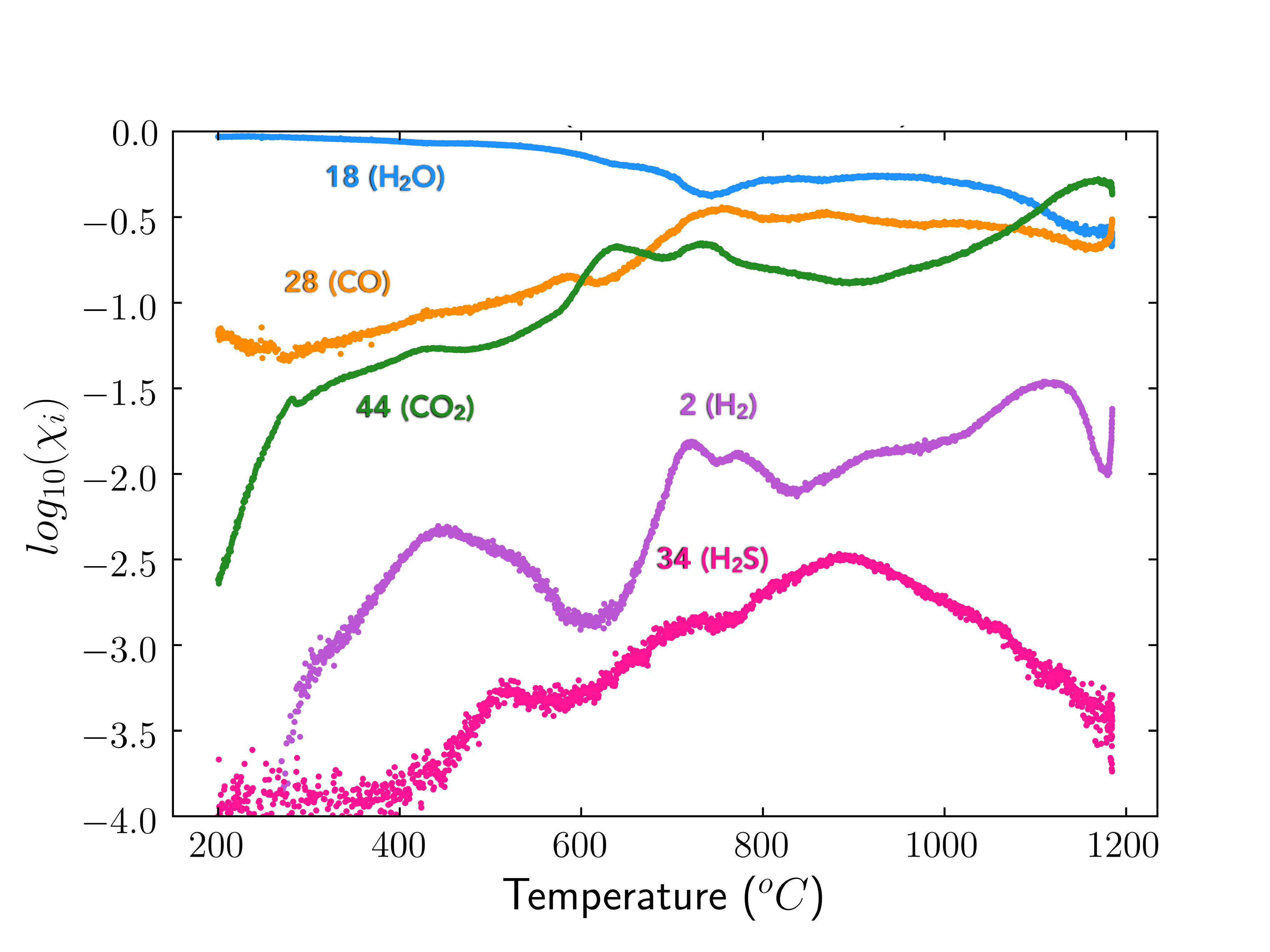}
        \caption{\small Average of 3 CM Chondrite Samples with the Signal at 32 amu Separated into Sulfur and O\textsubscript{2} Components}
        \label{fig:s6b}  
    \end{subfigure}
\caption{\textbf{Comparison between original results and results of separating the 32 amu signal into sulfur and O\textsubscript{2} components}. Figure (a) shows the outgassing abundances in which the signal at 32 amu is not separated into the sulfur and O\textsubscript{2} components (i.e., Figure 3 (b)). Figure (b) shows the results of separating the signal at 32 amu into its sulfur and O\textsubscript{2} abundances.}
\label{fig:supimage6}
\end{figure}

\begin{figure}[hbt!]
    \begin{subfigure}{0.5\textwidth}
        \centering
        \includegraphics[width=\textwidth]{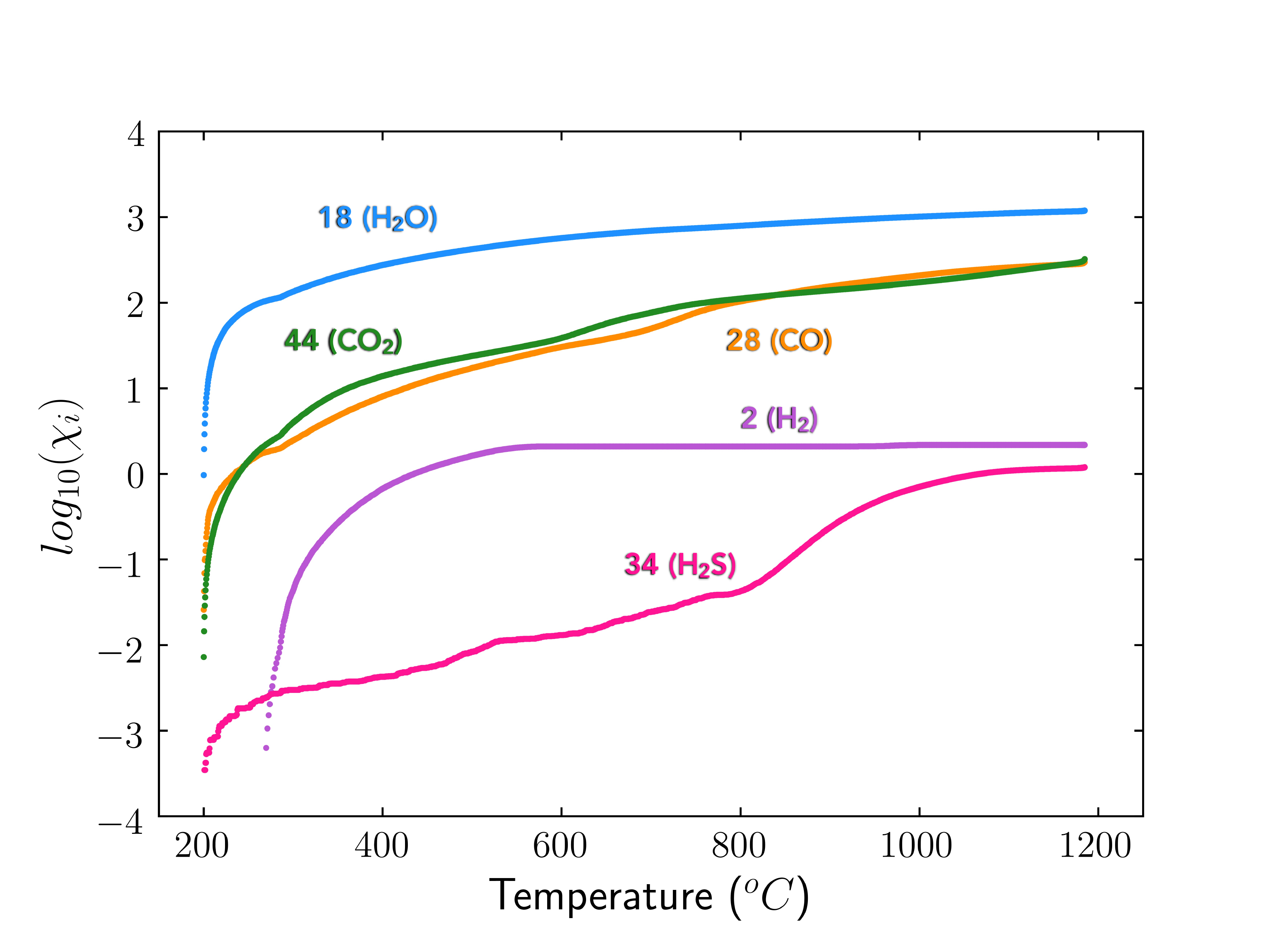}
        \caption{\small Murchison}  
        \label{fig:s7a} 
    \end{subfigure}
    \hfill
    \begin{subfigure}{0.5\textwidth}  
        \centering 
        \includegraphics[width=\textwidth]{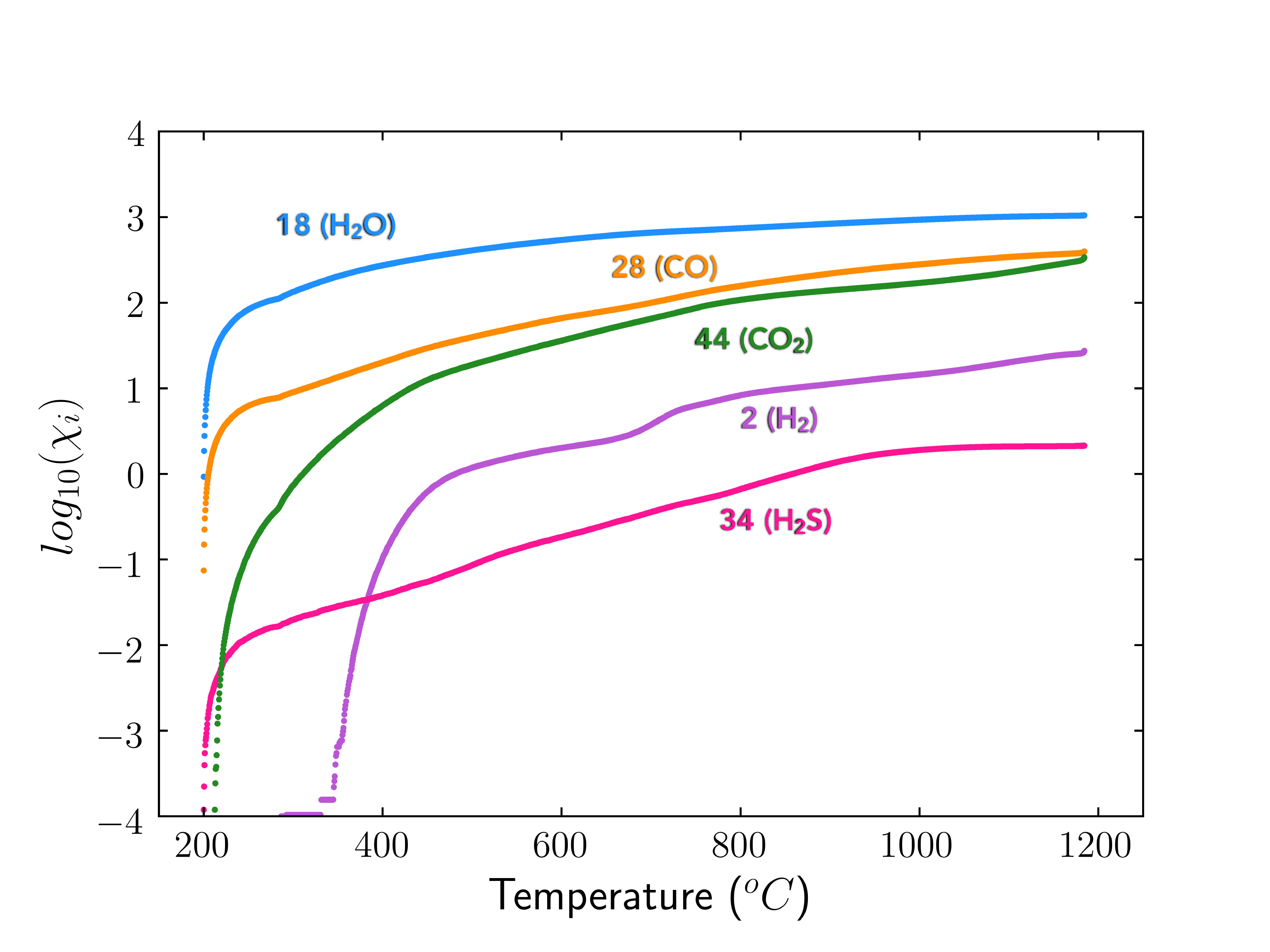}
        \caption{\small Winselwan}
        \label{fig:s7b}  
    \end{subfigure}
    \hfill
    \begin{subfigure}{0.5\textwidth}   
        \centering 
        \includegraphics[width=\textwidth]{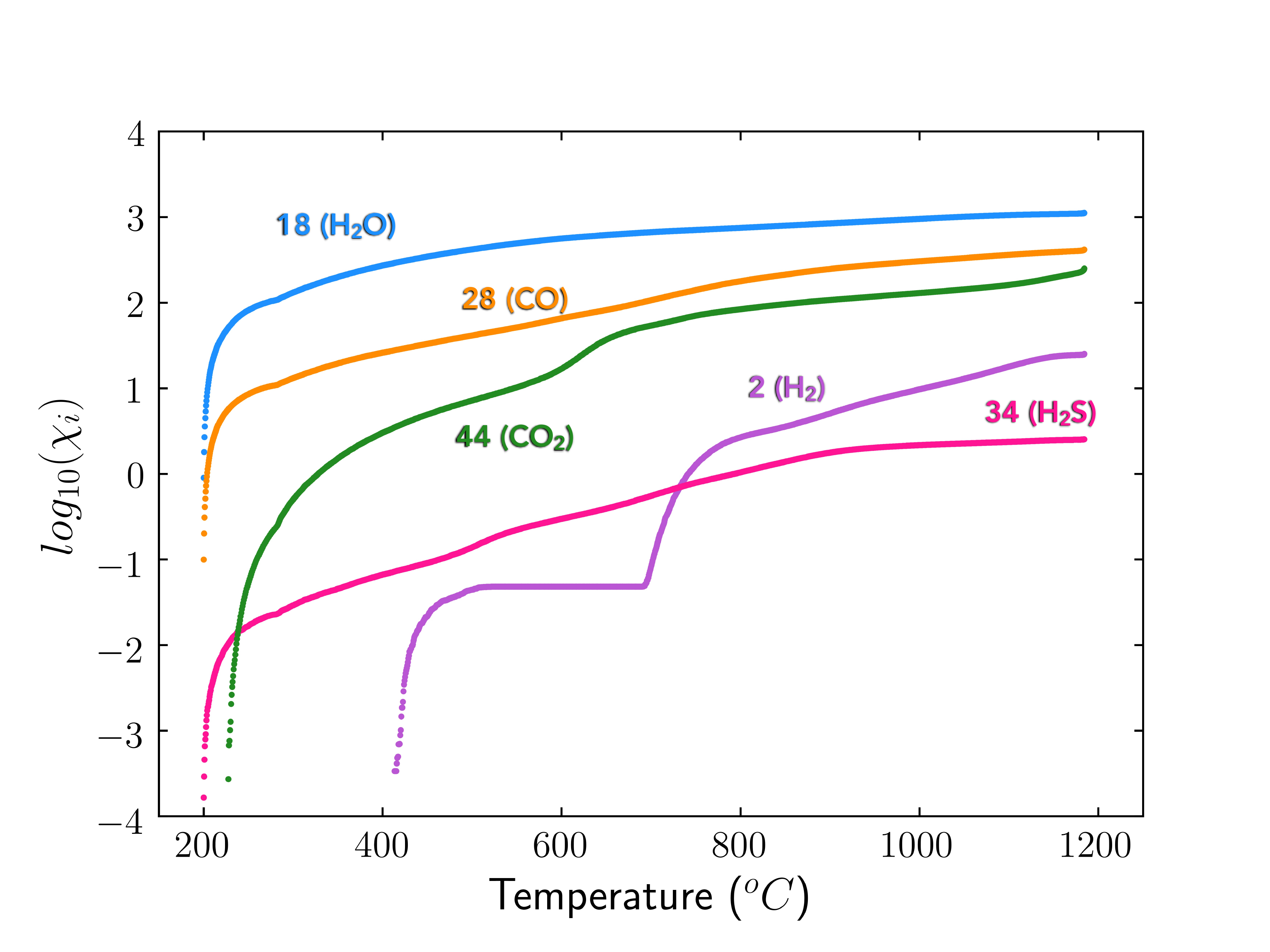}
        \caption{\small Aguas Zarcas}   
        \label{fig:s7c}
    \end{subfigure}
    \hfill
    \begin{subfigure}{0.5\textwidth}   
        \centering 
        \includegraphics[width=\textwidth]{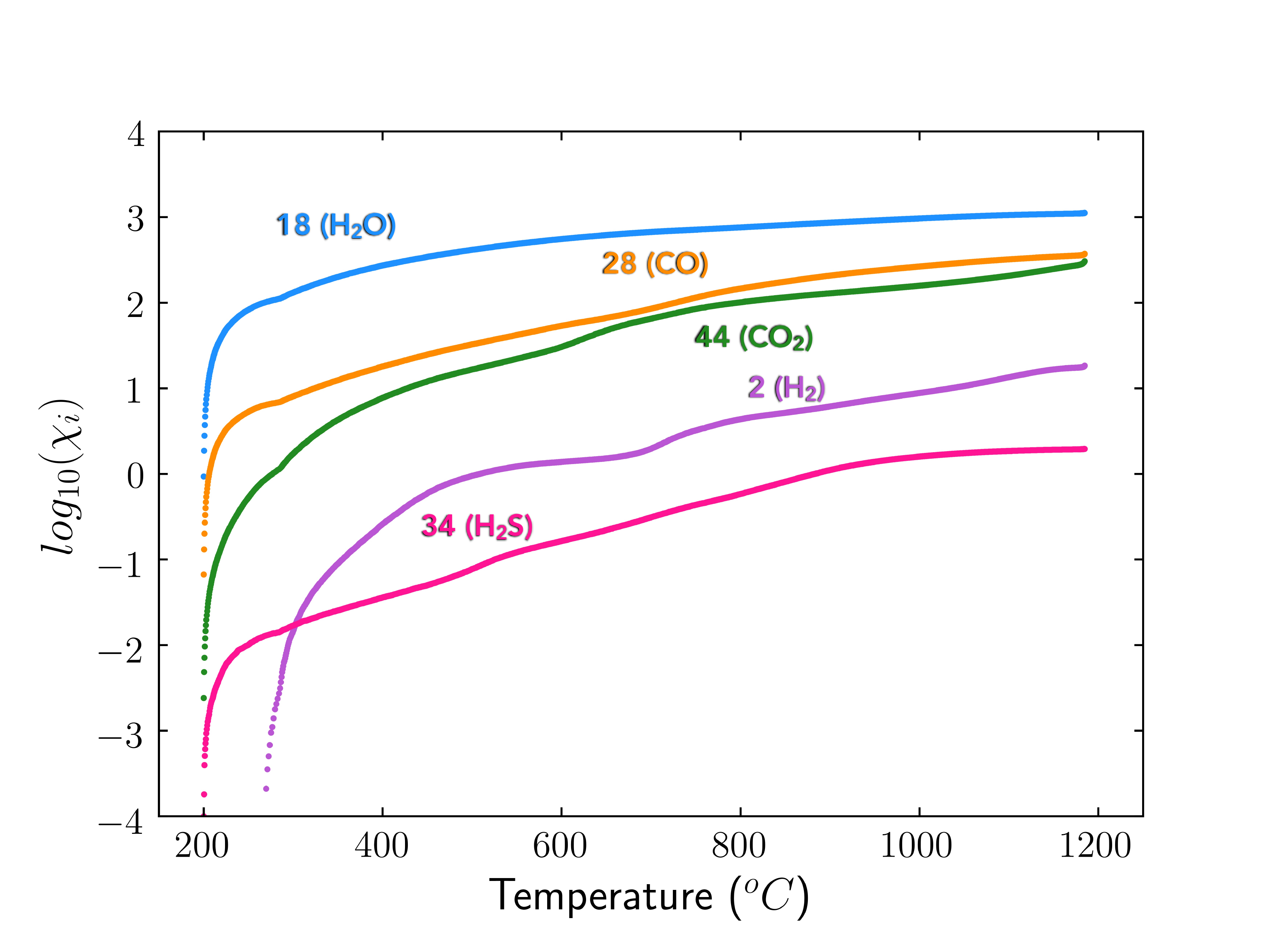}
        \caption{\small Average of 3 CM Chondrites}   
        \label{fig:s7d}
    \end{subfigure}
\caption{\textbf{Cumulative Outgassing Abundances}. The cumulative outgassing trends for samples of (a) Murchison, (b) Jbilet Winselwan, (c) Aguas Zarcas, and (d) the average of the three CM chondrite samples.}
\label{fig:supimage7}
\end{figure}


\begin{thebibliography}{10}
\expandafter\ifx\csname url\endcsname\relax
  \def\url#1{\texttt{#1}}\fi
\expandafter\ifx\csname urlprefix\endcsname\relax\def\urlprefix{URL }\fi
\providecommand{\bibinfo}[2]{#2}
\providecommand{\eprint}[2][]{\url{#2}}

\bibitem{Petigura2013a}
\bibinfo{author}{Petigura, E.~A.}, \bibinfo{author}{Marcy, G.~W.} \&
  \bibinfo{author}{Howard, A.~W.}
\newblock \bibinfo{title}{A plateau in the planet population below twice the
  size of earth}.
\newblock \emph{\bibinfo{journal}{The Astrophysical Journal}}
  \textbf{\bibinfo{volume}{770}}, \bibinfo{pages}{21} (\bibinfo{year}{2013}).

\bibitem{Petigura2013b}
\bibinfo{author}{Petigura, E.~A.}, \bibinfo{author}{Howard, A.~W.} \&
  \bibinfo{author}{Marcy, G.~W.}
\newblock \bibinfo{title}{Prevalence of earth-size planets orbiting sun-like
  stars}.
\newblock \emph{\bibinfo{journal}{Proceedings of the National Academy of
  Sciences}} \textbf{\bibinfo{volume}{110}}, \bibinfo{pages}{19273--19278}
  (\bibinfo{year}{2013}).

\bibitem{Dressing2015}
\bibinfo{author}{Dressing, C.~D.} \& \bibinfo{author}{Charbonneau, D.}
\newblock \bibinfo{title}{The occurrence of potentially habitable planets
  orbiting m dwarfs estimated from the full kepler dataset and an empirical
  measurement of the detection sensitivity}.
\newblock \emph{\bibinfo{journal}{The Astrophysical Journal}}
  \textbf{\bibinfo{volume}{807}}, \bibinfo{pages}{23} (\bibinfo{year}{2015}).

\bibitem{Sharp2017}
\bibinfo{author}{Sharp, Z.~D.}
\newblock \bibinfo{title}{Nebular ingassing as a source of volatiles to the
  terrestrial planets}.
\newblock \emph{\bibinfo{journal}{Chemical Geology}}
  \textbf{\bibinfo{volume}{448}}, \bibinfo{pages}{137--150}
  (\bibinfo{year}{2017}).

\bibitem{Schlichting2018}
\bibinfo{author}{Schlichting, H.~E.} \& \bibinfo{author}{Mukhopadhyay, S.}
\newblock \bibinfo{title}{Atmosphere impact losses}.
\newblock \emph{\bibinfo{journal}{Space Science Reviews}}
  \textbf{\bibinfo{volume}{214}} (\bibinfo{year}{2018}).

\bibitem{Wu2018}
\bibinfo{author}{Wu, J.} \emph{et~al.}
\newblock \bibinfo{title}{Origin of earth's water: Chondritic inheritance plus
  nebular ingassing and storage of hydrogen in the core}.
\newblock \emph{\bibinfo{journal}{Journal of Geophysical Research: Planets}}
  \textbf{\bibinfo{volume}{123}}, \bibinfo{pages}{2691--2712}
  (\bibinfo{year}{2018}).

\bibitem{Lammer2018}
\bibinfo{author}{Lammer, H.} \emph{et~al.}
\newblock \bibinfo{title}{Origin and evolution of the atmospheres of early
  venus, earth and mars}.
\newblock \emph{\bibinfo{journal}{The Astronomy and Astrophysics Review}}
  \textbf{\bibinfo{volume}{26}}, \bibinfo{pages}{72} (\bibinfo{year}{2018}).

\bibitem{Elkins-Tanton2008}
\bibinfo{author}{Elkins-Tanton, L.~T.} \& \bibinfo{author}{Seager, S.}
\newblock \bibinfo{title}{Ranges of atmospheric mass and composition of
  super-earth exoplanets}.
\newblock \emph{\bibinfo{journal}{The Astrophysical Journal}}
  \textbf{\bibinfo{volume}{685}}, \bibinfo{pages}{1237--1246}
  (\bibinfo{year}{2008}).

\bibitem{Ahrens1989}
\bibinfo{author}{Ahrens, T.~J.}, \bibinfo{author}{O'Keefe, J.~D.} \&
  \bibinfo{author}{Lange, M.~A.}
\newblock \emph{\bibinfo{title}{Origin and Evolution of Planetary and Satellite
  Atmospheres}}, chap. \bibinfo{chapter}{Formation of Atmospheres during
  Accretion of the Terrestrial Planets} (\bibinfo{publisher}{University of
  Arizona Press}, \bibinfo{year}{1989}).

\bibitem{LoddersFegley}
\bibinfo{author}{Lodders, K.} \& \bibinfo{author}{Fegley, B.}
\newblock \emph{\bibinfo{title}{The Planetary Scientist's Companion}}
  (\bibinfo{publisher}{Oxford University Press}, \bibinfo{year}{1998}).

\bibitem{Wasson1988}
\bibinfo{author}{Wasson, J.~T.} \& \bibinfo{author}{Kallemeyn, G.~W.}
\newblock \bibinfo{title}{Compositions of chondrites}.
\newblock \emph{\bibinfo{journal}{Philosophical Transactions of the Royal
  Society of London. Series A, Mathematical and Physical Sciences}}
  \textbf{\bibinfo{volume}{325}}, \bibinfo{pages}{535--544}
  (\bibinfo{year}{1988}).

\bibitem{Zahnle1988}
\bibinfo{author}{Zahnle, K.~J.}, \bibinfo{author}{Kasting, J.~F.} \&
  \bibinfo{author}{Pollack, J.~B.}
\newblock \bibinfo{title}{Evoluation of a steam atmosphere during earth's
  accretion}.
\newblock \emph{\bibinfo{journal}{Icarus}} \textbf{\bibinfo{volume}{74}},
  \bibinfo{pages}{62--97} (\bibinfo{year}{1988}).

\bibitem{Gaillard2014}
\bibinfo{author}{Gaillard, F.} \& \bibinfo{author}{Scaillet, B.}
\newblock \bibinfo{title}{A theoretical framework for volcanic degassing
  chemistry in a comparative planetology perspective and implications for
  planetary atmospheres}.
\newblock \emph{\bibinfo{journal}{Earth and Planetary Science Letters}}
  \textbf{\bibinfo{volume}{403}}, \bibinfo{pages}{307--316}
  (\bibinfo{year}{2014}).

\bibitem{SchaeferFegly2010}
\bibinfo{author}{Schaefer, L.} \& \bibinfo{author}{Fegley., B.~Jr.}
\newblock \bibinfo{title}{Chemistry of atmospheres formed during accretion of
  the earth and other terrestrial planets}.
\newblock \emph{\bibinfo{journal}{Icarus}} \textbf{\bibinfo{volume}{208}},
  \bibinfo{pages}{438--448} (\bibinfo{year}{2010}).

\bibitem{Herbort2020}
\bibinfo{author}{Herbort, O.}, \bibinfo{author}{Woitke, P.},
  \bibinfo{author}{Helling, C.} \& \bibinfo{author}{Zerkle, A.}
\newblock \bibinfo{title}{The atmospheres of rocky exoplanets i. outgassing of
  common rock and the stability of liquid water}.
\newblock \emph{\bibinfo{journal}{Astronomy \& Astrophysics}}
  \textbf{\bibinfo{volume}{636}}, \bibinfo{pages}{1--19}
  (\bibinfo{year}{2020}).

\bibitem{CourtSephton2009}
\bibinfo{author}{Court, R.~W.} \& \bibinfo{author}{Sephton, M.~A.}
\newblock \bibinfo{title}{Meteorite ablation products and their contribution to
  the atmospheres of terrestrial planets: An experimental study using
  pyrolysis-ftir}.
\newblock \emph{\bibinfo{journal}{Geochimica et Cosmochimica Acta}}
  \textbf{\bibinfo{volume}{73}}, \bibinfo{pages}{3512--3521}
  (\bibinfo{year}{2009}).

\bibitem{GoodingMuenow1977}
\bibinfo{author}{Gooding, J.~L.} \& \bibinfo{author}{Muenow, D.~W.}
\newblock \bibinfo{title}{Experimental vaporization of the holbrook chondrite}.
\newblock \emph{\bibinfo{journal}{Meteoritics}} \textbf{\bibinfo{volume}{12}},
  \bibinfo{pages}{401--408} (\bibinfo{year}{1977}).

\bibitem{LangeAhrens1982}
\bibinfo{author}{Lange, M.~A.} \& \bibinfo{author}{Ahrens, T.~J.}
\newblock \bibinfo{title}{The evolution of an impact-generated atmosphere}.
\newblock \emph{\bibinfo{journal}{Icarus}} \textbf{\bibinfo{volume}{51}},
  \bibinfo{pages}{96--120} (\bibinfo{year}{1982}).

\bibitem{Burgess1991}
\bibinfo{author}{Burgess, R.}, \bibinfo{author}{Wright, I.~P.} \&
  \bibinfo{author}{Pillinger, C.~T.}
\newblock \bibinfo{title}{Determination of sulphur-bearing components in c1 and
  c2 carbonaceous chondrites by stepped combustion}.
\newblock \emph{\bibinfo{journal}{Meteoritics}} \textbf{\bibinfo{volume}{26}},
  \bibinfo{pages}{55--64} (\bibinfo{year}{1991}).

\bibitem{Springmann2019}
\bibinfo{author}{Springmann, A.} \emph{et~al.}
\newblock \bibinfo{title}{Thermal alteration of labile elements in carbonaceous
  chondrites}.
\newblock \emph{\bibinfo{journal}{Icarus}} \textbf{\bibinfo{volume}{324}},
  \bibinfo{pages}{104--119} (\bibinfo{year}{2019}).

\bibitem{Tyburczy1986b}
\bibinfo{author}{Tyburczy, J.~A.}, \bibinfo{author}{Frisch, B.} \&
  \bibinfo{author}{Ahrens, T.~J.}
\newblock \bibinfo{title}{Shock-induced volatile loss from a carbonaceous
  chondrite: implications for planetary accretion}.
\newblock \emph{\bibinfo{journal}{Earth and Planetary Science Letters}}
  \textbf{\bibinfo{volume}{80}}, \bibinfo{pages}{201--207}
  (\bibinfo{year}{1986}).

\bibitem{Ikramuddin1977a}
\bibinfo{author}{Ikramuddin, M.}, \bibinfo{author}{Binz, C.~M.} \&
  \bibinfo{author}{Lipschutz, M.~E.}
\newblock \bibinfo{title}{Thermal metamorphism of primitive meteorites iii. ten
  trace elements in krymka l3 chondrite heated at 400-1000 c}.
\newblock \emph{\bibinfo{journal}{Geochimica et Cosmochimica Acta}}
  \textbf{\bibinfo{volume}{41}}, \bibinfo{pages}{393--401}
  (\bibinfo{year}{1977}).

\bibitem{Murchison1969}
\bibinfo{author}{Krinov, E.~L.}
\newblock \bibinfo{title}{Fall of murchison stone meteorite shower, australia}.
\newblock \emph{\bibinfo{journal}{The Meteoritical Bulletin (Meteoritics)}}
  \textbf{\bibinfo{volume}{5}}, \bibinfo{pages}{2} (\bibinfo{year}{1969}).

\bibitem{Winselwan2015}
\bibinfo{author}{Ruzicka, A.}, \bibinfo{author}{Grossman, J.},
  \bibinfo{author}{Bouvier, A.}, \bibinfo{author}{Herd, C. D.~K.} \&
  \bibinfo{author}{Agee, C.~B.}
\newblock \bibinfo{title}{The meteoritical bulletin, no 102}.
\newblock \emph{\bibinfo{journal}{Meteoritics and Planetary Science}}
  \textbf{\bibinfo{volume}{50}} (\bibinfo{year}{2015}).

\bibitem{AguasZarcas2019}
\bibinfo{author}{Gattacceca, J.}, \bibinfo{author}{McCubbin, F. M.},
  \bibinfo{author}{Bouvier, A.}, \& \bibinfo{author}{Grossman, J. N.}
\newblock \bibinfo{title}{Meteoritical bulletin no. 108}
\newblock \emph{\bibinfo{journal}{Meteoritics and Planetary Science}}
  \textbf{\bibinfo{volume}{55}} (\bibinfo{year}{2020}).

\bibitem{Nittler2004}
\bibinfo{author}{Nittler, L.~R.} \emph{et~al.}
\newblock \bibinfo{title}{Bulk element compositions of meteorites: a guide for
  interpreting remote-sensing geochemical measurements of planets and
  asteroids}.
\newblock \emph{\bibinfo{journal}{Antarctic Meteorite Research}}
  \textbf{\bibinfo{volume}{17}}, \bibinfo{pages}{231--251}
  (\bibinfo{year}{2004}).
  
\bibitem{Alexander2012}
\bibinfo{author}{Alexander, C.~M.~O'D.} \emph{et~al.}
\newblock \bibinfo{title}{The Provenances of Asteroids, and Their Contributions to the Volatile Inventories of the Terrestrial Planets}.
\newblock \emph{\bibinfo{journal}{Science}}
  \textbf{\bibinfo{volume}{337}}, \bibinfo{pages}{721}
  (\bibinfo{year}{2012}).

\bibitem{OBrienNielsen1959}
\bibinfo{author}{O'Brien, W.~J.} \& \bibinfo{author}{Nielsen, J.~P.}
\newblock \bibinfo{title}{Decomposition of gypsum investment in the presence of
  carbon}.
\newblock \emph{\bibinfo{journal}{Journal of Dental Research}}
  \textbf{\bibinfo{volume}{38}}, \bibinfo{pages}{541--547}
  (\bibinfo{year}{1959}).

\bibitem{Zhao2011}
\bibinfo{author}{Zhao, S.}, \bibinfo{author}{Jiang, J.} \&
  \bibinfo{author}{Zheng, J.}
\newblock \bibinfo{title}{Thermal analysis on the kinetics of thermal
  decomposition of FeS}.
\newblock \emph{\bibinfo{journal}{Journal of Chongqing University}}
  (\bibinfo{year}{2011}).

\bibitem{Gooding1987}
\bibinfo{author}{Gooding, J.~L.} \& \bibinfo{author}{Zolensky, M.~E.}
\newblock \bibinfo{title}{Thermal stability of tochilinite}.
\newblock \emph{\bibinfo{journal}{Lunar and Planetary Science Conference}}
  \textbf{\bibinfo{volume}{18}}, \bibinfo{pages}{343--344}
  (\bibinfo{year}{1987}).

\bibitem{Miller-Ricci2009}
\bibinfo{author}{Miller-Ricci, E.}, \bibinfo{author}{Seager, S.} \&
  \bibinfo{author}{Sasselov, D.}
\newblock \bibinfo{title}{The atmospheric signatures of super-earths: How to
  distinguish between hydrogen-rich and hydrogen-poor atmospheres}.
\newblock \emph{\bibinfo{journal}{The Astrophysical Journal}}
  \textbf{\bibinfo{volume}{690}}, \bibinfo{pages}{1056--1067}
  (\bibinfo{year}{2009}).

\bibitem{Fortney2013}
\bibinfo{author}{Fortney, J.~J.} \emph{et~al.}
\newblock \bibinfo{title}{A framework for characterizing the atmospheres of
  low-mass low-density transiting planets}.
\newblock \emph{\bibinfo{journal}{The Astrophysical Journal}}
  \textbf{\bibinfo{volume}{775}}, \bibinfo{pages}{13} (\bibinfo{year}{2013}).

\bibitem{Greene2016}
\bibinfo{author}{Greene, T.~P.} \emph{et~al.}
\newblock \bibinfo{title}{Characterizing transiting exoplanet atmospheres with
  jwst}.
\newblock \emph{\bibinfo{journal}{The Astrophysical Journal}}
  \textbf{\bibinfo{volume}{817}}, \bibinfo{pages}{22} (\bibinfo{year}{2016}).

\bibitem{Morley2017}
\bibinfo{author}{Morley, C.~V.}, \bibinfo{author}{Kreidberg, L.},
  \bibinfo{author}{Rustamkulov, Z.}, \bibinfo{author}{Robinson, T.} \&
  \bibinfo{author}{Fortney, J.~J.}
\newblock \bibinfo{title}{Observing the atmospheres of known temperate
  earth-sized planets with jwst}.
\newblock \emph{\bibinfo{journal}{The Astrophysical Journal}}
  \textbf{\bibinfo{volume}{850}}, \bibinfo{pages}{18} (\bibinfo{year}{2017}).

\bibitem{Bower2019}
\bibinfo{author}{Bower, D.~J.} \emph{et~al.}
\newblock \bibinfo{title}{Linking the evolution of terrestrial interiors and an
  early outgassed atmosphere to astrophysical observations}.
\newblock \emph{\bibinfo{journal}{Astronomy \& Astrophysics}}
  \textbf{\bibinfo{volume}{631}}, \bibinfo{pages}{18} (\bibinfo{year}{2019}).
  
\bibitem{SossiFegley2018}
\bibinfo{author}{Sossi, P.~A.} \& \bibinfo{author}{Jr., B.~F.}
\newblock \bibinfo{title}{Thermodynamics of element volatility and its application to planetary processes}.
\newblock \emph{\bibinfo{journal}{Reviews in Mineralogy \& Geochemistry}}
  \textbf{\bibinfo{volume}{84}}, \bibinfo{pages}{393--459}
  (\bibinfo{year}{2018}).\\
  
\noindent \textbf{Methods References}


\bibitem{RGAm}
\bibinfo{author}{Systems, S.~R.}
\newblock \emph{\bibinfo{title}{Operating Manual and Programming Reference:
  Models RGA100, RGA200, and RGA300 Residual Gas Analyzer}}.
\newblock \bibinfo{organization}{Stanford Research Systems},
  \bibinfo{address}{1290-D Reamwood Avenue, Sunnyvale, CA 94089},
  \bibinfo{edition}{1.8} edn. (\bibinfo{year}{2009}).

\bibitem{Okumura}
\bibinfo{author}{Okumura, F.} \& \bibinfo{author}{Mimura, K.}
\newblock \bibinfo{title}{Gradual and stepwise pyrolyses of insoluble organic
  matter from the murchison meteorite revealing chemical structure and isotopic
  distribution}.
\newblock \emph{\bibinfo{journal}{Geochimica et Cosmochimica Acta}}
  \textbf{\bibinfo{volume}{75}}, \bibinfo{pages}{7063--7080}
  (\bibinfo{year}{2011}).

\bibitem{NIST}
\bibinfo{author}{NIST, M. S. D.~C.} \& \bibinfo{author}{Wallace, W.~E.}
\newblock \emph{\bibinfo{title}{NIST Chemistry WebBook}}, chap.
  \bibinfo{chapter}{Mass Spectra}.
\newblock \bibinfo{number}{69} (\bibinfo{publisher}{NIST Standard Reference
  Database}, \bibinfo{year}{2018}).

\bibitem{GradyWright2003}
\bibinfo{author}{Grady, M.~M.} \& \bibinfo{author}{Wright, I.~P.}
\newblock \bibinfo{title}{Elemental and isotopic abundances of carbon and
  nitrogen in meteorites}.
\newblock \emph{\bibinfo{journal}{Space Science Reviews}}
  \textbf{\bibinfo{volume}{106}}, \bibinfo{pages}{231--248}
  (\bibinfo{year}{2003}).

\bibitem{SchaeferFegley2017}
\bibinfo{author}{Schaefer, L.} \& \bibinfo{author}{Fegley, B.~Jr.}
\newblock \bibinfo{title}{Redox states of initial atmospheres outgassed on
  rocky planets and planetesimals}.
\newblock \emph{\bibinfo{journal}{The Astrophysical Journal}}
  \textbf{\bibinfo{volume}{843}}, \bibinfo{pages}{18} (\bibinfo{year}{2017}).

\bibitem{Fegley2013}
\bibinfo{author}{Fegley, B.~Jr.}
\newblock \emph{\bibinfo{title}{Practical Chemical Thermodynamics for
  Geoscientists}}, chap. \bibinfo{chapter}{Chapter 10 Chemical Equilibria},
  \bibinfo{pages}{482} (\bibinfo{publisher}{Academic Press (Elsevier)},
  \bibinfo{year}{2013}).

\bibitem{CourtSephton2009_2}
\bibinfo{author}{Court, R.~W.} \& \bibinfo{author}{Sephton, M.~A.}
\newblock \bibinfo{title}{Investigating the contribution of methane produced by
  ablating micrometeorites to the atmosphere of mars}.
\newblock \emph{\bibinfo{journal}{Earth and Planetary Science Letters}}
  \textbf{\bibinfo{volume}{288}}, \bibinfo{pages}{382--385}
  (\bibinfo{year}{2009}).

\bibitem{Huss1996}
\bibinfo{author}{Huss, G.~R.}, \bibinfo{author}{Lewis, R.~S.} \&
  \bibinfo{author}{Hemkin, S.}
\newblock \bibinfo{title}{The "normal planetary" noble gas component in
  primitive chondrites: Compositions, carrier and metamorphic history}.
\newblock \emph{\bibinfo{journal}{Geochimica et Cosmochimica Acta}}
  \textbf{\bibinfo{volume}{60}}, \bibinfo{pages}{3311--3340}
  (\bibinfo{year}{1996}).

\bibitem{AbeMatsui1985}
\bibinfo{author}{Abe, Y.} \& \bibinfo{author}{Matsui, T.}
\newblock \bibinfo{title}{The formation of an impact-generated h2o atmosphere
  and its implications for the early thermal history of the earth}.
\newblock \emph{\bibinfo{journal}{Journal of Geophysical Research}}
  \textbf{\bibinfo{volume}{90}}, \bibinfo{pages}{C545--C559}
  (\bibinfo{year}{1985}).

\bibitem{Hasimoto2007}
\bibinfo{author}{Hasimoto, G.~L.}, \bibinfo{author}{Abe, Y.} \&
  \bibinfo{author}{Sugita, S.}
\newblock \bibinfo{title}{The chemical composition of the early terrestrial
  atmosphere: Formation of a reducing atmosphere from ci-like material}.
\newblock \emph{\bibinfo{journal}{Journal of Geophysical Research}}
  \textbf{\bibinfo{volume}{112}}, \bibinfo{pages}{12} (\bibinfo{year}{2007}).
  
\bibitem{SchaeferFegley2007}
\bibinfo{author}{Schaefer, L.} \& \bibinfo{author}{Fegley, B.~Jr.}
\newblock \bibinfo{title}{Outgassing of ordinary chondritic material and some
  of its implications for the chemistry of asteroids, planets, and satellites}.
\newblock \emph{\bibinfo{journal}{Icarus}} \textbf{\bibinfo{volume}{186}},
  \bibinfo{pages}{462--483} (\bibinfo{year}{2007}).

\bibitem{Lupu2014}
\bibinfo{author}{Lupu, R.~E.} \emph{et~al.}
\newblock \bibinfo{title}{The atmospheres of earthlike planets after giant
  impact events}.
\newblock \emph{\bibinfo{journal}{The Astrophysical Journal}}
  \textbf{\bibinfo{volume}{784}}, \bibinfo{pages}{19} (\bibinfo{year}{2014}).

\bibitem{MbarekKempton2016}
\bibinfo{author}{Mbarek, R.} \& \bibinfo{author}{Kempton, E. M.-R.}
\newblock \bibinfo{title}{Clouds in super-earth atmospheres: Chemical
  equilibrium calculations}.
\newblock \emph{\bibinfo{journal}{The Astrophysical Journal}}
  \textbf{\bibinfo{volume}{827}}, \bibinfo{pages}{10} (\bibinfo{year}{2016}).

\bibitem{Dorn2018}
\bibinfo{author}{Dorn, C.}, \bibinfo{author}{Noack, L.} \&
  \bibinfo{author}{Rozel, A.~B.}
\newblock \bibinfo{title}{Outgassing on stagnant-lid super-earths}.
\newblock \emph{\bibinfo{journal}{Astronomy \& Astrophysics}}
  \textbf{\bibinfo{volume}{614}}, \bibinfo{pages}{20} (\bibinfo{year}{2018}).

\bibitem{Muenow1995}
\bibinfo{author}{Muenow, D.~W.}, \bibinfo{author}{Keil, K.} \&
  \bibinfo{author}{McCoy, T.~J.}
\newblock \bibinfo{title}{Volatiles in unequilibrated ordinary chondrites:
  abundances, sources and implications for explosive volcanism on
  differentiated asteroids}.
\newblock \emph{\bibinfo{journal}{Meteoritics}} \textbf{\bibinfo{volume}{30}},
  \bibinfo{pages}{639--645} (\bibinfo{year}{1995}).

\bibitem{Gerasimov1998}
\bibinfo{author}{Gerasimov, M.~V.}, \bibinfo{author}{Ivanov, B.~A.},
  \bibinfo{author}{Yakovlev, O.~I.} \& \bibinfo{author}{Dikov, Y.~P.}
\newblock \bibinfo{title}{Physics and chemistry of impacts}.
\newblock \emph{\bibinfo{journal}{Earth, Moon and Planets}}
  \textbf{\bibinfo{volume}{80}} (\bibinfo{year}{1998}).

\bibitem{MKS}
\bibinfo{author}{MKS, Gas Analysis}
\newblock \bibinfo{title}{RGA Application Bulleton \#208 Spectra Reference}.
\newblock \emph{\bibinfo{journal}{Application Note}} \textbf{\bibinfo{volume}{03/02-2/11}}, \bibinfo{pages}{1--2} (\bibinfo{year}{2005}).

\bibitem{Fuchs1973}
\bibinfo{author}{Fuchs, L.~H.}, \bibinfo{author}{Olsen, E.} \&
  \bibinfo{author}{Jensen, K.~J.}
\newblock \bibinfo{title}{Mineralogy, mineral-chemistry and composition of the
  murchison (c2) meteorite}.
\newblock \emph{\bibinfo{journal}{Smithsonian Contributions to the Earth
  Sciences}} \bibinfo{pages}{1--39} (\bibinfo{year}{1973}).

\end{thebibliography}

\clearpage

\begin{thebibliography}{10}
\expandafter\ifx\csname url\endcsname\relax
  \def\url#1{\texttt{#1}}\fi
\expandafter\ifx\csname urlprefix\endcsname\relax\def\urlprefix{URL }\fi
\providecommand{\bibinfo}[2]{#2}
\providecommand{\eprint}[2][]{\url{#2}}


\bibitem{NIST2}
\bibinfo{author}{NIST, M. S. D.~C.} \& \bibinfo{author}{Wallace, W.~E.}
\newblock \emph{\bibinfo{title}{NIST Chemistry WebBook}}, chap.
  \bibinfo{chapter}{Mass Spectra}.
\newblock \bibinfo{number}{69} (\bibinfo{publisher}{NIST Standard Reference
  Database}, \bibinfo{year}{2018}).
  
\bibitem{MKS2}
\bibinfo{author}{MKS, Gas Analysis}
\newblock \bibinfo{title}{RGA Application Bulleton \#208 Spectra Reference}.
\newblock \emph{\bibinfo{journal}{Application Note}} \textbf{\bibinfo{volume}{03/02-2/11}}, \bibinfo{pages}{1--2} (\bibinfo{year}{2005}).

\bibitem{Alexander2012_2}
\bibinfo{author}{Alexander, C.~M.~O'D.} \emph{et~al.}
\newblock \bibinfo{title}{The Provenances of Asteroids, and Their Contributions to the Volatile Inventories of the Terrestrial Planets}.
\newblock \emph{\bibinfo{journal}{Science}}
  \textbf{\bibinfo{volume}{337}}, \bibinfo{pages}{721}
  (\bibinfo{year}{2012}).

\bibitem{Nittler2004_2}
\bibinfo{author}{Nittler, L.~R.} \emph{et~al.}
\newblock \bibinfo{title}{Bulk element compositions of meteorites: a guide for
  interpreting remote-sensing geochemical measurements of planets and
  asteroids}.
\newblock \emph{\bibinfo{journal}{Antarctic Meteorite Research}}
  \textbf{\bibinfo{volume}{17}}, \bibinfo{pages}{231--251}
  (\bibinfo{year}{2004}).

\bibitem{Fuchs1973_2}
\bibinfo{author}{Fuchs, L.~H.}, \bibinfo{author}{Olsen, E.} \&
  \bibinfo{author}{Jensen, K.~J.}
\newblock \bibinfo{title}{Mineralogy, mineral-chemistry and composition of the
  murchison (c2) meteorite}.
\newblock \emph{\bibinfo{journal}{Smithsonian Contributions to the Earth
  Sciences}} \bibinfo{pages}{1--39} (\bibinfo{year}{1973}).
  
\end{thebibliography}
\end{document}